\documentclass[12pt]{iopart}
\usepackage{iopams}  
\usepackage{setstack}  
\usepackage[pdftex]{graphicx}
\usepackage{bbm}
\eqnobysec

\begin{document}

\newcommand{\ket}[1]{\left|#1\right\rangle }
\newcommand{\bra}[1]{\left\langle #1\right|}
\newcommand{\braket}[2]{\left\langle#1\middle|#2\right\rangle}
\newcommand{\unity}{\mathbbm{1}}
\newcommand{\komma}{\:,}
\newcommand{\punkt}{\:.}
\newcommand{\rhoC}{\tilde{\rho}^{C}}
\newcommand{\rhoCp}{\tilde{\rho}'{}^{C}}
\newcommand{\atil}{\tilde{\alpha}}
\newcommand{\btil}{\tilde{\beta}}
\newcommand{\Op}{\hat{O}}
\newcommand{\fmat}[2]{{f ^{#1,#2}}}
\newcommand{\ftmat}[2]{{\tilde{f} ^{#1,#2}}}
\newcommand{\kF}{k_{\mathrm{F}}}
\newcommand{\Kbar}{\bar{K}}
\newcommand{\CD}{CD }
\newcommand{\CDe}{CD}
\newcommand{\CDf}{charge-density }
\newcommand{\CDfe}{charge-density}
\newcommand{\CDfu}{Charge-density }
\newcommand{\CDfue}{Charge-density}
\newcommand{\FL}{1P }
\newcommand{\FLe}{1P}
\newcommand{\FLf}{one-particle }
\newcommand{\FLfe}{one-particle}
\newcommand{\FLfu}{One-particle }
\newcommand{\FLfue}{One-particle}
\newcommand{\SC}{2P }
\newcommand{\SCe}{2P}
\newcommand{\SCf}{two-particle }
\newcommand{\SCfe}{two-particle}
\newcommand{\SCfu}{Two-particle }
\newcommand{\SCfue}{Two-particle}
\newcommand{\rms}{\textit{r.m.s.} net correlations }
\newcommand{\rmse}{\textit{r.m.s.} net correlations}

\title[Variational matrix product state approach]{Correlation density matrices for 1- dimensional quantum chains based on the density matrix renormalization group}

\author{W M\"under$^1$, A Weichselbaum$^1$, A Holzner$^1$, Jan von Delft$^1$ and C L Henley$^2$}
\address{$^1$Physics Department, Arnold Sommerfeld Center for Theoretical Physics and Center for NanoScience, Ludwig-Maximilians-Universit\"at, 80333 Munich, Germany}
\address{$^2$Laboratory of Atomic and Solid State Physics, Cornell University, Ithaca, New York, 14853-2501}
\ead{wolfgang.muender@physik.uni-muenchen.de}

\begin{abstract}
A useful concept for finding numerically the dominant correlations of a given 
ground state in an interacting quantum lattice system in an unbiased 
way is the correlation density matrix. For two disjoint, separated clusters, it 
is defined to be the density matrix of their union minus the direct 
product of their individual density matrices and contains all correlations 
between the two clusters. We show how to extract from the correlation 
density matrix a general overview of the correlations as well as detailed 
information on the operators carrying long-range correlations and the spatial 
dependence of their correlation functions. 
To determine the correlation density matrix, we calculate the ground state 
for a class of spinless extended Hubbard models using the density 
matrix renormalization group. This numerical method is based on matrix 
product states for which the correlation density matrix can be obtained 
straightforwardly. In an appendix, we give a detailed tutorial introduction to 
our variational matrix product state approach for ground state calculations 
for 1- dimensional quantum chain models. We show in detail how matrix product 
states overcome the problem of large Hilbert space dimensions in these 
models and describe all techniques which are needed for handling them in 
practice.
\end{abstract}

\pacs{02.70.-c, 05.10.Cc, 03.65.Fd, 01.30.Rr, 71.10.Pm, 71.10.Hf}

\submitto{\NJP}
\maketitle

\tableofcontents{} 

\section{Introduction \label{sec:Introduction}}
\markboth{Introduction}{Introduction}

In an interacting quantum lattice model the ground 
state may have several kinds of correlations, such 
as long-range order, power-law, or exponentially 
decaying correlations. In the numerical treatment 
of such a model it is not clear \textit{a priori} what 
kind of correlation will be dominant and what kind 
of operators corresponds to these correlations. 
Before calculating correlation functions, one typically 
chooses in advance which operators to consider, 
using prior knowledge and making initial assumptions. 
The need to make such choices introduces a certain 
bias into the investigation, which can be somewhat 
unsatisfying, especially when hidden or exotic 
correlations are present.

\subsection{The correlation density matrix}\label{sub:CDM}

The correlation density matrix (CDM) \cite{CheongPhysRevB79} 
has been proposed as an unbiased tool to discover the dominant 
kind of correlations between two separated clusters, given the 
density matrix for their union (obtained by tracing out the rest of the 
system). For two disjoint, separated clusters $A$ and $B$ the 
CDM is defined to be the density matrix of their union minus 
the direct product of their respective density matrices to get rid 
of trivial correlations, 
\begin{equation}
	\hat{\rho}^{C}	\equiv \hat{\rho}^{A \cup B}  - \hat{\rho}^{A} \otimes \hat{\rho}^{B}
	\komma \label{eq:CDM}
\end{equation}
which is completely unbiased except for the specification of 
the clusters. If the two clusters were not correlated at all, this 
would imply $\hat{\rho}^{AB} = \hat{\rho}^{A} \otimes \hat{\rho}^{B}$ 
and therefore $\hat{\rho}^{C} = 0$. The CDM encodes all possible 
correlations between the clusters $A$ and $B$, as can be seen 
from the fact that 
\begin{eqnarray}
	\tr \left( \hat{\rho}^{C} \Op^{A} \otimes \Op'^{B} \right	)	
			& = &	\tr \left( \hat{\rho}^{A \cup B} \left( \Op^{A} \otimes \Op'^{B} \right) \right)
						- \tr \left( \left( \hat{\rho}^{A} \Op^{A} \right) \otimes \left( \hat{\rho}^{B} \Op'^{B} \right) \right)
						\nonumber \\
			& = &	\langle \Op^{A} \Op'^{B} \rangle - \langle \Op^{A} \rangle \langle \Op'^{B} \rangle
						\equiv C_{\Op\Op'}
						\komma \label{CDMcorrs}
\end{eqnarray}
where $\Op^{A}$ and $\Op'^{B}$ are operators acting on clusters 
$A$ and $B$, respectively.

\subsection{Lessons from Luttinger liquid theory}\label{sub:LLtheory}

To extract useful information from the CDM, it will be helpful 
to develop some intuition for its general structure. To this end, 
let us recall some fundamental facts from one-dimensional 
critical fermion systems. They are described by 
Luttinger liquid theory, in which one of the key parameters 
is the Fermi wave vector $\kF$.  The asymptotic behavior 
of any kind of correlation or Green's function is typically an 
oscillation inside a power-law envelope, 
\begin{equation}
	C \left( r \right) \sim \cos \left( m \kF r +\phi \right) / r^{\gamma}
	\komma \label{eq:lutt-asymptotic}
\end{equation}
for some exponent $\gamma$, where $m$ is some integer. 
For the particular model to be used in this study, a nontrivial 
mapping is known to a free fermion chain \cite{CheongSFL}, 
a special case of Luttinger liquid.

Renormalization group theory \cite{WilsonRevModPhys47} 
quite generally implies the existence of {\it scaling operators} 
in any critical system such as a Luttinger liquid. They are 
eigenvectors of the renormalization transformation and 
consequently their correlations are purely of a form like 
\eref{eq:lutt-asymptotic} for all $r$, not just asymptotically. 
The scaling operators usually have complicated forms. The 
correlation of a simple operator (e.g. fermion density $n(x)$ 
at position $x$ along a chain) has overlap with various 
scaling operators, and correspondingly the correlation function 
of that simple operator is a linear combination of contributions 
like \eref{eq:lutt-asymptotic} from those scaling operators.

Our aim is to discover the leading scaling operators 
numerically. The leading scaling operator encodes all 
the local fluctuations that are correlated with faraway 
parts of the system. Intuitively, for a given cluster $A$, 
that operator does not depend significantly on the 
exact position of the (distant) cluster $B$. That is 
particularly obvious in a one dimensional system: any 
correlation at distances $r'>r$ must be propagated 
through some sort of correlation at $r$, so we expect 
the same operators from cluster $A$ to be involved in 
$\hat{\rho}^{C} \left( r \right)$, irrespective of the distance $r$.

This suggests an \textit{ansatz} for  leading contributions in the CDM:
\begin{equation}
	\hat{\rho}^{C} \left( r \right) = \sum_{s} \Op^{A,s} \otimes \Op^{B,s} c_{s} \frac{ e^{i k_{s} r} }{ r^{\gamma_{s}} }
	\punkt \label{eq:ansatz}
\end{equation}
Here $\Op^{A,s}$ and $\Op^{B,s}$ are 
a pair of (distance-independent) scaling operators acting 
on clusters $A$ and $B$, respectively, $k_{s}$ is the 
characteristic wave vector for oscillations in their correlation, 
and $\gamma_{s}$ is the corresponding scaling exponent. 
When $k_{s} \neq 0$, the operator pairs must themselves 
come in pairs, labelled, say, by $s$ and $s+1$, with 
$k_{s+1} = -k_{s}$, $c_{s+1} = c_{s}^{*}$, and 
$\gamma_{s+1} = \gamma_{s}$, so that $\hat{\rho}^{C}$ 
is hermitian. The scaling operators for each cluster form an 
orthonormal set. We expect that only a few terms in the 
sum in \eref{eq:ansatz} capture most of the weight. 
Correspondingly, it may be feasible to truncate the complete 
basis sets $\Op^{A,s}$ and $\Op^{B,s}$ to a smaller set 
of ``dominant operators", whose correlators carry the dominant 
correlations of the system. The ansatz \eref{eq:ansatz} will 
guide our steps in the ensuing analysis; at the end, we shall 
check how well it is satisfied by the actual CDMs calculated 
for the model studied in this paper (see \sref{sub:CDfmatrix}). 

Notice that although a particular correlation function 
may have nodes, see \eref{eq:lutt-asymptotic}, for 
a CDM of the form \eref{eq:ansatz} the norm, 
\begin{equation}
	\Vert \hat{\rho}^{C} \left( r \right) \Vert^{2} = \sum_{s} \frac{ \vert c_{s} \vert^{2} }{ r^{2 \gamma_{s}} }
	\komma \label{CDMnorm}
\end{equation}
is monotonically decaying with $r$. This expresses 
the fact that information can only be lost with increasing 
distance, never restored, in a one-dimensional system.

\subsection{Operator basis and f-matrix}\label{sub:Opsfmatrix}

In \cite{CheongPhysRevB79} the operators entering 
the dominant correlation were found by a kind of 
singular value decomposition (SVD), which was 
done independently for each separation. However, 
the operators obtained from the SVD will in general 
be different for different separations $r$. This does not 
correspond to the form \eref{eq:ansatz}, where the 
operators are \textit{distance-independent} and only 
the coefficients are $r$-dependent. Therefore, we shall 
explore in this paper a new scheme to decompose the 
CDMs for all separations in concert, so as to obtain a 
small set of scaling operators characterizing the 
dominant correlations at any (sufficiently large) 
separation. We decompose $\hat{\rho}^{C}$ in the form
\begin{equation}
	\hat{\rho}^{C} \left( r \right) =		\sum_{S_{i}} \left( \sum_{\mu \mu'}
												\fmat{\mu}{\mu'} \left( r \right)
												\Op^{A, \mu} \otimes \Op^{B, \mu'}
												\right)_{S_{i}}
												\komma \label{fmatrix}
\end{equation}
where the $S_{i}$ represent the symmetry-sectors of 
the discrete, Abelian symmetries of the Hamiltonian 
(see \sref{sub:Symm}). The subscript of the brackets 
indicates that the decomposition within the brackets is 
done for each symmetry-sector individually. This 
decomposition is possible for any two complete, 
$r$-independent operator sets $\Op^{A, \mu}$ and 
$\Op^{B, \mu'}$ acting on the part of the Hilbert space 
of clusters $A$ and $B$, respectively, which correspond 
to the symmetry sector $S_{i}$. The goal is to find two 
operator sets $\Op^{A, \mu}$ and $\Op^{B, \mu'}$ 
such that these operator sets may be truncated to a 
small number of operators each, while still bearing the 
dominant correlations of the system. The distance 
dependence of the CDM is then only contained in the 
matrix $\fmat{\mu}{\mu'} \left( r \right)$. Then, all analysis 
concerning the distance-dependence of correlations can 
be done in terms of this f-matrix.

\subsection{Ground state calculation with DMRG}\label{sub:DMRG}

The CDM in \cite{CheongPhysRevB79} was calculated 
using the full ground state obtained from exact 
diagonalization. This limits the system size, so that the 
method was appropriate mainly in cases of rapidly 
decaying, or non-decaying correlations -- not for critical 
or slowly decaying ones. In the present work, we use 
the density matrix renormalization group (DMRG) 
\cite{WhitePhysRevLett69} (see the excellent review by 
U. Schollw\"ock \cite{SchollwockRevModPhys77}) to 
compute the ground state for a ladder system which is 
known to have algebraic correlations \cite{CheongSFL}. 
We use the matrix product state (MPS) formulation of 
DMRG \cite{VerstraetePhysRevLett93} in which an 
efficient variational procedure is used to obtain the 
ground state.

\subsection{Structure of the paper}\label{sub:structure}

The structure of the main body of the paper is as follows: 
in \sref{sec:Model} we introduce the model to be 
considered for explicit calculations. In \sref{sec:Calculation} 
we show how the CDM is defined, how to calculate it, 
and explain how a first overview of the relative strengths 
of various types of correlations can be obtained. In 
\sref{sec:OperatorBasis} we show how to analyze the 
CDM and its distance dependence. Sections 
\ref{sec:Numerics} to \ref{sec:Comparison} present our 
numerical results, and \sref{sec:Conclusions} our conclusions. 
In an extended appendix we offer a tutorial introduction to 
the MPS formulation of DMRG, and also explain how it can 
be used to efficiently calculate the CDM.

\section{Model}\label{sec:Model}
\markboth{Model}{Model}

To be concrete in the following analysis of the 
CDM, we begin by introducing the model for 
which we did our numerical calculations. This 
model contains rich physics and its treatment 
below can readily be generalized to other models.

\subsection{Definition of the model}\label{sub:ModelDef}

We analyze the CDM for a class of spinless extended 
Hubbard models for fermions, which was intensely studied 
by Cheong and Henley \cite{CheongSFL}. They computed correlation 
functions up to separations of about $r = 20$, using nontrivial 
mappings to free fermions and hardcore bosons. The correlation 
functions are calculated with an intervening-particle expansion 
\cite{CheongSFL}, which expresses the correlation functions in 
terms of one-dimensional Fermi-sea expectation values (an evaluation 
of the CDM for that model has also been done by Cheong and Henley 
\cite{CheongPhysRevB79}, using exact diagonalization, but the 
system sizes are too short to be conclusive).
For spinless fermions on a two-leg ladder with length $N$, we use the 
following Hamiltonian:
\begin{eqnarray}
	H	& = &	- t_{\parallel} \sum_{a=1}^{2} \sum_{x=1}^{N-1} (\hat{c}_{a,x}^{\dagger} \hat{c}_{a,x+1} + \mathrm{h.c.})
					- t_{\perp} \sum_{x=1}^{N} (\hat{c}_{1,x}^{\dagger} \hat{c}_{2,x} + \mathrm{h.c.})
					\nonumber \\
		&  &		- t_{c} \sum_{x=2}^{N-1} (\hat{c}_{1,x-1}^{\dagger} \hat{n}_{2,x} \hat{c}_{1,x+1}
					+ \hat{c}_{2,x-1}^{\dagger} \hat{n}_{1,x} \hat{c}_{2,x+1} + \mathrm{h.c.})
					\nonumber \\
		&  &		+ V \sum_{a=1}^{2} \sum_{x=1}^{N-1} \hat{n}_{a,x} \hat{n}_{a,x+1} + V \sum_{x=1}^{N} \hat{n}_{1,x} \hat{n}_{2,x}
					\komma \label{eq:Hamiltonian}
\end{eqnarray}
where $\hat{c}_{a,x}$ destroys a spinless fermion on leg 
$a$ and rung $x$, and $\hat{n}_{a,x} = \hat{c}_{a,x}^{\dagger} \hat{c}_{a,x}$ 
is the corresponding number operator. Effectively, the model 
corresponds to a one-dimensional pseudo-spin chain, where 
the $a=1$ leg is denoted by spin $\uparrow$ and the $a=2$ 
leg by spin $\downarrow$. Hence, in the following sections 
which generally apply to quantum chain models we will treat 
this model as a quantum chain consisting of $N$ sites and 
return to view the system as a ladder model in the sections 
where we discuss our results.

We will focus on infinite nearest-neighbour repulsion 
$V \rightarrow \infty$, which we treat differently along the 
legs and the rungs in our numerical calculations. In the 
pseudo-spin description we can enforce the nearest-neighbour 
exclusion along rungs by removing double 
occupancy from the local Hilbert space of the pseudo-spin 
sites. The nearest-neighbour exclusion along the legs cannot 
be implemented so easily and we mimic $V \rightarrow \infty$ 
by a value of $V$ which is much larger than all the other 
energies in the Hamiltonian (typically $V/t_{\parallel} = 10^{4}$). 

\begin{figure}
	\begin{centering}
	\includegraphics{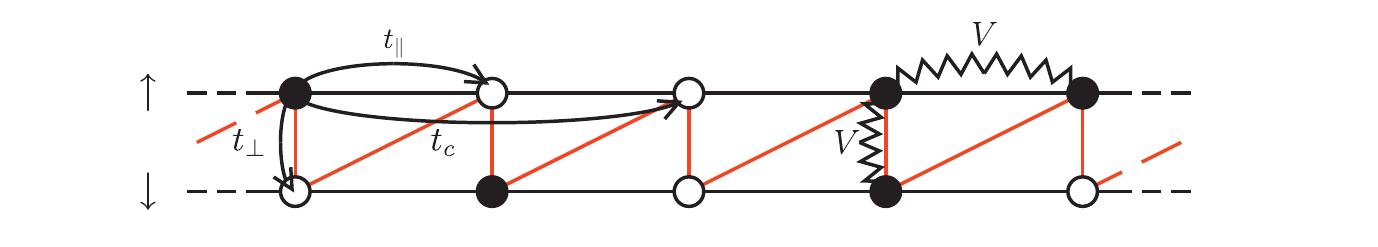}
	\par
	\end{centering}
	\caption{Ladder model with the terms 
		of the Hamiltonian in \eref{eq:Hamiltonian}. 
		Fermions are depicted 	by black circles and 
		empty lattice positions by white circles. The 
		ordering used for our Jordan-Wigner transformation 
		of fermionic creation and annihilation 
		operators is depicted by the red line.}
	\label{fig:Order}
\end{figure}

For fermionic systems, the fermionic sign due to the 
anti-commutation relations of the fermionic creation- and 
annihilation-operators needs to be taken into account. 
Specifically, we have to choose an order in which 
we pick the Fock basis, where we have to keep 
in mind that this choice produces a so called 
Jordan-Wigner-string of the form 
$\sum_{x'' = x+1}^{x'-1} e^{i \pi \hat{n}_{x''}}$ 
when evaluating correlators $\langle \hat{c}_{x} \hat{c}_{x'}^{\dagger} \rangle$ 
at distance $r = \vert x - x' \vert$. In the present system 
it is convenient to choose this order such that the operators of 
the two sites of a rung are succeeding each other (see \fref{fig:Order}), 
as this choice yields the shortest Jordan-Wigner strings.

\subsection{Expectations for simple limiting cases}\label{sub:ModelLimits}

Setting $t_{\parallel} \equiv 1$ as a reference 
scale, we are left with two parameters in the Hamiltonian: 
the rung hopping $t_{\perp}$ and the correlated hopping 
$t_{c}$. The physics of the system is governed by the 
competition of $t_{\perp}$ to localize the fermions on the 
rungs and $t_{c}$ to pair the fermions. There are three 
limiting cases which have been studied in detail by Cheong 
and Henley \cite{CheongPhysRevB79,CheongSFL}.

\begin{enumerate}
\item[(i)]
The paired limit, $t_{c} \gg t_{\parallel},t_{\perp}$ (we used 
$t_{c}/t_{\parallel} = 10^{2}$ and $t_{\perp} = 0$ for our 
calculations). In this limit the fermions form tight pairs which 
behave similar to hardcore bosons \cite{CheongSFL}. For two 
given rungs $x$ and $x+1$, there are two possibilities to create 
a pair of fermions, due to infinite nearest-neighbour repulsion: 
$\hat{c}_{\uparrow x}^{\dagger} \hat{c}_{\downarrow x+1}^{\dagger}$ 
and $\hat{c}_{\downarrow x}^{\dagger} \hat{c}_{\uparrow x+1}^{\dagger}$. 
It has been shown in \cite{CheongSFL} that, based on these two 
bound pairs, one may classify the bound pairs in two flavours 
along the ladder and that the ground state has only one definite 
flavour, causing a twofold symmetry breaking in the ground 
state. This symmetry breaking introduces complications that will 
be addressed below. The dominant correlations are expected to 
be \CDf correlations at short distances and \SCf at long distances. 
These \CDf and \SCf correlations decay as power laws, oscillating with $k = 2\kF$, 
where the Fermi wavelength $\kF$ is related to the filling as 
$\kF = 2\nu$ \cite{CheongSFL}. In this system, the \FLf 
correlations are suppressed and are expected to decay exponentially, 
as a nonzero expectation value depends on a local fluctuation 
completely filling the rungs between the clusters (as elaborated 
in \sref{sub:FL}).

\item[(ii)]
The two-leg limit, $t_{\perp} \ll t_{\parallel}$, $t_{c}=0$. 
In this limit the two legs are decoupled with respect to 
hopping, but still the infinite nearest-neighbour repulsion 
introduces correlations between the two legs. At large 
distances, power-law \CDf correlations dominate, while 
\SCf correlations show much faster power-law decay 
and \FLf correlations decay exponentially.

\item[(iii)]
The rung-fermion limit, $t_{\perp} \gg t_{\parallel}$, $t_{c}=0$. 
In this limit the particles are delocalized along the rungs. 
For fillings smaller than quarter-filling, \CDf, \FLf and 
\SCf correlations all decay as power laws where 
\CDf correlations dominate at large distances.
\end{enumerate}

Our analysis in this paper is limited to the case (i), 
where DMRG also showed best performance.

\subsection{Smooth boundary conditions}\label{sub:SBC}

For a ladder of length $N$ (treated as a pseudo-spin 
chain), we have attempted to reduce effects from the 
boundaries by implementing \textit{smooth boundary conditions}, 
adopting a strategy proposed in \cite{VekicPhysRevLett71} 
for a spin chain to our present fermionic system. 
(Alternatively, it is possible to use periodic boundary 
conditions \cite{VerstraetePhysRevLett93}. However, 
this leads to some difficulties, since it is not possible 
to work with orthonormal basis sets describing the 
left or right part of the chain with respect to a given 
site.) Smooth boundary conditions are open boundary 
conditions together with an artificial decay of all terms 
of the Hamiltonian over the last $M$ rungs at each 
end of the chain. We shall calculate expectation 
values only of operators located in the central part of 
the system (sites $x$, with $M < x \leq N-M$), thus 
the system's effective length is $N' = N - 2M$.

For both smooth and open boundary conditions 
the average site filling strongly decreases near the 
boundaries. To determine the average filling $\nu$, 
which influences the system's correlations in an 
important manner, we thus use only the central 
$N'$ sites: 
\begin{equation}
	\nu = \sum_{x = M+1}^{N-M} \left( \langle \hat{n}_{\uparrow x} \rangle + \langle \hat{n}_{\downarrow x} \rangle \right) / (2N')
	\punkt \label{eq:Filling}
\end{equation}
Due to the infinite nearest neighbour repulsion, 
this implies that $\nu \in [0,0.5]$.

\section{Calculation of the CDM}\label{sec:Calculation}
\markboth{Calculation}{Calculation}

Throughout the paper we will use the Frobenius 
inner product and norm for any matrices $M_{ij}$ 
and $M_{ij}'$ of matching dimension, 
\begin{eqnarray}
	\langle M , M' \rangle	& \equiv &	\sum_{i j} M_{i j}^{*} M_{i j}'
													= \tr \left( M^{\dagger} M' \right) 
													\label{eq:FroIP} \\
	\Vert M \Vert				& \equiv &	\langle M , M \rangle^{1/2}
													\punkt \label{eq:FroNorm}
\end{eqnarray}

\subsection{Definition of the CDM}\label{sub:DefinitionCDM}

We take two disjoint, separated clusters $A$ and $B$ of 
equal size from a one-dimensional quantum chain, i.e. two 
sets of adjacent sites ${ x_{1}^{A} , \dots , x_{n}^{A} }$ and 
${ x_{1}^{B} , \dots , x_{n}^{B} }$ where $n$ is the size of 
the clusters and all the indices $x$ are distinct from each other. 
The local Hilbert spaces of clusters $A$ and $B$ with dimension 
$d^{n}$ are described in terms of sets of basis states 
$\ket{\alpha}$ and $\ket{\beta}$, which are product states of the 
local states of each site in the cluster. The CDM of the two clusters, 
defined by \eref{eq:CDM}, can be expanded in this basis as
\begin{equation}
	\hat{\rho}^{C}	 = \rho^{C}_{\alpha \beta \alpha' \beta'} \ket{\alpha} \ket{\beta} \bra{\alpha'} \bra{\beta'}
	\punkt \label{eq:CDMdetail}
\end{equation}
For processing the CDM we fuse the two indices of each cluster \cite{CheongPhysRevB79}:
\begin{equation}
	\rhoC_{\atil \btil} \equiv \rhoC_{(\alpha \alpha') (\beta \beta')} \ket{\alpha} \bra{\alpha'} \ket{\beta} \bra{\beta'}
\end{equation}
with $\tilde{\alpha} = \left( \alpha \alpha' \right)$ and 
$\tilde{\beta} = \left( \beta \beta' \right)$, and denote the 
reshaped object $\rhoC$ itself by an extra tilde.
This corresponds to a partial transpose of the CDM (note 
that $\rhoC$ is no longer a symmetric tensor). For the 
CDM expressed in the indices $\tilde{\alpha}$ and 
$\tilde{\beta}$, we may use the Frobenius inner 
product \eref{eq:FroIP} and norm \eref{eq:FroNorm}.

To study the distance dependence of the correlations, we vary the 
position of the clusters $A$ and $B$, resulting in a position-dependent 
CDM $\rhoC \left( x^{A}_{1} , x^{B}_{1} \right)$. If the system is 
translationally invariant, this object depends only on the distance 
$r = \vert x_{1}^{A} - x_{1}^{B} \vert$ (the minimal distance for two 
adjacent clusters is equal to the cluster size $n$). For a finite system, 
though, $\rhoC$ will also depend on $\frac{1}{2} \left( x_{1}^{A} + x_{1}^{B} \right)$, 
at best weakly if the system is long. Strategies for minimizing the 
dependence on $\frac{1}{2} \left( x_{1}^{A} + x_{1}^{B} \right)$ by 
taking suitable averages will be discussed in \sref{sub:TranslationalInvariance}.

\subsection{DMRG-calculation of the CDM}\label{sub:CalculationDMRG}

The fact that the Hamiltonian in \eref{eq:Hamiltonian} 
is a one-dimensional pseudo-spin chain allows us 
to calculate ground state properties with the density 
matrix renormalization group (DMRG) \cite{WhitePhysRevLett69}. 
Using the variational matrix product state formulation 
of that method (see appendix for a detailed description), 
we calculated the ground state of the Hamiltonian in 
\eref{eq:Hamiltonian} for several values of $t_{\perp}$ 
and $t_{c}$. The framework of MPS also allows the CDM 
to be calculated efficiently (see \sref{sub:MPSReducedDensityMatrix} 
for details). Limiting ourselves to the case $t_{\perp} = 0$ 
in this paper, we have calculated the CDM derived from the 
ground state for distances up to $40$ rungs, which is 
significantly larger than in previous approaches.

\subsection{Symmetry sectors}\label{sub:Symm}

All the symmetries of the Hamiltonian are reflected in the 
CDM, making the CDM block-diagonal, where each block 
can be labeled uniquely by a set of quantum numbers that 
are conserved by the Hamiltonian. This means for Abelian 
symmetries (which are the only ones we are considering in 
practice), that the CDM in the original form 
$\rho^{C}_{\alpha \beta , \alpha' \beta'}$ fulfills 
$Q_{\alpha} + Q_{\beta} = Q_{\alpha'} + Q_{\beta'}$, 
where $Q_{\alpha}$ corresponds to the quantum numbers 
of state $\ket{\alpha}$, etc. The rearrangement of the CDM 
into $\rhoC_{\atil \btil}$ then implies 
$\Delta Q_{\atil} = - \Delta Q_{\btil}$ with 
$\Delta Q_{\atil} \equiv Q_{\alpha} - Q_{\alpha'}$ and 
$\Delta Q_{\btil} \equiv Q_{\beta} - Q_{\beta'}$. Since 
$\hat{\rho}^{AB}$ is hermitian, for every block of the CDM 
involving $\Delta Q_{\atil}$ ($\Delta Q_{\btil}$) there has to 
be a block involving $- \Delta Q_{\atil}$ ($- \Delta Q_{\btil}$), 
respectively. Therefore, it is convenient to sort the various parts 
of the CDM in terms of their change in quantum numbers 
$\Delta Q \equiv \vert \Delta Q_{\atil} \vert = \vert \Delta Q_{\btil} \vert$ 
and to analyze each symmetry sector individually.

To obtain a general classification of the CDM we sort the various 
contributions of the CDM according to the conserved quantum 
number(s) $Q$. In the case of the Hamiltonian in \eref{eq:Hamiltonian}, 
we consider particle conservation ($Q = \hat{N}_{\mathrm{tot}}$) 
which breaks the CDM into blocks with well-defined particle transfer 
$\Delta N \equiv \vert \Delta N_{\atil} \vert = \vert \Delta N_{\btil} \vert$ 
between clusters $A$ and $B$. The following \rms then is a 
measure for the correlations with transfer of $\Delta N$ particles 
between $A$ and $B$ (with $\Delta N =0,1,2$):
\begin{equation}
	w_{\Delta N}^{2} \left( r \right) = \sum_{\atil \btil \in {\cal S}_{\Delta N}} \left( \rhoC_{\atil \btil} \left( r \right) \right)^{2}
	\komma \label{eq:WeightN}
\end{equation}
where $\sum_{\Delta N=0}^{2} w_{\Delta N}^{2} \left( r \right) = \Vert \rho^{C} ( r ) \Vert^{2}$. 
Here the notation $\atil \equiv \left( \alpha \alpha' \right) \in {\cal S}_{\Delta N}$ 
indicates that only pairs of states $\left( \alpha \alpha' \right)$ 
are considered which differ by $\Delta N$ in particle number (similarly 
for $\btil \equiv \left( \beta \beta' \right) \in {\cal S}_{\Delta N}$). In the 
following we will call correlations involving $\Delta N = 0,1,2$ particles 
\textit{\CDfe} correlations (\CDe), \textit{\FLfe} correlations (\FLe), and 
\textit{\SCfe} correlations (\SCe), respectively. The following analysis is 
done for each symmetry sector individually. Depending on the decay of 
the \rms \eref{eq:WeightN}, some symmetry sectors may become 
irrelevant with increasing distance.

\subsection{``Restoration" of numerically broken symmetries}\label{sub:TranslationalInvariance}

Although we have tried to minimize the effect of boundaries, 
our numerical methods for calculating the ground state and 
CDM do not produce strictly translationally invariant results. 
(In contrast, analyses based on exact diagonalization start 
from a ground state wavefunction in which the symmetry (in 
a finite system) is restored, even if there is a symmetry 
breaking in the thermodynamic limit.) Therefore, we construct 
the CDM $\rhoC \left( r \right)$ for a given distance $r$ from 
an average over several CDMs $\rhoC \left( x,x' \right)$ with 
constant $r = \vert x - x' \vert$, where $x$ and $x'$ give the 
position of the first site of clusters $A$ and $B$, respectively. 

Moreover, if the exact ground state is degenerate under a 
discrete symmetry, we expect that DMRG breaks this 
symmetry unless it is implemented explicitly in the code. As 
mentioned in \sref{sub:ModelLimits} for the specific models 
of this paper we expect a discrete symmetry under interchange 
of legs for some parameter regimes. Since we did not implement 
this symmetry explicitly in our code, we also average the CDM 
by interchanging the legs of the ladder. Thus, all the data 
analysis presented in subsequent sections will be based on 
using the following ``symmetry-restored" form of the CDM, 
\begin{equation}
	\rhoC \left( r \right) = \frac{1}{\mathcal{N}} \sum_{xx' , \vert x - x' \vert = r} \left( \rhoC \left( x,x' \right) + \rhoCp \left( x,x' \right) \right)
	\komma \label{eq:CDMaverage}
\end{equation}
where $\rhoCp$ is obtained from $\rhoC$ by interchanging the 
legs of the ladder, and $\mathcal{N}$ is some normalization factor.

One might argue that it is not sufficient to average over the broken 
symmetry \textit{w.r.t.} leg-interchange on the level of the density 
matrix, but that instead the symmetry should be restored on the 
level of the ground state wave function. Specifically, for a ground 
state $\ket{\psi_{1}}$ (however it is calculated) which breaks this 
symmetry, we could restore the symmetry in the following way, 
\begin{equation}
	\ket{\psi^{+}} = \frac{1}{\sqrt{2}} \left( \ket{\psi_{1}} + \ket{\psi_{2}} \right)
	\komma \label{eq:PSIaverage}
\end{equation}
where $\ket{\psi_{2}} = \hat{S} \ket{\psi_{1}}$ and $\hat{S}$ 
describes the action of interchanging the legs. This would lead 
to a total density matrix 
\begin{equation}
	\ket{\psi^{+}} \bra{\psi^{+}} = \frac{1}{2}
	\left( \ket{\psi_{1}} \bra{\psi_{1}} + \ket{\psi_{2}} \bra{\psi_{2}}
	+ \ket{\psi_{1}} \bra{\psi_{2}} + \ket{\psi_{2}} \bra{\psi_{1}} \right)
	\punkt \label{eq:RHOaverage}
\end{equation}
Now, for two clusters $A$ and $B$, the first two terms on 
the \textit{r.h.s.} yield the CDM of \eref{eq:CDMaverage}, 
while the last two terms turn out to be negligible when 
traced out over all sites except for the two \textit{local} 
clusters $A$ and $B$. This follows from $\ket{\psi_{1}}$ 
and $\ket{\psi_{2}}$ being orthogonal, hence 
$\tr (\ket{\psi_{1}} \bra{\psi_{2}}) = \braket{\psi_{2}}{\psi_{1}} = 0$, 
implying that for a long chain with local clusters $A$ and 
$B$, the reduced density matrix 
$\hat{\rho}^{AB,12} \equiv \tr_{x \notin A,B} (\ket{\psi_{1}} \bra{\psi_{2}})$ 
will be very close to zero due to the orthogonality of the 
wave functions on the sites outside of clusters $A$ and 
$B$. Consequently, it is sufficient to retain only the first 
two terms of \eref{eq:RHOaverage}, i.e. to restore the 
broken symmetry on the level of the density matrices 
only, as done in \eref{eq:CDMaverage}.

\section{Finding a distance-independent operator basis}\label{sec:OperatorBasis}
\markboth{Operator basis}{Operator basis}

The goal of this section is to extract a (likely) small set of 
operators from the CDM, which will describe the dominant 
correlations in the system as a function of distance. We 
will assume in this section that the CDM does not include 
any broken symmetries as indicated in 
\sref{sub:TranslationalInvariance}.

\subsection{Need for operator bases for clusters $A$ and $B$}\label{sub:ClusterBasis}

As already mentioned, the CDM (obtained from \eref{eq:CDMaverage}) 
may be investigated by applying a singular value decomposition (SVD) 
for each distance individually \cite{CheongPhysRevB79}: 
\begin{equation}
	\rhoC_{\atil \btil} = \sum_{s} w^{s} O^{A,s}_{\atil} \otimes O^{B,s}_{\btil}
	\komma \label{CDMusv}
\end{equation}
or, in operator notation: 
\begin{equation}
	\hat{\rho}^{C} = \sum_{s} w^{s} \Op^{A,s} \otimes \Op^{B,s}
	\komma \label{CDMusvOperator}
\end{equation}
where $\Op^{A,s}$ and $\Op^{B,s}$ act on clusters $A$ 
and $B$, respectively. Here the singular values $w^{s}$ 
are strictly positive real numbers. By construction, $\Op^{A,s}$ 
and $\Op^{B,s}$ form orthonormal sets in their corresponding 
Hilbert spaces, i.e. $O^{A,s}_{\atil} = O^{A,s}_{\alpha \alpha'}$ 
and $O^{B,s}_{\btil} = O^{B,s}_{\beta \beta'}$ form a 
complete set in the operator space of clusters $A$ and $B$,
respectively, using the inner product as in \eref{eq:FroIP}. 
The set includes operators with $w_{s} = 0$, such as the 
identity operator, since these will be produced by the SVD. 
The SVD \eref{CDMusvOperator} yields for each specific 
distance $r$ a set of operators $\Op^{A,s} \left( r \right)$ 
and $\Op^{B,s} \left( r \right)$ acting on clusters $A$ and 
$B$, respectively. 

However, the dominant operators so obtained, i.e. the ones 
with large weight from the SVD of $\rhoC \left( r \right)$, are 
likely not the same as each other for different distances and 
hence not convenient for characterizing the ``dominant 
correlations" of the system. What is needed, evidently, is a 
strategy for reducing the numerous sets of operators 
$\Op^{A,s} \left( r \right)$ and $\Op^{B,s} \left( r \right)$ to 
two ``basis sets of operators" for clusters $A$ and $B$, 
respectively, say $\Op^{A,\mu}$ and $\Op^{B,\mu}$, which 
are $r$-independent and whose correlators yield the dominant 
correlations in the system in the spirit of \eref{eq:ansatz}. (For 
a translationally invariant system the two sets have to be 
equal for both clusters $A$ and $B$, but we will treat them 
independently in the analysis.) Following the ansatz 
\eref{eq:ansatz} from the Luttinger liquid theory, these operators 
ought to be distance-independent, carrying common correlation 
content for all distances. Thus we seek an expansion of 
$\rhoC \left( r \right)$ of the form \eref{fmatrix}, in which \textit{only} 
the coefficients, not the operators, are $r$-dependent.

\subsection{Construction of operator bases}\label{sub:OperatorBases}

We have explored a number of different strategies for extracting operators from 
the CDM which carry common information for all distances. We 
will discuss in detail only one of these, which is rather simple to 
formulate and reliably yields operator sets with the desired properties. 
(Several other strategies yielded equivalent results, but in a somewhat 
more cumbersome fashion.) 

The simplest possible strategy one may try is to average over all 
the CDMs at different distances and to singular-value decompose
the resulting crude ``average CDM". However, since the elements 
for the CDM are expected to be oscillating functions of $r$, such 
a crude average can cancel out important contributions of the 
CDM. Thus we need a procedure that avoids such possible 
cancellations. To this end, we construct the following operators, 
bilinear in the CDM:
\numparts
\begin{eqnarray}
	\hat{K}^{A} \left( r \right)	& \equiv &		\tr_{B} \left( \hat{\rho}^{C \dagger} \left( r \right)  \hat{\rho}^{C} \left( r \right) \right)
															/ \Vert \hat{\rho}^{C} \Vert^{2}
															\label{eq:bilinearA} \\
	\hat{K}^{B} \left( r \right)	& \equiv &		\tr_{A} \left( \hat{\rho}^{C} \left( r \right)  \hat{\rho}^{C \dagger} \left( r \right) \right)
															/ \Vert \hat{\rho}^{C} \Vert^{2}
															\komma \label{eq:bilinearB}
\end{eqnarray}
\endnumparts
with matrix elements
\numparts
\begin{eqnarray}
	K_{\tilde{\alpha} \tilde{\alpha}'}^{A} \left( r \right)	& = &	\sum_{\tilde{\beta}}
																					\rhoC_{\tilde{\alpha} \tilde{\beta}} \left( r \right)
																					\rhoC_{\tilde{\alpha}' \tilde{\beta}}{}^{*} \left( r \right)
																					/ \Vert \tilde{\rho}^{C} \left( r \right) \Vert^{2}
																					\label{eq:KA} \\
	K_{\tilde{\beta} \tilde{\beta}'}^{B} \left( r \right)		& = &	\sum_{\tilde{\alpha}}
																					\rhoC_{\tilde{\alpha} \tilde{\beta}} \left( r \right)
																					\rhoC_{\tilde{\alpha} \tilde{\beta}'}{}^{*} \left( r \right)
																					/ \Vert \tilde{\rho}^{C} \left( r \right) \Vert^{2}
																					\punkt \label{eq:KB}
\end{eqnarray}
\endnumparts
We normalize by $\Vert \tilde{\rho}^{C} \left( r \right) \Vert^{2}$ 
in order to treat the operator correlations of $\rhoC \left( r \right)$ 
for different distances on an equal footing. Note that the eigenvalue 
decomposition on the hermitian matrices $K^{A} \left( r \right)$ 
and $K^{B} \left( r \right)$ (in short K-matrices) yields the same 
operators $\Op^{A} \left( r \right)$ and $\Op^{B} \left( r \right)$ 
as the SVD of $\rhoC \left( r \right)$, with eigenvalues being 
equal to singular values squared, up to the additional 
normalization factor $\Vert \tilde{\rho}^{C} \left( r \right) \Vert^{2}$. 
(Reason: for a matrix of the form $M = usv^{\dagger}$ we have 
$MM^{\dagger} = us^{2}u^{\dagger}$ and $M^{\dagger}M = vs^{2}v^{\dagger}$.) 

The object $\hat{K}^{X}$ (for $X = A,B$) is positive-definite and 
according to ansatz \eref{eq:ansatz}, it is expected to have the form 
\begin{equation}
	\hat{K}^{X} \left( r \right) = \mathcal {N}_{K}^{-1} 
			\sum_{s} \frac{ \vert c_{s} \vert^{2} }{ r^{2 \gamma_{s}} } \Op^{X} \Op^{X \dagger}
			\punkt \label{eq:bilinear2}
\end{equation}
In particular, it no longer contains any oscillating parts 
(in contrast to \eref{eq:ansatz}), and hence is suitable 
for being averaged over $r$.

Summing up the $K^{X}$-matrices over a range $R$ of 
distances ($r \in R$, where $R$ will be specified below) 
gives a mean $\Kbar^{X}$-matrix for cluster $X$ ($ = A,B$), 
namely $\Kbar^{X,R} \equiv \sum_{r \in R} \hat{K}^{X} \left( r \right)$. 
We do not divide the latter expression by the number of 
terms in the sum (as would be required for a proper 
mean), as at this stage we are only interested in the 
operator eigendecomposition, 
\begin{equation}
	\Kbar^{X,R} = \sum_{\mu} w^{R,\mu} \left( \Op^{X,R,\mu} \otimes \Op^{X,R,\mu} {}^{\dagger} \right)
	\komma \label{eq:KbarEig}
\end{equation}
with the operators normalized such that $\Vert \Op^{X,R,\mu} \Vert = 1$. 
The operator set $\Op^{X,R,\mu}$ gives an orthonormal, 
$r$-independent basis for cluster $X$. In practice, however, 
many of the $w^{R,\mu}$ (which turn out to be the same 
for $X=A$ or $B$) will be very small. Thus, it will be sufficient 
to work with a truncated set of these operators having 
significant weight.

To explore the extent to which $\Kbar^{X}$ depends on 
the summation range, we shall study several such ranges: 
$R_{\rm{all}}$ includes all distances, $R_{\rm{short}}$ short 
distances (first third of distances analyzed), $R_{\rm{int}}$ 
intermediate distances (second third) and $R_{\rm{long}}$ 
long distances (last third). The resulting (truncated) sets of 
operators can be compared via their mutual overlap matrix 
$O_{\mu\mu'}^{RR'} = \tr ( \Op^{R,X,\mu} \Op^{R',X,\mu'} )$, 
or more simply, by the single number 
$O^{RR'} = \sum_{\mu\mu'} (O_{\mu\mu'}^{RR'})^{2}$, 
which may be interpreted as the dimension of the common 
subspace of the two operator sets. The value of $O^{RR'}$ 
ranges from $0$ to $dim({\Op^{R,X,\mu}})$. By comparing 
$O^{RR'}$ for the different distance ranges, additional clues 
can be obtained about how the relative weight of correlations 
evolves from short to long distances. (Such a comparison is 
carried out in \tref{tab:OpsCD} below.)

\subsection{Definition of f-Matrix}\label{sub:fmatrix}

Once a convenient basis of operators $\Op^{A, \mu}$ 
and $\Op^{B, \mu}$ has been found, the correlation density 
matrix can be expanded in terms of this basis as in \eref{fmatrix}, 
\begin{equation}
	\rhoC_{\atil \btil} \left( r \right) 
		= \sum_{\mu \mu'} \fmat{\mu}{\mu'} \left( r \right)
		O^{A, \mu}_{\atil} O^{B, \mu'}_{\btil}
		\komma \label{eq:fmatrix}
\end{equation}
with matrix elements 
\begin{equation}
	\fmat{\mu}{\mu'} \left( r \right)
		\equiv \sum_{\atil \btil} \rhoC_{\atil \btil} \left( r \right) 
		O^{A, \mu}_{\atil} O^{B, \mu'}_{\btil}
		\punkt \label{fmatrixentries}
\end{equation}
For complete operator 
spaces $\Op^{A, \mu}$ and $\Op^{B, \mu'}$, by definition, 
the set of amplitudes squared sum up to the norm of the CDM:
\begin{equation}
	\sum_{\mu \mu'} 
	\vert \fmat{\mu}{\mu'} \left( r \right) \vert^{2} 
	= \Vert \rhoC \left( r \right) \Vert^{2}
	\punkt \label{fmatrixsum}
\end{equation}
However, as alluded to above, we expect that the dominant 
correlators can be expressed in terms of a \emph{truncated} 
set of dominant operators. If the sum on the left hand side of 
\eref{fmatrixsum} is restricted to this truncated set, its deviation 
from the right hand side gives an estimate of how well $\rhoC$ 
is represented by the truncated set of operators. It will turn out 
that only a handful of dominant operators (typically 4 or 6) are 
needed, implying very significant simplifications in the analysis. 
Thus, the data analysis will be done in terms of the matrices 
$\fmat{\mu}{\mu'} \left( r \right)$ (in short ``f-matrix") for this 
truncated set of dominant operators. 

\subsection{Fourier-analysis and decay of f-matrix}\label{sub:ffmatrix}

According to the expectations expressed in \eref{eq:ansatz}, 
the elements of the f-matrix are expected to be products of 
oscillating and decaying functions of $r$. The corresponding 
dominant wave vectors can be identified via Fourier transform 
on each element of the f-matrix. For an oscillating function 
times a monotonically decaying envelope, the peaks of the 
Fourier spectrum of the oscillating function will be broadened 
by the presence of the envelope. To minimize this unwanted 
broadening, we introduce a rescaled f-matrix (denoted by a tilde), 
$\ftmat{\mu}{\mu'} \left( r \right) = u \left( r \right) \fmat{\mu}{\mu'} \left( r \right)$, 
where the positive weighting-function $u \left( r \right)$ is chosen 
such that all values of $\vert \ftmat{\mu}{\mu'} \left( r \right) \vert$ 
are of the  same order, and Fourier decompose the rescaled 
$\tilde{f}$-matrix as $\ftmat{\mu}{\mu'} \left( k \right) = \sum_{r} e^{-ikr} \ftmat{\mu}{\mu'} \left( r \right)$. 
Its norm $\Vert \tilde{f} \left( k \right) \Vert^2 = \sum_{\mu \mu'} \vert \ftmat{\mu}{\mu'} \left( r \right) \vert^2$, 
plotted as a function of $k$, will contain distinct peaks that 
indicate which wave vectors characterize the dominant 
correlations. Subsequently, the elements of the f-matrix, 
can be fitted to the forms 
\begin{equation}
	\fmat{\mu}{\mu'} \left( r \right) = \sum_{j} A_{\mu,\mu'}^{[j]} e^{i k_{j} r} f_{j} \left( r \right)
	\komma \label{eq:Fit}
\end{equation}
where $A_{\mu,\mu'}^{[j]}$ are complex amplitudes, $f_{j} (r)$ 
describes the decay with distance (e.g. $f_{j} (r) = r^{-\gamma_{j}}$ 
or $e^{- r / r_{j}}$ for power-law or exponential decay, respectively), 
and $k_{j}$ is a set of dominant wave vectors. The latter appear 
pairwise in combinations $\left( +k ; -k \right)$, since $\fmat{\mu}{\mu'} \in \mathbbm{R}$, 
which implies $A_{\mu,\mu'}^{[i]} = A_{\mu,\mu'}^{[j]*}$ for 
$k_{i} = - k_{j}$. The results of such a fit for each pair of dominant 
operators $\Op^{A,\mu}$ and $\Op^{B,\mu'}$, is the final outcome 
of our analysis, since it contains the information needed to check the 
applicability of ansatz \eref{eq:lutt-asymptotic}.

\section{Numerical results: general remarks}\label{sec:Numerics}
\markboth{Numerical results}{Numerical results}

In this section, we illustrate the analysis proposed 
above for the model introduced in \sref{sec:Model}. 
We will focus on the limiting case of large $t_{c}$, 
which we expect to have the most complex 
behavior among all three limiting cases introduced 
in \cite{CheongPhysRevB79} and \cite{CheongSFL}. 
After some preliminary analysis, we will discuss in 
\sref{sec:NumericsSymm} each of the three symmetry 
sectors (\CDe, \FLe, and \SCe) characterized by the 
operators' fermion number, and in \sref{sec:Comparison} 
compare our results to those found by \cite{CheongSFL} 
using a different method.

\subsection{Specification of the clusters $A$ and $B$}\label{sub:Ops}

\begin{figure}
	\begin{centering}
	\includegraphics{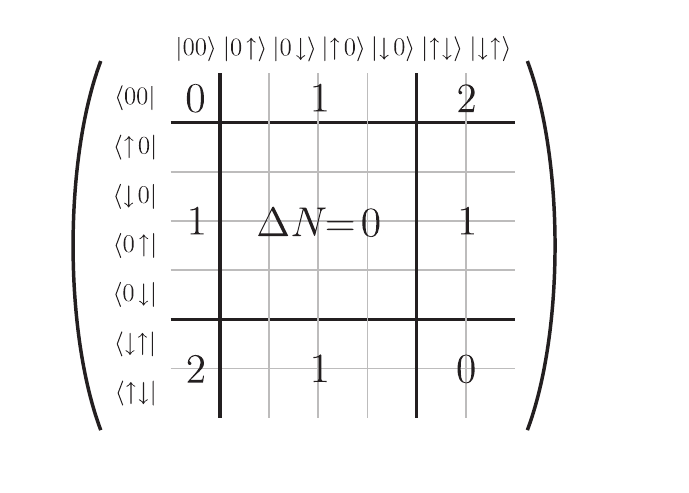}
	\par
	\end{centering}
	\caption{The symmetry sectors of 
		an operator acting on a cluster of 
		two rungs in the basis
		$\ket{0 0}$, $\ket{0 \uparrow}$, $\ket{0 \downarrow}$, 
		$\ket{\uparrow 0}$, $\ket{\downarrow 0}$, 
		$\ket{\uparrow \downarrow}$, $\ket{\downarrow \uparrow}$ 
		in pseudo-spin notation.
		}
	\label{fig:operator}
\end{figure}

For the following analysis it is convenient to take the 
size of the clusters $A$ and $B$ to be two rungs, 
because clusters of at least that size allow for up to 
two particles in one cluster (due to infinite nearest-neighbour 
repulsion). Thus, correlations involving $\Delta N = 0,1,2$ 
are possible, i.e \CDe, \FLe, and \SC correlations, 
respectively. Note that larger clusters can be studied, 
but would significantly increase numerical costs. 
Taking into account the infinite nearest-neighbour 
repulsion, clusters of size two have a seven-dimensional 
Hilbert space spanned by the kets 
$\ket{0 0}$, $\ket{0 \uparrow}$, $\ket{0 \downarrow}$, 
$\ket{\uparrow 0}$, $\ket{\downarrow 0}$, 
$\ket{\uparrow \downarrow}$, $\ket{\downarrow \uparrow}$, 
where the first (second) entry corresponds to the first 
(second) rung, $0$ represents an empty rung and 
$\uparrow$ and $\downarrow$ a fermion on the 
upper and lower leg in pseudo-spin notation (recall 
that we are dealing with spinless fermions). The space 
of operators acting on a cluster has dimension $7^2 = 49$, 
where the subspaces for $\Delta N = 0$, 1 or 2 have 
dimensions 21, 24 and 4, respectively, as depicted 
schematically in \fref{fig:operator}.

\subsection{Average site occupation}\label{sub:NumericsOcc}

As a first check of the influence of the boundaries, 
we investigate the average site occupation on the 
ladder. It is expected to be uniform in a translationally 
invariant system. However, there are two ways in 
which our calculation breaks translational symmetry, 
which cause residual oscillations in the density of 
particles along the ladders.

Firstly, there is the spontaneous breaking of the pair 
flavor symmetry described in \sref{sub:ModelLimits}. 
In the ground state produced by DMRG, all pairs 
have the same flavor, so only one of the two 
sublattices actually has any fermions on it. Thus a 
strong alternation in the density is observed between 
one leg for even rungs and the other leg for odd rungs; 
this can be taken care of by the symmetrization with 
respect to legs (as in \eref{eq:CDMaverage}).

\begin{figure}
	\begin{centering}
	\includegraphics{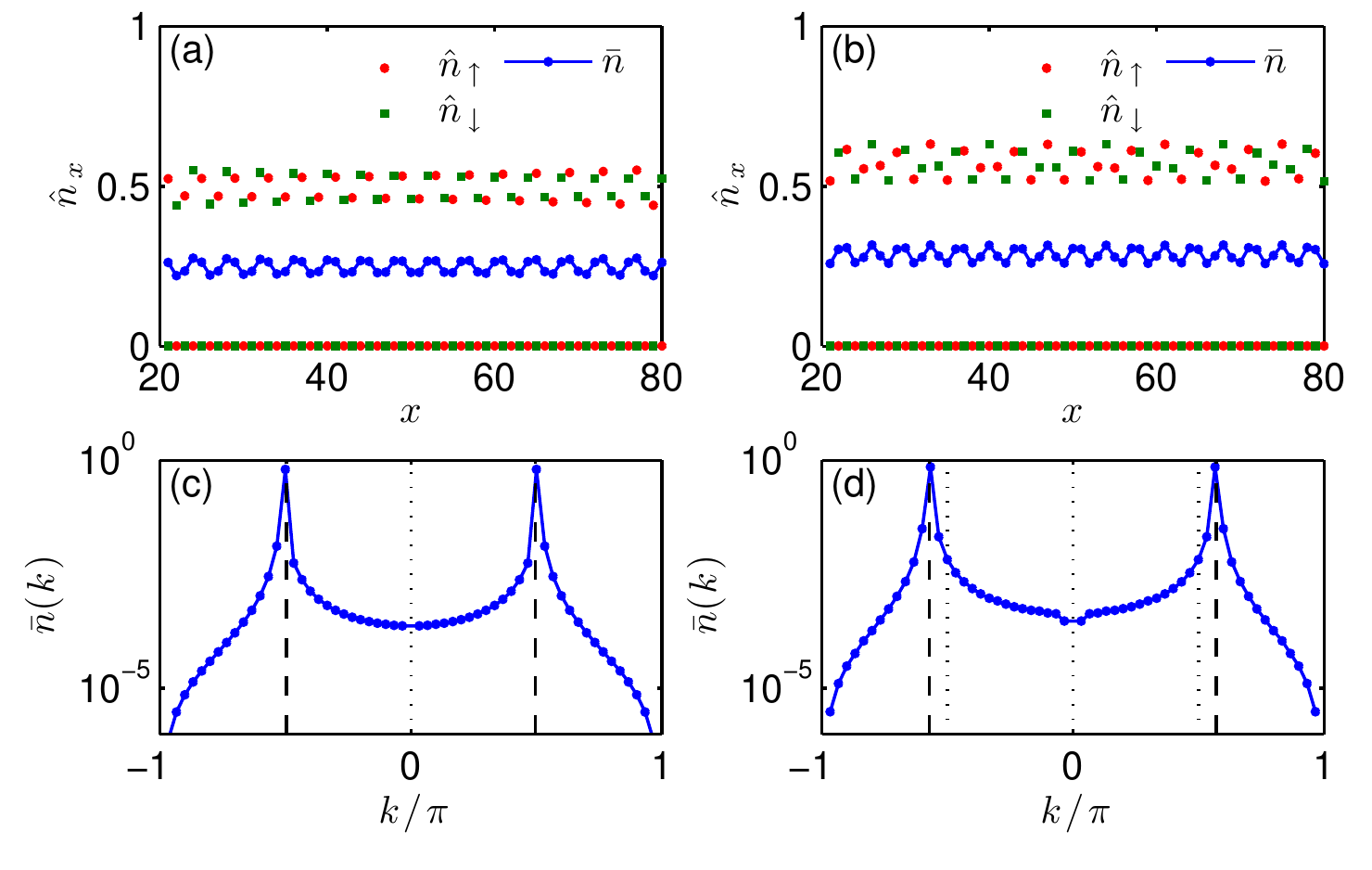}
	\par
	\end{centering}
	\caption{The average occupation along the legs of 
		the ladder for a filling of $\nu = 0.248$ (panels a,c) 
		and a filling of $\nu = 0.286$ (panels b,d). Panels 
		(a) and (b) show the average occupation 
		$\hat{n}_{\uparrow}$ on the upper (red) and 
		$\hat{n}_{\downarrow}$ on the lower (green) leg, 
		with every second value being zero. The end regions 
		$i = 1 \dots 20$ and $i = 81 \dots 100$ were skipped 
		in the figures and also in the analysis as these are 
		affected by the smooth open boundary condition. The 
		leg symmetrized occupation 
		$\bar{n} = \frac{1}{2} ( \hat{n}_{\uparrow} + \hat{n}_{\downarrow} )$ 
		(blue, same for upper and lower leg) eliminates this 
		strong even odd alternation but still shows small 
		modulations. This can be seen in detail in the Fourier 
		transform of the symmetrized occupation in panels (c) 
		and (d). There is a clear peak at $k = \pm 2\kF$ 
		(dashed vertical lines).
		}
	\label{fig:Occ}
\end{figure}

Secondly, translational symmetry is broken due to finite 
size in the DMRG calculation. This induces oscillations 
in the average occupation as a function of $x$ (see 
\fref{fig:Occ}), whose period is clearly dependent on the 
filling. In fact, their period is $2\kF$, so they may be 
interpreted as Friedel-like oscillations caused by the 
boundaries. Although the amplitude of density oscillation 
appears rather flat in the central portion of the system, 
it does have a minimum there; so we expect that the 
amplitude in the center of the system would vanish in 
a sufficiently large system.

Although the intent of the smooth boundary conditions is 
to minimize effects such as these oscillations, in fact, their 
amplitude appeared to be of about the same strength 
independent of whether we used smooth or plain open 
boundary conditions. We suspect, however, that the 
amplitude could be reduced by further careful optimization 
(not attempted here) of the parameters of the smooth 
boundary conditions.

\subsection{\rms $w_{\Delta N} \left( r \right)$}\label{sub:rmsnetcorrelations}

\begin{figure}
	\begin{centering}
	\includegraphics{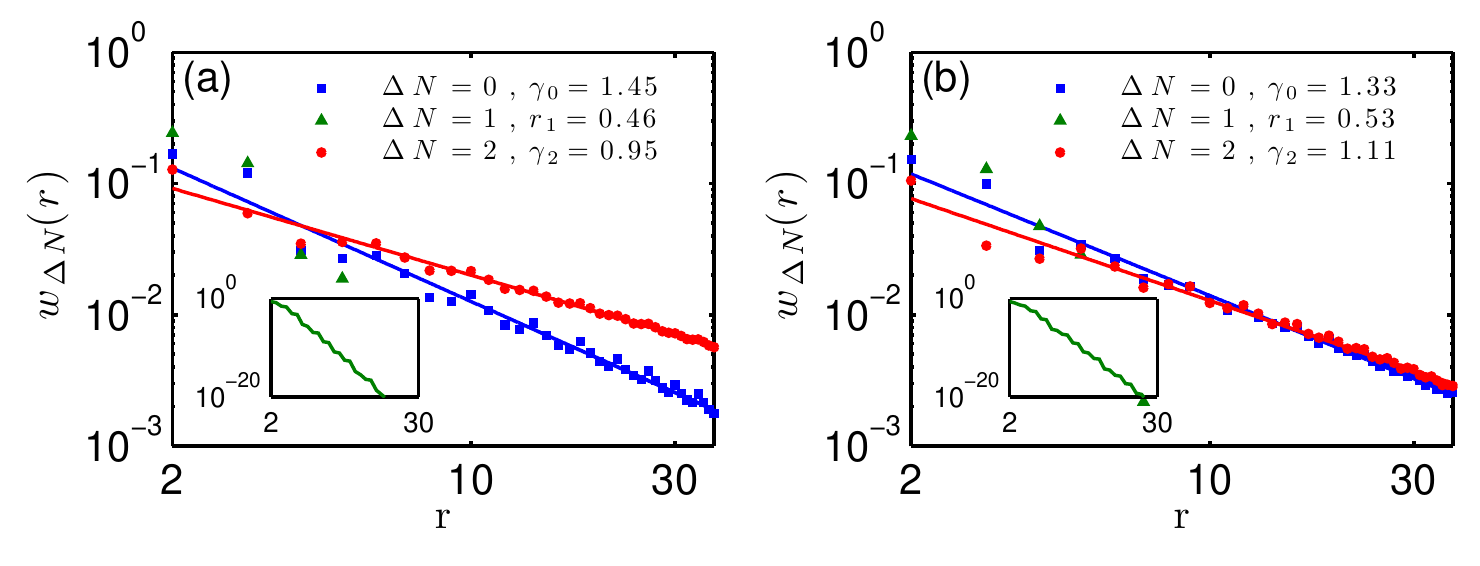}
	\par
	\end{centering}
	\caption{The \rms of \eref{eq:WeightN}, plotted
		as a function of distance for (a) a filling of $\nu = 0.248$ 
		and (b) a filling of $\nu = 0.286$. The symmetry 
		sectors are $\Delta N = 0$ (blue, no particle transfer, \CDe), 
		$\Delta N = 1$ (green, transfer of one particle, \FLe) 
		and $\Delta N = 2$ (red, transfer of two particles, \SCe). 
		We see that \CD and \SC correlations decay as 
		power-laws ($r^{-\gamma}$, blue and red solid lines)
		with small residual oscillations at $k=2\kF$, 
		while the \FL correlations show exponential 
		decay ($e^{-r / r_{1}}$, see semi-logarithmic plot in the inset). 
		The value $r_{1} \simeq 0.5$ for both fillings is reasonable 
		as we would expect a value of the order of one, which 
		is the size of the bound pairs.
		}
	\label{fig:rms}
\end{figure}

The next basic step is to identify the leading correlations 
in terms of the \rms $w_{\Delta N}$ defined in \eref{eq:WeightN}. 
These reveal which sectors of correlations dominate at 
large distances. The results (see \fref{fig:rms}) show 
that the \rms decay exponentially in the \FL sector, 
whereas they decay algebraically in both the \CD and 
\SC sectors, consistent with \cite{CheongSFL}. The 
latter two correlations are comparable in size over a 
significant range of distances, but for the fillings we 
investigated, \SC correlations ultimately dominate over 
CD correlations at the largest distances. Both the \CD 
and \SC \rms can be fitted to power laws, with the 
exponent dependent on the filling. The \rms in each 
sector are monotonic and only weakly modulated, even 
though the dominant correlation functions and the dominant 
parts of the CDM itself are oscillating (as will be discussed 
in more detail in \sref{sub:numericsCD}, see, e.g., 
\fref{fig:fitCD}). This implies that the correlations in each 
sector can be represented by a linear combination of 
correlation functions (associated with different operators) 
which oscillate out of phase, in such a way that in the 
sum of their squared moduli the oscillations more or less 
average out, resulting in an essentially monotonic decay 
with $r$, as expected according to \eref{CDMnorm}. 

We will next apply the analysis proposed in \sref{sub:OperatorBases} 
to the respective symmetry sectors (which will provide 
more exact fits of the exponents of the power-law decays).
The analysis in any sector consists of two stages. First, 
following \sref{sub:OperatorBases}, we try to find an optimal 
truncated basis which describes best the dominant correlations. 
Second, we examine the f-matrix of \sref{sub:fmatrix} 
(i.e. represent the CDM in the truncated basis) to see the nature 
of its $r$ dependence, and to fit this to an appropriate form, 
following \sref{sub:ffmatrix}.

\section{Numerical results: symmetry sectors}\label{sec:NumericsSymm}
\markboth{Numerical results}{Numerical results}

\subsection{\CDfu correlations}\label{sub:numericsCD}

\subsubsection{Operator basis}\label{sub:CDopbasis}

First we calculated the mean K-matrices $\bar{K}^{A,R}$ 
and $\bar{K}^{B,R}$ from $\bar{\rho}^{C}_{R}$ defined in 
\eref{eq:bilinearA} and \eref{eq:bilinearB}, and obtained 
operator sets from their eigenvalue decomposition, 
using various distance ranges.  

In order to decide how many operators to include in the 
truncated basis, we used the diagnostic described in 
\sref{sub:OperatorBases}. In presenting the results, we 
limit ourselves to cluster $A$ as the results for cluster $B$ 
are completely analogous. The operator set 
$\Op^{A,R_{\rm{all}},\mu}$ corresponding to the full range 
of distances $R_{\rm{all}}$ (specified in section 
\sref{sub:OperatorBases}) is used as a reference set to 
be compared with the operator sets obtained from 
$R_{\rm{short}}$, $R_{\rm{int}}$ and $R_{\rm{long}}$. The 
results are given in \tref{tab:OpsCD}. We see that, for 
intermediate or long distances, the effective dimension 
($O^{R_{\rm{all}}R_{\rm{int}}}$ and $O^{R_{\rm{all}}R_{\rm{long}}}$) 
of the common operator space shared between the operator 
set $\Op^{A,R_{\rm{all}},\mu}$ and the operator sets 
$\Op^{A,R_{\rm{int}},\mu}$ and $\Op^{A,R_{\rm{long}},\mu}$, 
respectively, saturates at six even if a larger operator 
space is allowed. Similarly, also  the short-distance 
operator set $\Op^{A,R_{\rm{short}},\mu}$ agrees best 
with the other three operator sets at dimension six: a 
further increase of the number of operators, however, adds 
only operators in the short range sector of the CDM. Hence 
we truncate to a six-dimensional operator basis. Within 
this reduced operator space, all dominant correlations are 
well-captured, as can be seen from the relative weights 
of \tref{tab:OpsCD}. For the resulting truncated basis set 
equation \eref{fmatrixsum} holds up to a relative deviation 
of the order of $\mathcal{O} \left( 10^{-5} \right)$.

\Table{\label{tab:OpsCD}
	Comparison of the operator sets on cluster $A$ 	for a filling 
	of $\nu = 0.286$. (The results for  $\nu = 0.248$ are similar, with only minor differences.)
	The first and second column of the table give the number
	of operators kept and the corresponding smallest singular value of  
	the set of operators $\Op^{A,R_{\rm{all}},\mu}$ obtained from the full range of 
	distances $R_{\rm{all}}$. The other three columns show $O^{R_{\rm{all}}R_{\rm{short}}}$, 
	$O^{R_{\rm{all}}R_{\rm{int}}}$ and $O^{R_{\rm{all}}R_{\rm{long}}}$	for the given number of operators.}
	\begin{tabular}{@{}lllll}
	\br
	number of   & $w^{R_{\rm{all}},\mu}/w^{R_{\rm{all}},1}$ & $O^{R_{\rm{all}}R_{\rm{short}}}$  & $O^{R_{\rm{all}}R_{\rm{int}}}$ & $O^{R_{\rm{all}}R_{\rm{long}}}$ \\
	operators & & (short) \,\,\, & (intermediate) & (long) \\
	\mr
	\01 & 1 & \01 & 0.99 & 1 \\
	\02 & 0.784122 & \01.99 & 2 & 2 \\
	\03 & 0.579242 & \02.99 & 3 & 3 \\
	\04 & 0.176043 & \03.99 & 4 & 4 \\
	\05 & 0.011250 & \05 & 5 & 4.99 \\
	\06 & 0.003040 & \06 & 6 & 5.99 \\
	\07 & 0.000004 & \07 & 6 & 6 \\
	\08 & 0.000001 & \08 & 6 & 6 \\
	\09 & 0.000001 & \09 & 6 & 6 \\
	10 & 0.000001 & 10 & 6 & 6 \\
	\br
	\end{tabular}
\endTable

Investigating the six-dimensional set of operators 
in more detail reveals that they can be classified 
with respect to their symmetry with respect to interchanging 
the legs of the ladder, i.e. they obey 
$\hat{S} \Op^{A,R_{\rm{all}},\mu} = \pm \Op^{A,R_{\rm{all}},\mu}$, 
with $\hat{S}$ describing the action of interchanging 
legs. The set breaks into two subsets 
of three operators each, which have positive or negative parity 
with respect to $\hat{S}$, respectively. It turns
out that all six operators are linear combinations 
of operators having matrix elements on the diagonal 
only, in the representation of \fref{fig:operator}. Moreover, together with the 
unit matrix they span the full space of diagonal operators 
(therefore the dimension of $6 = 7 - 1$). Explicitly, the 
symmetric operators are given by 
\numparts
\begin{eqnarray}
\fl	\Op^{1}		&=&	\textstyle{\frac{1}{\sqrt{12}}} \left(
							- \hat{n}_{0,x} \hat{n}_{\uparrow,x+1}
							- \hat{n}_{\uparrow,x} \hat{n}_{0,x+1} 
							+ 2 \hat{n}_{\uparrow,x} \hat{n}_{\downarrow,x+1}
							+ \rm{leg~symmetrized} \right) \\
\fl	\Op^{2}		&=&	\textstyle{\frac{1}{2}} \left(
							\hat{n}_{0,x} \hat{n}_{\uparrow,x+1}
							- \hat{n}_{\uparrow,x} \hat{n}_{0,x+1}
							+ \rm{leg~symmetrized} \right) \\
\fl	\Op^{3}		&=&	\textstyle{\frac{1}{\sqrt{42}}}  \left[
							- 6 \hat{n}_{0,x} \hat{n}_{0,x+1} + \left(
							\hat{n}_{0,x} \hat{n}_{\uparrow,x+1}
							+ \hat{n}_{\uparrow,x} \hat{n}_{0,x+1}
							+ \hat{n}_{\uparrow,x} \hat{n}_{\downarrow,x+1}
							+ \rm{leg~symmetrized} \right) \right]
\end{eqnarray}
\endnumparts
and the antisymmetric operators by 
\numparts
\begin{eqnarray}
\fl	\Op^{4}		&=&	\textstyle{\frac{1}{\sqrt{2}}}
							\hat{n}_{0,x} \left( \hat{n}_{\uparrow,x+1}
							- \hat{n}_{\downarrow,x+1} \right) \\
\fl	\Op^{5}		&=&	\textstyle{\frac{1}{\sqrt{2}}}
							\left( \hat{n}_{\uparrow,x} - \hat{n}_{\downarrow,x} \right)
							\hat{n}_{0,x+1} \\
\fl	\Op^{6}		&=&	\textstyle{\frac{1}{\sqrt{2}}} 
							\left( \hat{n}_{\uparrow,x} \hat{n}_{\downarrow,x+1}
							- \hat{n}_{\downarrow,x} \hat{n}_{\uparrow,x+1} \right)
\end{eqnarray}
\endnumparts
where $\hat{n}_{0} = \left( 1 - \hat{n}_{\uparrow} - \hat{n}_{\downarrow} \right)$.
We use this operator basis for both cluster $A$ and 
cluster $B$. If we calculate the f-matrix \eref{eq:fmatrix} based on these 
operators we see that it breaks into two blocks 
corresponding to their symmetry with respect to leg interchange.

\subsubsection{f-matrix elements: oscillations and decay}\label{sub:CDfmatrix}

\begin{figure}
	\begin{centering}
	\includegraphics{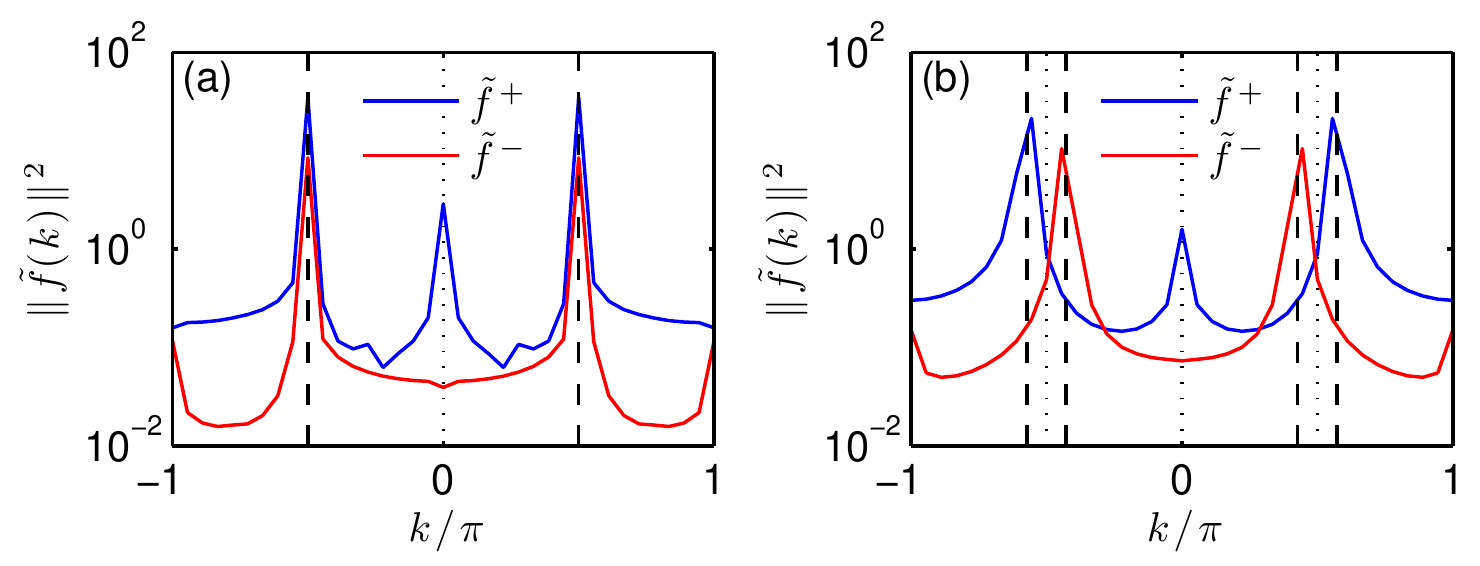}
	\par
	\end{centering}
	\caption{Fourier transform of the rescaled f-matrix $\tilde{f}$ 
		for \CD correlations based on operators chosen from a 
		reduced six-dimensional operator space, for a filling of 
		(a) $\nu = 0.248$ and (b) $\nu = 0.286$. We obtain 
		these Fourier spectra from the rescaled f-matrix 
		$\tilde{f}^{\mu , \mu'} \left( r \right) = r^{\gamma''} \fmat{\mu}{\mu'} \left( r \right)$, 
		with $\gamma''$ extracted from a power-law fit on 
		$\vert \fmat{\mu}{\mu'} \left( r \right) \vert$. The Fourier 
		spectrum breaks up into a contribution coming from the 
		operators symmetric or antisymmetric under leg-interchange, 
		labelled $\tilde{f}^{+}$ (blue) and $\tilde{f}^{-}$ (red), 
		respectively. The spectrum of $\tilde{f}^{+}$ shows strong 
		peaks at $k=\pm 2\kF$ (dashed lines) and a smaller peak 
		at $k=0$ with $\kF/\pi = \nu$. The spectrum of $\tilde{f}^{-}$, 
		having peaks at $k=\pm 2\kF + \pi$ (dashed lines) and 
		$k = \pi$, is shifted w.r.t. $\tilde{f}^{+}$ by $\pi$. For a filling 
		close to $\frac{1}{4}$ the dominant peaks of $\tilde{f}^{\pm}$, 
		at $k=\pm 2\kF$ and $k=\pm 2\kF + \pi$. are nearly at the same position.
		}
	\label{fig:ffmatrixCD}
\end{figure}

We now turn to extracting the distance-dependence of the dominant 
correlation in this symmetry sector, which is now visualizable since 
we drastically reduced the operator space to six dimensions. All 
relevant information is contained in the f-matrix and its Fourier transform. The 
first step is to identify the oscillation wave vector(s) $k$ to be used 
as initial guesses in the fit.  A general method is to plot the Fourier 
spectrum $\Vert \tilde{f} (k) \Vert$ of the rescaled f-matrix (\fref{fig:ffmatrixCD}). 
When using a logarithmic scale for the vertical axis, 
even sub-leading contributions show up clearly. We find that 
the spectra belonging to the symmetric and anti-symmetric 
operators are shifted against each other by $\pi$. This relative 
phase shift implies a trivial additional distance dependence of 
$e^{i \pi r}$ of $f^{-}(r)$ with respect to $f^{+}(r)$, reflecting 
the different parity under leg interchange of the two operator sets. 
We have found it convenient to undo this shift by redefining 
$f^{-} (r)$, the part of the f-matrix belonging to the anti-symmetric 
operators, to $e^{i \pi r} f^{-} (r)$. The resulting combined Fourier 
spectrum for $f^{+}$ and $e^{i \pi r}f^{-}$ has strong peaks at 
$k = 2\kF$ and a smaller peak at $k = 0$, in agreement with 
the result from \cite{CheongSFL}.

Based on the Fourier spectrum, we rewrite the fitting form 
\eref{eq:Fit} as 
\begin{equation}
	\fmat{\mu}{\mu'} \left( r \right) = A_{\mu\mu'} r^{-\gamma} \cos \left( kr + \phi_{\mu\mu'} \right) + B_{\mu\mu'} r^{-\gamma'}
	\komma \label{eq:FitCD}
\end{equation}
with real numbers $A_{\mu\mu'} > 0$ and $B_{\mu\mu'}$, 
where we expect $\gamma' > \gamma$, 
due to the relative sharpness of the peaks in the 
Fourier spectrum. The non-linear fitting over the full 
range of distances is done in several steps to also 
include the decaying part at long distances on an 
equal footing. First, the data is rescaled by 
$r^{+\gamma''}$, where we obtained $\gamma''$ 
from a simple power-law fit, in order to be able to 
fit the oscillations for all distances with comparable 
accuracy. Then we fit the rescaled data to 
\eref{eq:FitCD}, where initially we use the information 
from the Fourier spectrum in keeping $k$ fixed to 
$k = 2\kF$, but finally also release the constraint 
on $k$. This procedure showed best results, with 
relative error bounds up to $2\%$. The uncertainties 
are largest for the second term in \eref{eq:FitCD} 
as it acts mainly on short distances, having 
$\gamma' > \gamma$. 

\begin{figure}
	\begin{centering}
	\includegraphics{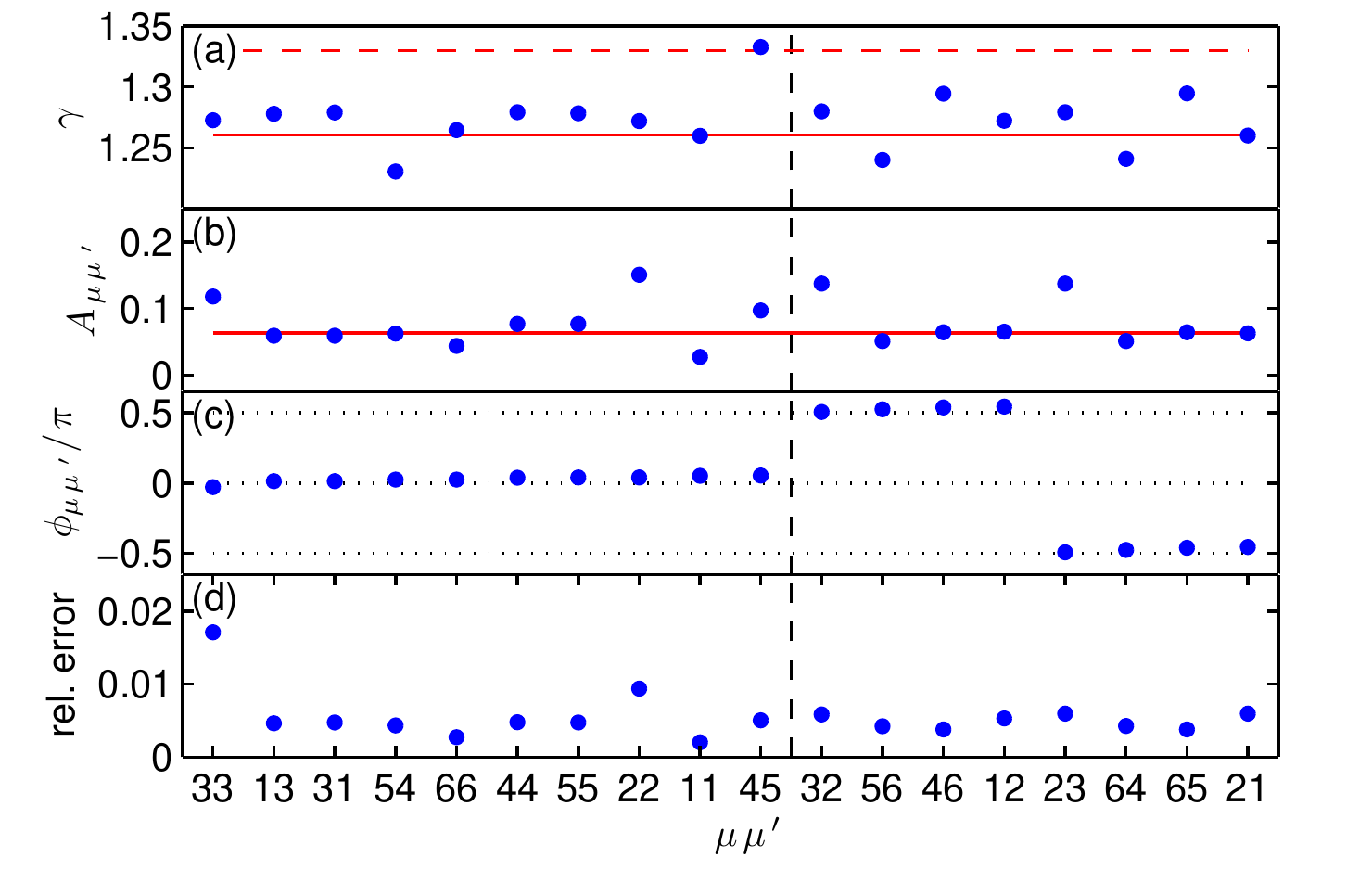}
	\par
	\end{centering}
	\caption{The results of the fit in \eref{eq:FitCD} to the 
		18 independent elements $f^{\mu,\mu'}$ of the f-matrix, 
		labeled along the horizontal axis by the index pair 
		$\mu \mu'$, for $\Delta N = 0$ at filling $\nu = 0.286$. 
		The results for $\nu = 0.248$ are qualitatively the same. 
		Panel (a) shows $\gamma$, panel (b) $A_{\mu\mu'}$, 
		panel (c) $\phi_{\mu\mu'}$ and panel (d) the error of the 
		fitting $\epsilon$, defined by	
		$\epsilon^2 = \sum_{r} ( f^{\mu\mu'} (r) - f^{\rm{fit}} (r) )^{2} / r^{-2\gamma''}$, 
		where $r^{-\gamma''}$ is the power-law we used to 
		rescale the data before Fourier-transforming. The red, 
		dashed line in the first panel shows the power-law 
		exponent obtained from the \rmse, $\gamma_{0} = 1.33$. 
		The phase $\phi_{\mu\mu'}$ is defined such that it is 
		in the interval $[-\pi,\pi]$. The matrix elements have 	been 
		grouped according to their relative phases $\phi_{\mu\mu'}$ 
		(separated by the black, dashed line), which clearly indicate 
		$\cos$ and $\sin$ behaviour for $\phi_{\mu\mu'}=0$ and 
		$\phi_{\mu\mu'}=\pm\frac{\pi}{2}$, respectively. The solid 
		red lines in panels (a) and (b) show the exponent 
		$\gamma_{0}$ and the amplitude $A$, respectively, from 
		the single fit \eref{eq:FitCDfinal}.}
	\label{fig:fittingCD}
\end{figure}

The results of this fitting procedure are depicted in \fref{fig:fittingCD}, 
for all 18 nonzero elements of the f-matrix. We see that the leading 
power-law exponents deviate from the fit to the \rms in \eref{eq:WeightN} 
(compare \fref{fig:rms}) by about $5\%$. The $k$-vectors from the 
non-linear fit are close to $k = 2\kF$ and deviate by less than $1\%$. 
The fit to the sub-leading second term in \eref{eq:FitCD} is not reliable, 
so we do not show the results for $\gamma'$ here, but note that every 
fit satisfied $\gamma' > \gamma$.

Since most of the exponents $\gamma$ and amplitudes 
$A_{\mu\mu'}$ are of comparable size, we fit the f-matrix 
elements to a \textit{single} $\gamma_{0}$ and $A$ (as 
well as a single $\gamma_{0}'$ and $B$ for the second 
term) for all the f-matrix elements, using the Ansatz:
\begin{equation}
\fl
f \left( r \right)
	= A r^{-\gamma_{0}}
	\left( {\scriptsize \arraycolsep0pt	
		\begin{array}{cc}
		\arraycolsep4pt
		\left( \begin{array}{ccc}
		\cos (kr) & \sin (kr) & \cos (kr) \\
		-\sin (kr) & \cos (kr) & -\sin (kr) \\
		\cos (kr) & \sin (kr) & \cos (kr) \\
		\end{array} \right) & 0 \\
		\arraycolsep4pt
		0 & e^{i \pi r} \left( \begin{array}{ccc}
		\cos (kr) & \cos (kr) & \sin (kr) \\
		\cos (kr) & \cos (kr) & \sin (kr) \\
		-\sin (kr) & -\sin (kr) & \cos (kr) \\
		\end{array} \right)
		\end{array}
	} \right)
	+ B r^{-\gamma_{0}'}
	\punkt \label{eq:FitCDfinal}
\end{equation}
The form of the matrices in the two blocks was obtained 
by inserting into \eref{eq:FitCD} the explicit values of the 
phases $\phi_{\mu\mu'}$ determined from the previous fit 
and summarized in \fref{fig:fittingCD}. Fitting to \eref{eq:FitCDfinal} 
gives an error of about $10\%$, with largest errors arising 
for the f-matrix elements where $A_{\mu\mu'}$ deviates 
strongly from $A$ (see \fref{fig:fittingCD}). For the filling 
$\nu = 0.286$ we find $\gamma_{0} = 1.26$ and $A = 0.06$. 
The values of $\gamma_{0}'$ and $B$ are unreliable in 
that the results from several fittings differ by about $30\%$, 
but still it holds that $\gamma_{0}' > \gamma_{0}$.

\begin{figure}
	\begin{centering}
	\includegraphics{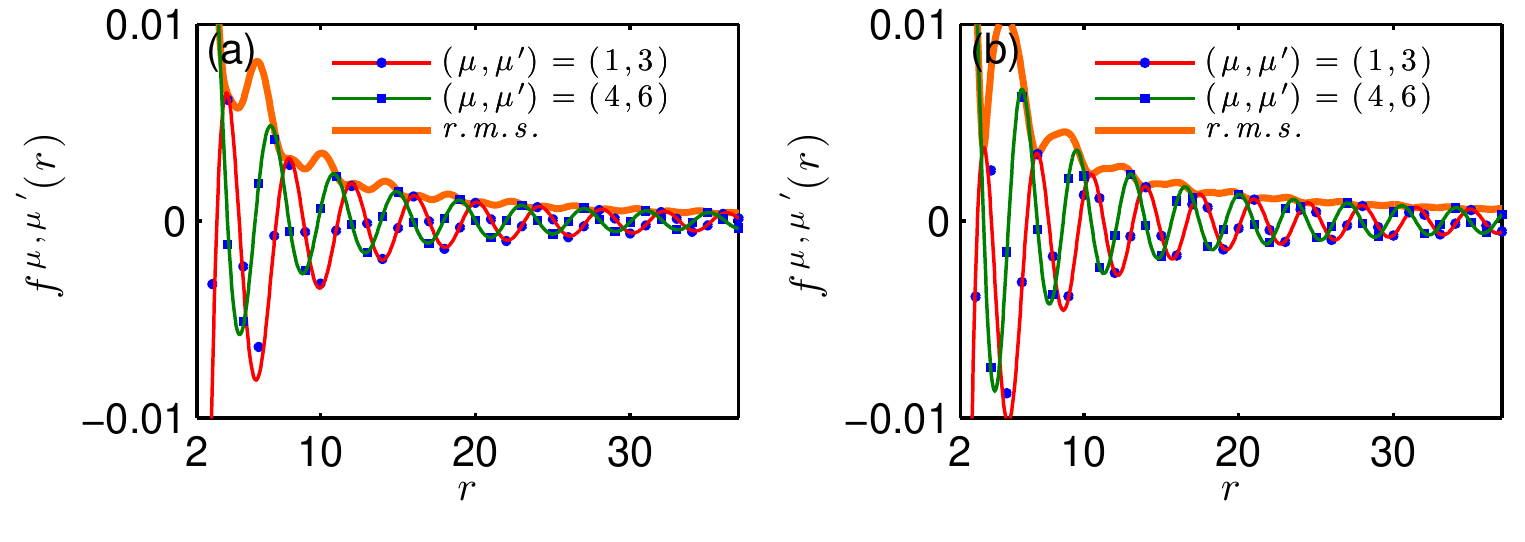}
	\par
	\end{centering}
	\caption{Two entries of the f-matrix for (a) $\nu=0.248$ 
		and (b) $\nu=0.286$ fitted to the form in \eref{eq:FitCD}. 
		The single points (blue circles and squares) are data points 
		from the f-matrix and the lines (red and green) are 
		the result of the fitting. They evidently oscillate with 
		a relative phase of $\Delta\phi = \pi/2$. As a result, 
		their contribution to the \rms, 
		$( \vert \fmat{1}{3} \vert^{2} + \vert \fmat{4}{6} \vert^{2} )^{\frac{1}{2}}$, 
		shown by the thick orange curve, has only small 
		oscillations at large distances.
		}
	\label{fig:fitCD}
\end{figure}

The form of \eref{eq:FitCDfinal} allows us to understand 
why the \rms displayed in \fref{fig:rms} show some 
residual oscillations, instead of decaying completely 
smoothly, as anticipated in \sref{sub:LLtheory}. The 
reason is that \eref{eq:FitCDfinal} contains 10 $\cos(kr)$ 
terms but only 8 $\sin (kr)$ terms. Although any two such 
terms oscillate out of phase, as illustrated in \fref{fig:fitCD}, 
the cancellation of oscillations will thus not be complete. 
Instead, the \rms contain a factor $[8 + 2 \cos^2(kr)]^{\frac{1}{2}}$ 
(compare to \eref{eq:WeightN}), which produces relative 
oscillations of about $10\%$, in accord with \fref{fig:rms}. 
(The fact that the total number of $\cos(kr)$ and $\sin(kr)$ 
terms is not equal is to be expected: the total operator 
Hilbert space per cluster is limited, and its symmetry 
subspaces might have dimensions not a multiple of 4.)

For each pair of wave vectors $\pm k$ in each parity 
sector, the effective operator basis per cluster can be 
reduced even further, from 3 operators to one conjugate 
pair of operators. This can be seen by rewriting 
\eref{eq:FitCDfinal} as follows: 
\begin{equation}
	f \left( r \right)
	=	A r^{-\gamma_{0}} 
		\left[ e^{i k r} \left( \begin{array}{cc} f_{+} & 0 \\ 0 & e^{i \pi r} f_{-} \\ \end{array} \right) + \rm{c.c.} \right]
		+ B r^{-\gamma_{0}'}
	\komma \label{eq:FitCDfinal2}
\end{equation}
with the matrices $f_{+}$ and $f_{-}$ defined as 
\begin{equation}
	f_{+} = \frac{1}{2}
		\left( \begin{array}{ccc}
		1 & -i & 1 \\
		i & 1 & i \\
		1 & -i & 1 \\
		\end{array} \right)
	\komma \:
	f_{-} = \frac{1}{2}
		\left( \begin{array}{ccc}
		1 & 1 & -i \\
		1 & 1 & -i \\
		i & i & 1 \\
		\end{array} \right)
	\punkt \label{eq:FitCDfinal2detail}
\end{equation}
Note that both $f_{+}$ and $f_{-}$ are matrices of rank 
one with eigenvalues $\frac{3}{2}$, $0$ and $0$. The 
eigenvectors with eigenvalue $\frac{3}{2}$ are 
$\frac{1}{\sqrt{3}} (1,i,1)$ and $\frac{1}{\sqrt{3}} (1,1,i)$, 
respectively. Thus, by transforming to an operator basis 
in which $f_{\pm}$ is diagonal, one finds that in both 
the even and the odd sector, the dominant correlations 
are actually carried by only a pair of operators, namely 
$\frac{1}{\sqrt{3}} ( \Op^{1} + i \Op^{2} + \Op^{3} )$ 
and its hermitian conjugate, and
$\frac{1}{\sqrt{3}} ( \Op^{4} + \Op^{5} + i \Op^{6} )$ 
and its hermitian conjugate, respectively. This result, 
whose precise form could hardly have been anticipated 
a priori, is a pleasing illustration of the power of a CDM 
analysis to uncover nontrivial correlations.

\subsection{\FLfu correlations}\label{sub:FL}

The correlations in the \FL sector are exponentially 
decaying, as already mentioned in 
\sref{sub:rmsnetcorrelations}. The reason for this 
was given in \cite{CheongPhysRevB79} and is the 
key to understanding the operators and correlations 
in this sector. In the limit where the fermions are all 
paired, the only possible way to annihilate one at 
$x$ and create one at $x'>x$ , such that the initial 
and final states are both paired, is that every rung 
in the interval $(x,x')$ has a fermion (necessarily on 
alternating legs). These fermions can be grouped as 
pairs in two different ways: $(x,x+1)$, $(x+2,x+3$), 
\dots, $(x'-2,x'-1)$ in the initial state, but $(x+1,x+2)$,
\dots, $(x'-1,x')$ in the final state. (Notice this requires 
that $x$ and $x'$ have the same parity.) 
\cite{CheongPhysRevB79} showed that the probability 
of such a run of filled sites decays exponentially with 
its length.

\Table{\label{tab:OpsFL}
	Comparison of the \FL operator sets on cluster $A$ for a filling 
	of $\nu = 0.286$, using the same conventions as for \tref{tab:OpsCD}.}
	\begin{tabular}{@{}lllll}
	\br
	number of   & $w^{R_{\rm{all}},\mu}/w^{R_{\rm{all}},1}$ & $O^{R_{\rm{all}}R_{\rm{short}}}$  & $O^{R_{\rm{all}}R_{\rm{int}}}$ & $O^{R_{\rm{all}}R_{\rm{long}}}$ \\
	operators & & (short) \,\,\, & (intermediate) & (long) \\
	\mr
	\04 & 1 & \04 & \04 & \04 \\
	\08 & 0.297162 & \08 & \08 & \08 \\
	12 & 0.014661 & 12 & 12 & 12 \\
	16 & 0.000402 & 16 & 16 & 16 \\
	20 & 0.000001 & 19.97 & 19.95 & 19.31 \\
	\br
	\end{tabular}
\endTable

\begin{figure}
	\begin{centering}
	\includegraphics{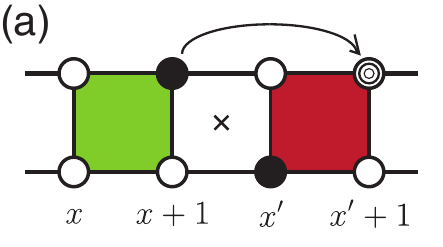}
	\includegraphics{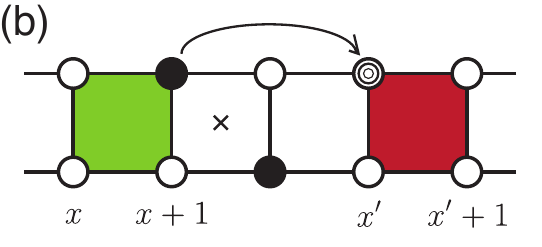}
	\includegraphics{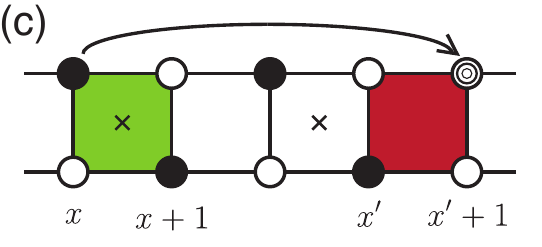}
	\par
	\end{centering}
	\caption{Three configurations of bound pairs 
	contributing to \FL correlations for a distance 
	(a) $r=2$ and (b),(c) $r=3$. Clusters $A$ and $B$ 
	are depicted by the green and 	red squares, 
	respectively. Fermions are depicted by black 
	circles, empty lattice positions by white 	circles 
	and the position where a fermion will be created 
	is depicted by concentric circles. The crosses show 
	the center of mass of the bound pairs. In 
	configuration (a) 	we have a correlation between 
	an operator corresponding to the first four eigenvalues 
	and an operator corresponding to the second four 
	eigenvalues in clusters $A$ and $B$, respectively. 
	In contrast, configuration (b) shows a correlation 
	between operators corresponding to the 
	largest eigenvalue only and configuration (c) a 
	correlation between operators corresponding to the 
	second eigenvalue only. }
	\label{fig:pairconfigs}
\end{figure}

Applying the operator analysis in this sector using 
the eigenvalue decomposition in \eref{eq:KbarEig} 
gives a series of fourfold degenerate eigenvalues 
for both clusters, see \tref{tab:OpsFL} for cluster $A$.
The table for cluster $B$ is exactly the same. For 
a specific eigenvalue, also the operators for cluster 
$B$ (residing at rungs $(x',x'+1)$) are the same as 
for cluster $A$ (residing at rungs $(x,x+1)$), but 
with mirrored rungs, i.e. an operator acting on rungs 
$(x,x+1)$ acts in the same fashion on rungs 
$(x'+1,x')$. 

\begin{figure}
	\begin{centering}
	\includegraphics{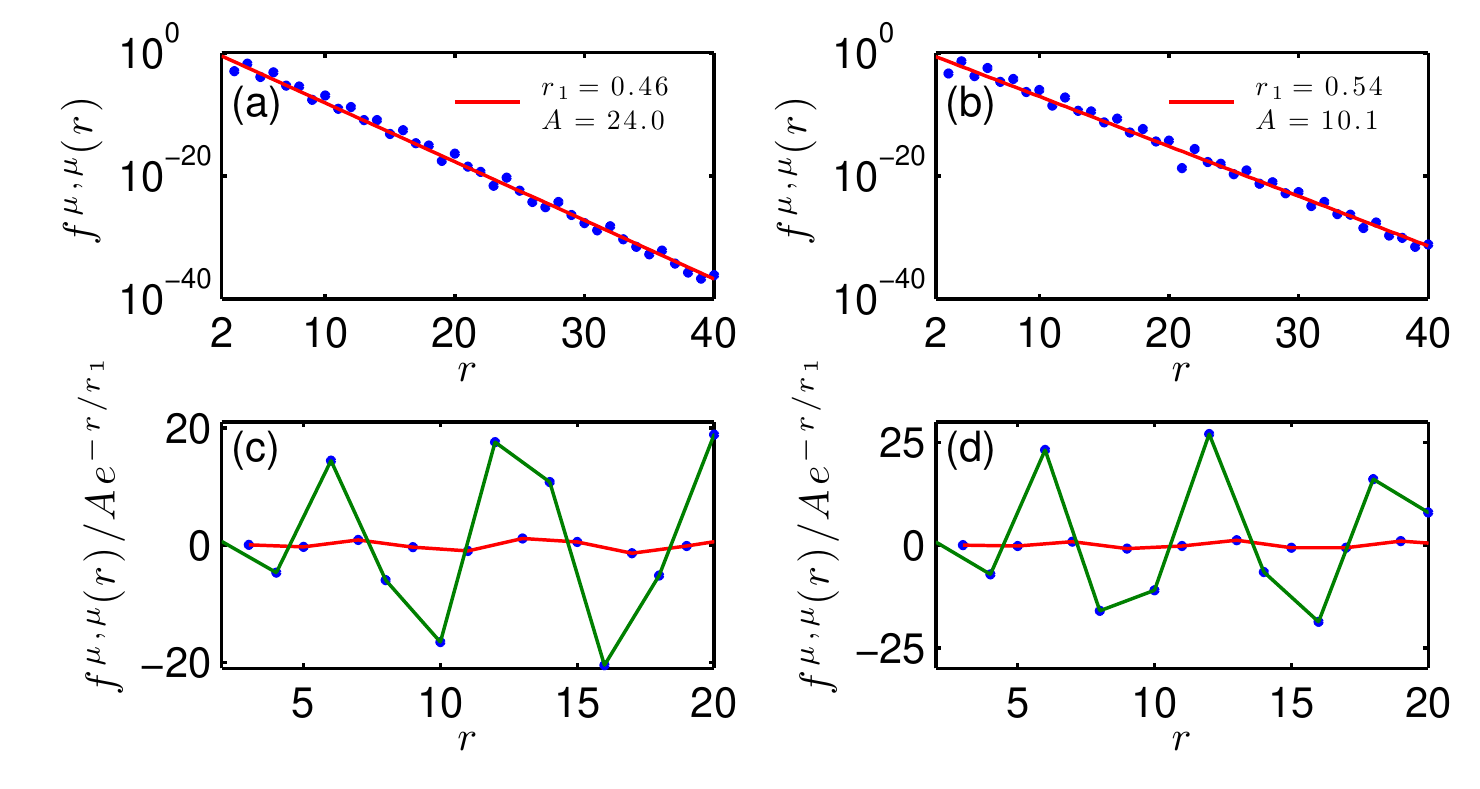}
	\par
	\end{centering}
	\caption{The \FL correlations for a filling of 
		(a),(c) $\nu = 0.248$ and (b),(d) $\nu = 0.286$. 
		Panels (a) and (b) show the \FL correlations 
		(blue symbols) together with a fit of the form 
		$A e^{-r/r_{1}}$ (red line). Panels (b) and (d) 
		show the rescaled correlator 
		$\fmat{\mu}{\mu} ( r ) / A e^{- r / r_{1}}$ 
		(blue symbols) for distances up to $r=20$. 
		(Larger distances are omitted, because for these
		$\fmat{\mu}{\mu} ( r ) < 10^{-16}$, which is the 
		maximal computer precision.) We 
		see a strong oscillation (green curve) and a 
		weak oscillation (red curve).
}
	\label{fig:fitFL}
\end{figure}

Looking more closely, the first four operators annihilate 
or create a particle on rungs $x+1$ or  $x'$, respectively, 
thereby breaking or regrouping bound pairs residing on 
$(x+1,x+2)$ or $(x'-1,x')$, respectively. The second set 
of four operators annihilates or creates a particle on 
rungs $x$ or $x'+1$, thereby breaking or regrouping 
bound pairs residing on rungs $(x,x+1)$ or $(x',x'+1)$. 
For a given odd separation $x'-x$, the combination of 
$x+1$ with $x'$ requires the smallest number of pairs 
to be present in between the two clusters. The alternative 
combination is $x$ with $x'+1$, which requires an 
additional pair in between (see \fref{fig:pairconfigs}). 
We could estimate their weights since the relative probability 
of an extra pair is the factor associated with increasing the 
separation by two. Since the correlations decay roughly as 
$\sim 10^{-r}$ (see \fref{fig:fitFL}), we predict two orders 
of magnitude. Similarly, when $x'-x$ is even, we get at 
mixture of the first and second four operators (see 
\fref{fig:pairconfigs}). This explains the difference in the 
weights of the two operator sets.

Thus, it turns out that for the \FL correlations a 
cluster size of one rung would already have 
been large enough to reveal the dominant 
correlations. We will hence use as operator basis
\numparts
\begin{eqnarray}
\fl	\Op^{A,\pm}	&=&	\unity_{x} \otimes \frac{1}{\sqrt{2}} \left( \hat{c}_{\uparrow,x+1} \pm \hat{c}_{\downarrow,x+1} \right)
								\\
\fl	\Op^{B,\pm}	&=&	\frac{1}{\sqrt{2}} \left( \hat{c}_{\uparrow,x} \pm \hat{c}_{\downarrow,x} \right) \otimes \unity_{x+1} \komma
\end{eqnarray}
\endnumparts
together with their hermitian conjugates. (The fact that our 
operator basis consists only of operators acting on a single 
rung implies that it would have been sufficient to use 
single-rung clusters. However, for the sake of consistency 
with the rest of our analysis, we retain two-rung clusters here, too.)

\begin{figure}
	\begin{centering}
	\includegraphics{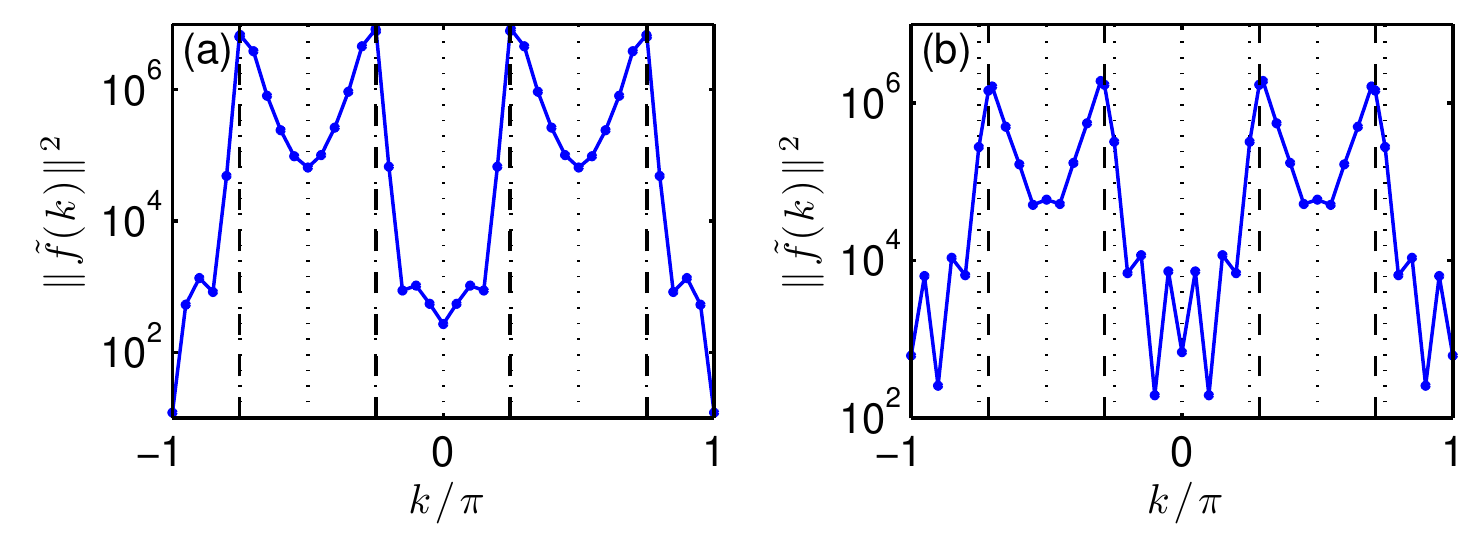}
	\par
	\end{centering}
	\caption{Fourier transform of the f-matrix obtained similarly 
		as \fref{fig:ffmatrixCD}, for \FL correlations based on 
		the four operators per cluster for (a) a filling of 
		$\nu = 0.248$ and (b) a filling of $\nu = 0.286$. 	We find 
		peaks at about $k = \pm \kF$ and $k = \pm \kF + \pi$ 
		(dashed black lines).}
	\label{fig:ffmatrixFL}
\end{figure}

The f-matrix based on these four operators (per cluster) 
is diagonal with equal entries for a given distance $r$. 
Its Fourier transform (see \fref{fig:ffmatrixFL}) gives a 
result distinct from the Fourier transform for \CD and \SC 
correlations. The dominant wave vectors are $k = \pm \kF$ 
and $k = \pi \pm \kF$, where the latter is the product 
of an oscillation with $k = \pi$ and an oscillation with 
$k = \pm \kF$. In total we have an oscillation in the 
correlations of the form $(1+(-1)^{r})e^{\pm i \kF r}$, i.e. 
an oscillation with $k = \pm \kF$, and every second term 
being close to zero. The dominant wave vector $k = \pm \kF$ i
s consistent with the usual behaviour of \FL Green's functions. 

The reason for every second term being essentially zero is 
that the dominant hopping in the system, the correlated 
hopping, always changes the position of a particle by two 
rungs, so every second position is omitted. The small but finite 
value for hopping onto intermediate rungs is related to the finite 
$t_{\parallel}/t_{c} = 10^{-2}$ that we use. It results in a second 
oscillation at $k = \pm \kF$ located at intermediate rungs, whose 
relative strength compared to the dominant one is about 
$10^{-2}$, which is consistent with the ratio $t_{\parallel}/t_{c}$ 
that we used (see \fref{fig:fitFL}). 

We fit the one independent f-matrix element $\fmat{\mu}{\mu}$ 
to an exponential decay of the form $A e^{-r/r_{1}}$ (see 
\fref{fig:fitFL}), but apart from this we were not able to fit the 
exact functional dependence on $r$, especially the oscillations 
with $k = \pm \kF$. The reason for this is the existence of two 
oscillations where one is zero on every second rung, and that 
the data range for which reasonable \FL correlations are still 
present is too small and thus makes it susceptible to numerical 
noise. This can be seen already in the Fourier spectrum, where 
we find relatively broad peaks, as a result of the influence of 
the exponential envelope and the relatively short distance range 
available.

\subsection{\SCfu correlations}\label{sub:SC}

The operator subspace for \SC ($\Delta N = 2$), in 
a cluster including two rungs has the comparatively 
small dimension of four due to the infinite nearest-neighbour 
repulsion (see \fref{fig:operator}). These are 
$\hat{c}_{\uparrow,x} \hat{c}_{\downarrow,x+1}$, 
$\hat{c}_{\downarrow,x} \hat{c}_{\uparrow,x+1}$ and 
their hermitian conjugates. In the present case of 
dominating $t_{c}$, these operators represent the 
creation- and annihilation-operators of bound pairs 
\cite{CheongSFL}. The operator analysis yields 
exactly the same four operators with degenerate 
weight for all distance regimes for both cluster $A$ 
and $B$. The four operators are 
$1/\sqrt{2} \left( \hat{c}_{\uparrow,x} \hat{c}_{\downarrow,x+1} \pm \hat{c}_{\downarrow,x} \hat{c}_{\uparrow,x+1} \right)$ 
together with their hermitian conjugates, and they 
already represent the symmetric and antisymmetric 
combinations of the operators mentioned above.

The f-matrix \eref{eq:fmatrix} is diagonal in the basis of the 
four operators, with equal strength of correlations for a fixed 
distance apart from a possible sign. This may be expected, 
given the similar structure of the operators. 

\begin{figure}
	\begin{centering}
	\includegraphics{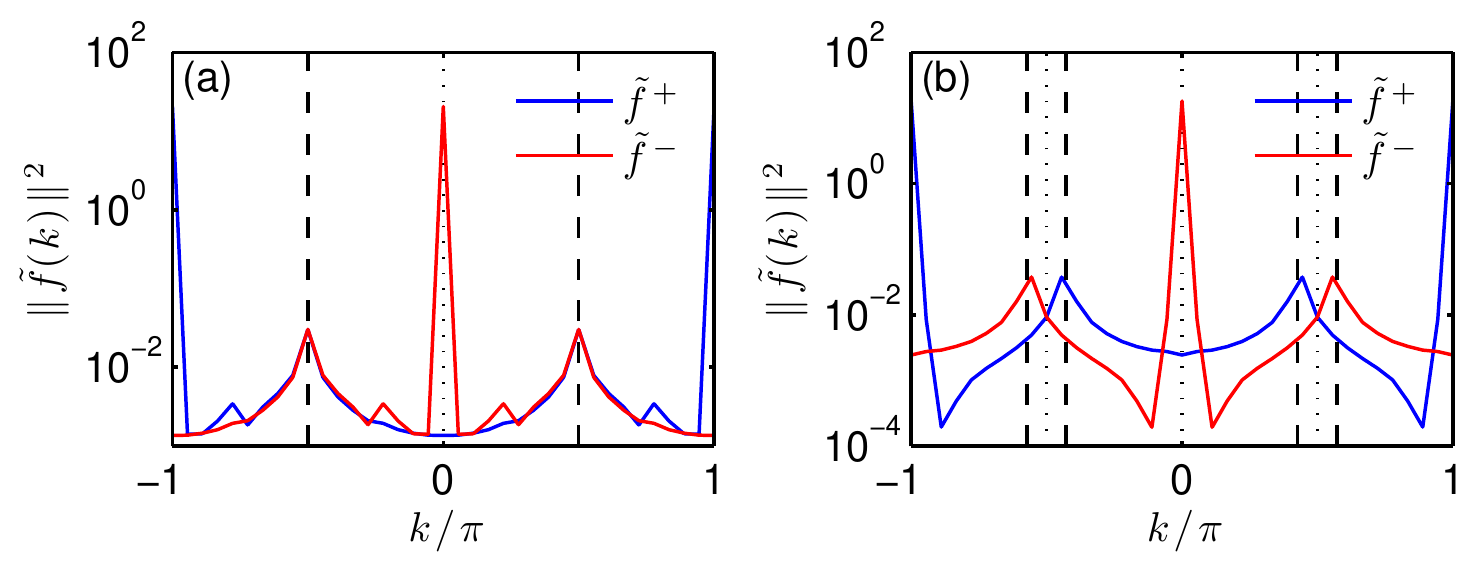}
	\par
	\end{centering}
	\caption{Fourier transform of the f-matrix for \SC 
		correlations based on the operators chosen 
		from the four-dimensional operator space for 
		(a) a filling of $\nu = 0.248$ and (b) a filling 
		of $\nu = 0.286$. For a detailed description 
		see \fref{fig:ffmatrixCD}.}
	\label{fig:ffmatrixSC}
\end{figure}

\begin{figure}
	\begin{centering}
	\includegraphics{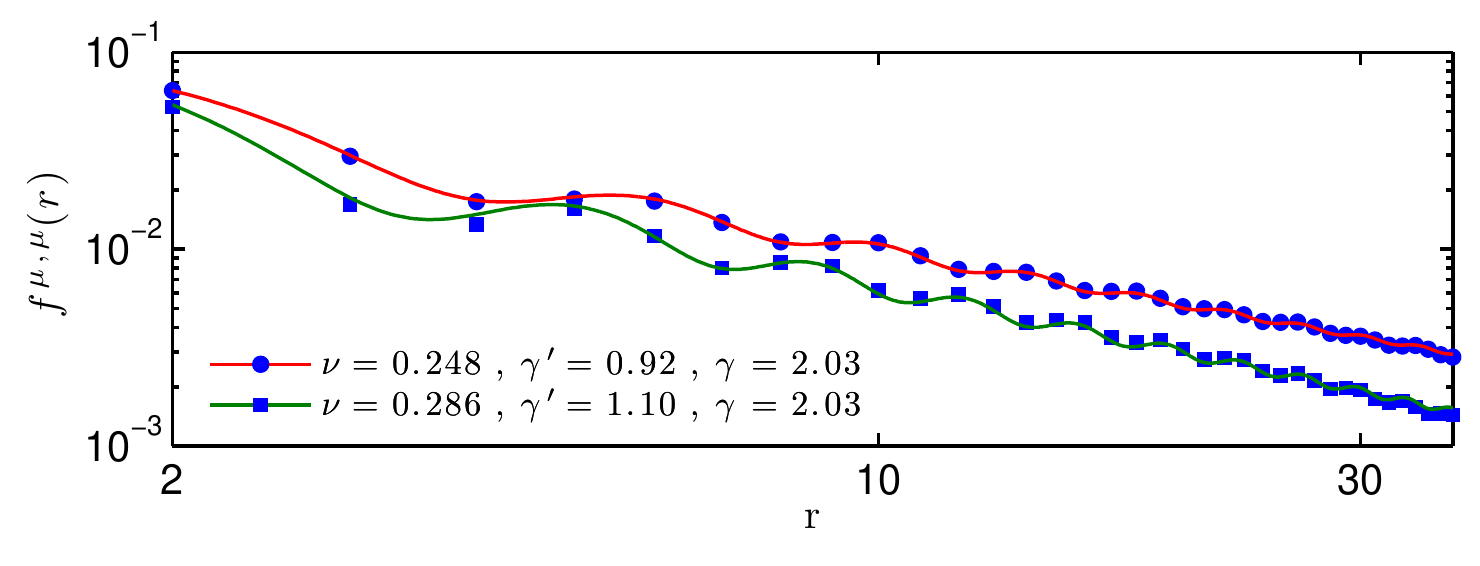}
	\par
	\end{centering}
	\caption{Fitting the \SC correlations to the form in \eref{eq:FitCD} 
		for a filling of $\nu = 0.248$ and $\nu = 0.286$. 
		The single points (blue circles and squares) are data points 
		from the f-matrix and the lines (red and green) are 
		the result of the fitting.
		}
	\label{fig:fitSC}
\end{figure}

As for the \CD correlations ($\Delta N = 0$), we apply 
a Fourier transform on the f-matrix (see \fref{fig:ffmatrixSC}) 
to identify the dominant wave vectors. Again, we find two 
spectra of similar form but shifted by $\pi$ with respect to 
each other. Consequently we redefine $f^{+}$ to 
$e^{i \pi r} f^{+}$, the part of the f-matrix belonging to the 
symmetric operators. Thus, we obtain one leading peak 
at $k = 0$ and sub-leading peaks at $k = 2\kF$. Given 
the similar structure of the Fourier spectrum to that of the 
\CD correlations, we fit the elements of the f-matrix to the 
form \eref{eq:FitCD}, but now expect $\gamma' < \gamma$ 
from the relative sharpness of the peaks. Already at the 
level of the f-matrix elements we find an overall leading 
decay with residual oscillations, whose relative magnitude 
becomes smaller at large distances (since $\gamma' < \gamma$). 
Since all matrix elements are the same after redefining 
$f^{+}$, it is sufficient to fit $\vert \fmat{\mu}{\mu} \vert$ for 
a given $\mu$, which will have dominant $k$-vectors $k=0$ 
and $k = \pm 2\kF$. The fit has errors of less than $5\%$ 
throughout, with results as shown in \fref{fig:fitSC}. The 
overall behaviour is very similar to the one already found 
from the \rms of this sector (see \fref{fig:rms}), up to the 
oscillatory part from the second term in \eref{eq:FitCD}. We 
see that the oscillations clearly decay more strongly than 
the actual strength $\vert \fmat{\mu}{\mu} \vert$, in accord 
with $\gamma' < \gamma$.

In contrast to the \CD correlations (see \fref{sub:CDfmatrix}), 
for the \SC correlations we do not find correlations which 
oscillate with phases shifted by $\Delta\phi = \pm \pi/2$ . 
This may come from the fact that clusters with the size of 
two rungs have the minimal possible size to capture \SC 
correlations. The corresponding operator space has dimension 
four and the four possible operators are very similar in structure. 
We expect that for larger clusters and hence a larger operator 
space, we would find correlations which also oscillate out of 
phase such that their oscillations cancel in the \rms, in accord 
with \eref{eq:ansatz}.

\section{Comparison to previous results}\label{sec:Comparison}
\markboth{Comparison}{Comparison}

We are now ready to compare our CDM-based results 
with those obtained in \cite{CheongSFL} by Cheong and 
Henley (CH) from fitting simple correlation functions. The 
latter were computed exactly in \cite{CheongSFL} for 
accessible separations after mapping the large $t_{c}$ 
model onto a hard-core bosonic system, but the functional 
forms of the $r$ dependencies were inferred from a purely 
numerical fitting procedure.

Overall, our results for the Hamiltonian \eref{eq:Hamiltonian} 
in the strongly correlated hopping regime agree with 
\cite{CheongSFL}, in that (i) \SC correlations and \CD 
correlations show power-law behaviour, (ii) the \SC 
correlations dominate at large distances for the fillings 
we were investigating, (iii) \FL correlations are 
exponentially decaying and are negligible over all but 
very short distances, and (iv) the dominating $k$-vector, 
for either \SC or \CD sectors, is $2\kF$.

\Table{\label{tab:Comparison}
	Comparison of the power-law exponents, which we 
	extracted from our numerical data, with those predicted 
	in \cite{CheongSFL}.}
	\begin{tabular}{@{}lllll}
	\br
	 & \centre{2}{\CD} & \centre{2}{\SC}\\
	filling & \cite{CheongSFL} & $\gamma_{0}$ & \cite{CheongSFL} & $\gamma_{2}$ \\
	\mr
	0.248 & 1.13 & 1.45 & 0.5 & 0.95 \\
	0.286 & 1.04 & 1.33 & 0.5 & 1.11 \\
	\br
	\end{tabular}
\endTable

However, the power-law exponents obtained from fitting 
f-matrix elements to \eref{eq:Fit} and summarized in 
\tref{tab:Comparison}, clearly deviate from the results in 
\cite{CheongSFL} by CH. For the \CD correlations, in 
\cite{CheongSFL} the dependence of $\gamma_{0}$ on 
the filling $\nu$ was given by the exponent 
$\gamma_{0}^{\rm{CH}} = \frac{1}{2} + \frac{5}{2} \left( \frac{1}{2} - \nu \right)$, 
from which our results deviate (see \fref{fig:rms} a,b) by 
about $25\%$. Nevertheless, our results for $\gamma_0$ 
agree qualitatively with this prediction, in that we also find 
$\gamma_0$ to decrease linearly with increasing filling. 

The \SC correlations deviate more strongly. For the dominant 
\SC correlations, CH predicted a constant power-law 
exponent of $\gamma_{2}^{\rm{CH}} = \frac{1}{2}$ 
independent of filling, coming from a universal correlation 
exponent for a chain of tightly-bound spinless fermion pairs 
\cite{EfetovLarkin76}. In contrast, we obtain a larger exponent 
(see \fref{fig:rms} a,b) for given fillings. Our result for 
$\gamma_{2}$ linearly decreases as the filling gets smaller 
and appears to approach $\frac{1}{2}$ only in the limit 
$\nu \to 0$. We also explicitly calculated the same correlation 
function as investigated in \cite{CheongSFL} but found a 
stronger decay than the $r^{-\frac{1}{2}}$ suggested there. 
We do not know whether the deviation is an artifact of the 
boundaries of our finite system, or whether the mapping used 
in \cite{CheongSFL} to a set of hardcore bosons might have 
omitted an important contribution. 

\begin{figure}
	\begin{centering}
	\includegraphics{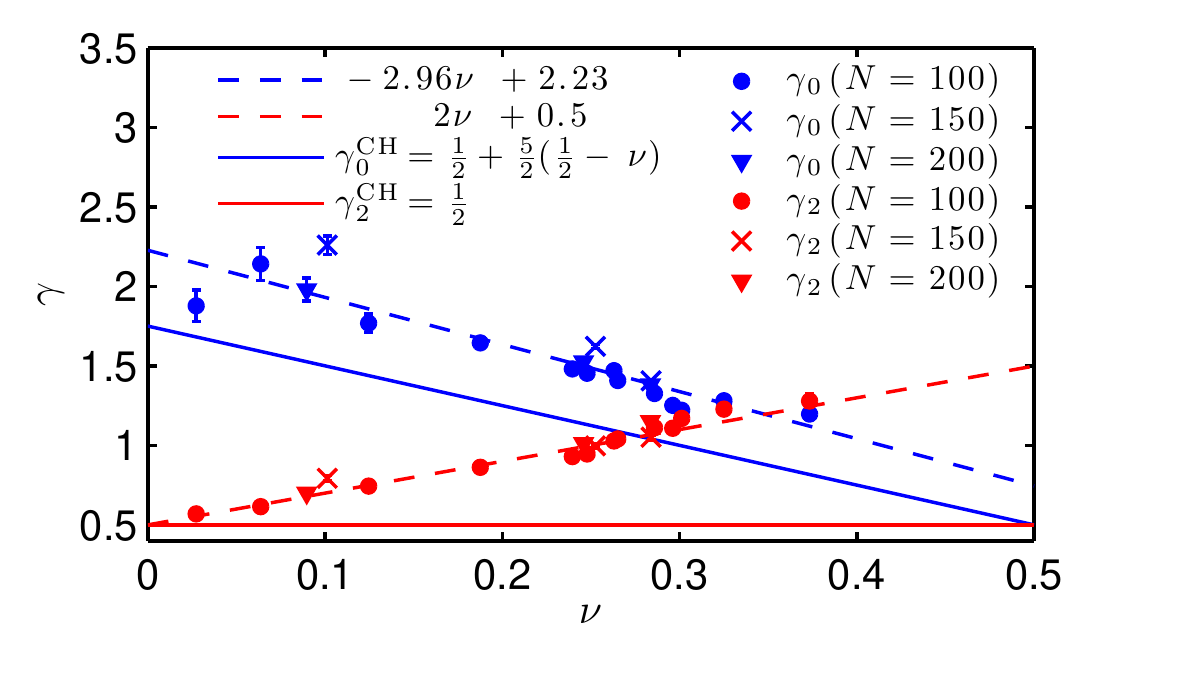}
	\par
	\end{centering}
	\caption{The power-law exponents for \CD correlations 
		($\gamma_{0}$, blue symbols) and \SC correlations 
		($\gamma_{2}$, red symbols) obtained from the \rms 
		for several fillings $\nu$. We used chain lengths of 
		$N=100$ (circles), $N=150$ (crosses), and $N=200$ 
		(triangles). The dashed blue and red lines are linear 
		fits to our numerical data for $\gamma_0$ and 
		$\gamma_2$, respectively. The solid blue and red lines 
		show the corresponding predictions of Cheong and 
		Henley \cite{CheongSFL}. For the \SC correlations, 
		our data implies a linear $\nu$-dependence going 
		from $\frac{1}{2}$ for $\nu=0$ to $\frac{3}{2}$ for 
		$\nu=\frac{1}{2}$. This crossover from $\frac{1}{2}$ 
		to $\frac{3}{2}$ is predicted by Cheong and Henley 
		as a sub-leading contribution, without giving an 
		explicit functional dependence on $\nu$. The two 
		linear $\nu$-dependencies imply that for large fillings 
		\CD correlations should become dominant over \SC 
		correlations. Unfortunately, we do not have been able 
		to obtain reliable data in that regime, because the \rms 
		showed strong oscillations here, contrary to our 
		expectations from \sref{sub:LLtheory}.}
	\label{fig:comparison}
\end{figure}

Moreover, it may be noted that by extrapolating the 
exponents in a linear fashion towards large fillings 
($\nu \to \frac{1}{2}$), it appears that for fillings larger 
than $\sim 0.35$ eventually the \CD correlations 
dominate over \SC correlations (see \fref{fig:comparison}). 
This conclusion has also been found in \cite{Micnas88} 
which similarly addresses diatomic real space pairing in 
the context of superconductivity. Their discussion, however, 
is not specifically constrained to one-dimensional systems, 
and one may wonder how the specific choice of parameters 
compare.

As the filling approaches $0.5$ in an excluded-fermion 
chain, it is appropriate to think about the degrees of 
freedom as impurity states or holes in the crystalline 
matrix of pairs \cite{Micnas88}. Then the natural length 
scale is the spacing between holes. The longer that 
spacing gets (it diverges as $\nu \to 0.5$), the larger 
also the system under investigation must be in order to 
reach the asymptotic limit. In other words, to see proper 
scaling behavior in a uniform way, the system size 
should increase proportional to $1/(0.5-\nu)$. In our 
case the data became unreliable for $\nu \gtrsim 0.4$ 
(see \fref{fig:comparison}). On the other hand, for certain 
fillings $\nu \lesssim 0.4$, we calculated the power-law 
exponents for \CD and \SC correlations for ladders of 
length $N=150$ and $N=200$ (this data is also included 
in \fref{fig:comparison}) and did not find different behaviour 
compared to out original data for ladders of length $N=100$.

\section{Conclusions}\label{sec:Conclusions}
\markboth{Conclusions}{Conclusions}

Summarizing, we found that the CDM is a useful tool 
to detect dominant correlations in a quantum lattice 
system. Starting from a ground state calculated with 
DMRG, we extracted all the important correlations 
present in our model system. We developed a method 
which, first, determines the distance-independent 
operators on each cluster that carry the dominant 
correlations of the system, and second, encodes the 
distance-dependence of the correlations in the f-matrix. 
The latter  is then analyzed in terms of decaying and 
oscillatory terms to extract the long-range behaviour of 
the correlations.

We saw that the size of the clusters $A$ and $B$ is a 
limitation of the method as it constrains the analysis to 
local operators. For some kind of correlations, however, 
larger clusters are needed to capture the relevant physics. 
This is not too easily implemented as it requires significantly 
more resources. As a possible alternative and as an 
outlook for possible future work, one may think of using a 
different cluster structure: one cluster as before and one 
``super-cluster" representing a larger continuous part of the 
system including one boundary. As MPS introduces, for each 
site, effective left and right Hilbert spaces describing the part 
of the chain to the left and to the right of that site, the description 
of such a super-cluster should be straightforward. The resulting 
effective density matrix describing a large part of the system 
can be calculated accordingly.

Overall, DMRG is a suitable method to calculate the CDM. 
The latter is easily and efficiently calculated within the framework 
of the MPS. The explicit breaking of (i) translational invariance 
by using finite system DMRG and (ii) a discrete symmetry of 
the model, lead us to develop certain strategies to restore these 
broken symmetries. The smoothing of the boundaries can still be 
further optimized, or be replaced by periodic boundary conditions. 
However, we do not expect that this will have significant influence 
on the conclusions drawn.

\ack
We would like to thank S.-A. Cheong and A. L\"auchli for discussions 
and comments on the manuscript. This work was supported by DFG 
(SFB 631, SFB-TR 12, De-730/4-1 and De-730/4-2), CENS (Center 
for NanoScience, LMU) and NIM (Nanosystems Initiative Munich). C. 
L. Henley acknowledges NSF grant DMR-0552461 for support. This 
research was supported in part by the National Science Foundation 
under Grant No. NSF PHY05-51164. J. von Delft acknowledges the 
hospitality of the Kavli Institute for Theoretical Physics, UCSB, and of 
the Institute for Nuclear Theory, University of Washington, Seattle.

\appendix
\renewcommand\thesection{\Alph{section}}
\section{The variational matrix product state approach}\label{cha:MPS}
\markboth{Variational matrix product state approach}{Variational matrix product state approach}

This appendix offers a tutorial introduction to the variational 
formulation of DMRG for finding the ground state of a 
one-dimensional quantum lattice model, , based on matrix 
product states (MPS). It also explains how this approach 
can be used to efficiently calculate the CDM. We point out 
all the important properties of the MPS and explain how to 
perform basic quantum calculations such as evaluating 
scalar products and expectation values, as well as 
determining the action of local operators on the MPS and 
constructing a reduced density matrix. We explain how a 
given MPS can be  optimized in an iterative fashion to find 
an excellent approximation for the global ground state. We 
also indicate briefly how the efficiency of the method can be 
enhanced by using Abelian symmetries.

We would like to emphasize that we make no attempt below 
at a historical overview of the DMRG approach, or at a complete 
set of references, since numerous detailed expositions of this 
approach already exist in the literature (see the excellent review 
by U. Schollw\"ock \cite{SchollwockRevModPhys77}). Our aim 
is much more modest, namely to describe the strategy implemented 
in our code in enough detail to be understandable for interested 
non-experts.

\subsection{Introduction}\label{sec:IntroductionMPS}

Quantum many-body systems deal with very large 
Hilbert spaces even for relatively small system sizes. 
For example, a one- dimensional quantum chain of 
$N$ spin $\case{1}{2}$ particles forms a Hilbert space 
of dimension $2^N$, which is exponential in system 
size. For quantum lattice models in $1$D a very efficient 
numerical method is the density matrix renormalization 
group (DMRG), introduced by Steven R. White 
\cite{WhitePhysRevLett69}. The problem of large Hilbert 
space dimension is avoided by an efficient description 
of the ground state, which discards those parts of the 
Hilbert space which have negligible weight in the ground 
state. In this manner the state space dimension of the 
effective description becomes tractable, and it has been 
shown that this produces excellent results in many quasi 
one-dimensional systems. 

The algebraic structure of the ground state for one-dimensional 
systems calculated with DMRG is described in terms of 
matrix product states (MPS) 
\cite{OstlundPhysRevLett75,DukelskyEPL43,FannesCommMathPhys144,VerstraetePhysRevLett93,TakasakiJPhysSocJpn68}. 
The origin of this MPS structure can be understood as follows 
(a detailed description will follow later): pick any specific site of 
the quantum lattice model, say site $k$, representing a local degree 
of freedom whose possible values are labeled by an index 
$\sigma_k$ (e.g., for a chain of spinless fermions,  $\sigma_k = 0$ 
or 1 would represent an empty or occupied site). Any many-body 
state $\ket{\psi}$ of the full chain can be expressed in the form
\begin{equation}
	\ket{\psi} = \sum_{l_{k} r_{k} \sigma_{k}}
					A_{l_{k} r_{k}}^{[\sigma_{k}]}
					\ket{l_{k}} \ket{\sigma_{k}} \ket{r_{k}}
					\komma \label{eq:MPSfullMPS}
\end{equation}
where  $\ket{l_{k}}$ and $\ket{r_{k}}$ are sets of states (say 
$N_{l}$ and $N_{r}$ in number) describing the parts of the 
chain to the left and right of current site $k$, respectively, and 
for each $\sigma_{k}$, $A^{[\sigma_{k}]}$ is a matrix with matrix 
elements $A^{[\sigma_{k}]}_{l_{k} r_{k}}$ and dimension 
$N_{l} \times N_{r}$. Since such a description is possible for 
any site $k$, the state $\ket{\psi}$ can be specified in terms 
of the set of all matrices $A^{[\sigma_{k}]}$, resulting in a 
\emph{matrix product state} of the form 
\begin{equation}
	\ket{\psi} = \sum_{\sigma_{1} \dots \sigma_{N}}
					\left( A^{[\sigma_{1}]} \dots A^{[\sigma_{N}]} \right)_{l_{1} r_{N}}
					\ket{\sigma_{1}} \dots \ket{\sigma_{N}}
					\punkt \label{eq:MPSprodPsi}
\end{equation}
One may now seek to minimize the ground state energy within the 
space of all MPS, treating the matrix elements of the $A$-matrices 
as variational parameters to minimize the expectation value 
$\bra{\psi} H \ket{\psi}$. If this is done by sequentially stepping 
through all matrices in the MPS and optimizing one matrix at a time 
(while keeping the other matrices fixed), the resulting procedure is 
equivalent to a strictly variational minimization of the ground state energy 
within the space of all MPS of the form \eref{eq:MPSprodPsi} 
\cite{VerstraetePhysRevLett93,OstlundPhysRevLett75,DukelskyEPL43,FannesCommMathPhys144,TakasakiJPhysSocJpn68}. 
If instead the optimization is performed for two adjacent matrices at 
a time, the resulting (quasi-variational) procedure is equivalent to 
White's original formulation of DMRG
\cite{VerstraetePhysRevLett93,OstlundPhysRevLett75,DukelskyEPL43,FannesCommMathPhys144,TakasakiJPhysSocJpn68}. 
The MPS based formulation of this strategy has proven to be very 
enlightening and fruitful, in particular also in conjunction with concepts 
from quantum information theory \cite{VerstraetePhysRevLett93}.

In general, such an approach works for both bosonic and 
fermionic systems. However, to be efficient the method needs 
a local Hilbert space with finite and small dimension, limiting 
its applicability to cases where the local Hilbert space is 
finite dimensional a priori (e.g. fermions or hard-core bosons) 
or effectively reduced to a finite dimension, e.g. by interactions. 
For example, such a reduction is possible if there is a large 
repulsion between bosons on the same site such that only 
a few states with small occupation number will actually take 
part in the ground state. For fermions, on the other hand, the 
fermionic sign must be properly taken care of. The anti-commutation 
rules of fermionic creation and annihilation operators causes the 
action of an operator on a single site to be non-local because 
the occupations of the other sites have to be accounted for. To 
simplify the problem, a Jordan-Wigner transformation 
\cite{JordanWignerZPhys28} can be used to transform fermionic 
creation and annihilation operators to new operators that obey 
bosonic commutation relations for any two operators referring 
to different sites. This greatly simplifies the numerical treatment 
of these operators as fermionic signs can be (almost) ignored.

Before outlining in more detail  the above-mentioned 
optimization scheme for determining the ground state 
(see \sref{sec:MPSVariationalOptimizationScheme}), 
we present in \sref{sec:MPS} various technical 
ingredients needed when working with MPS.

\subsection{Matrix product states}\label{sec:MPS}

\subsubsection{Construction of matrix product states}\label{sub:MPSConstruction}

We consider a chain with open boundary conditions consisting 
of $N$ equal sites with a local Hilbert space dimension of $d$. 
A state $\ket{\psi}$ is described by
\begin{equation}
	\ket{\psi} = \sum_{\sigma_{1} \dots \sigma_{N}}
					\psi_{\sigma_{1} , \dots, \sigma_{N}}
					\ket{\sigma_{1}} \dots \ket{\sigma_{N}}
					\komma \label{eq:MPSfullPsi}
\end{equation}
where $\sigma_{i}=1 , \dots , d$ labels the local basis states of site
$i$. In general, the size of the coefficient space $\psi$ scales
with $\mathcal{O} ( d^{N} )$. This can be rewritten in a
matrix decomposition of the form \eref{eq:MPSprodPsi} 
with a set of $N$ times $d$ matrices $A^{[\sigma_{k}]}$ 
(see \sref{sub:MPSDetailsOfA} for details). Formally, this 
decomposition has two open indices, namely the first index 
of $A^{[\sigma_{1}]}$ and the second index of $A^{[\sigma_{N}]}$, 
as $A^{[\sigma_{1}]}$ and $A^{[\sigma_{N}]}$ are not 
multiplied onto a matrix to the left and to the right, respectively. 
For periodic boundary conditions these two indices would 
be connected by a trace over the matrix decomposition, 
giving a scalar. In the case of open boundary conditions, 
the two indices range only over one value (see 
\sref{sub:MPSDetailsOfA}), i.e. the matrix decomposition 
is a $1\times1$ matrix which is a scalar.

If these $A$-matrices are sufficiently large this decomposition
is formally exact, but since that would require $A$-matrices 
of exponentially large size, such an exact description is 
of academic interest only. The reason why the $A$-matrices 
are introduced is that they offer a very intuitive strategy for 
reducing the numerical resources needed to describe a given 
quantum state. This strategy involves limiting the dimensions 
of these matrices by systematically using singular-value 
decomposition and retaining only the set of largest singular 
values. The $A$-matrices can be chosen much smaller while still giving
a very good approximation of the state $\ket{\psi}$.

\begin{figure}[h]
	\begin{centering}
	\includegraphics{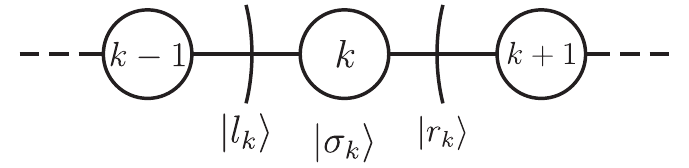} \par
	\end{centering}
	\caption{Current site with effective basis sets.}
	\label{fig:CurrentSite}
\end{figure}

Selecting a certain site $k$, the state can be rewritten in the form \eref{eq:MPSfullMPS}. 
The effective 'left' basis $\ket{l_{k}} = \sum_{\sigma_{1} \dots \sigma_{k-1}} 
A^{[\sigma_{1}]} \dots A^{[\sigma_{k-1}]} 
\ket{\sigma_{1}} \dots \ket{\sigma_{k-1}}$
describes the sites $j = 1, \dots, k-1$, the effective 'right' basis
$\ket{r_{k}}$ similarly describes the sites $j = k+1 , \dots , N$. 
Site $k$ is called the \emph{current} site, as the description of the 
state makes explicit only the $A$-matrix of this site (see \fref{fig:CurrentSite}).

So far \eref{eq:MPSfullPsi} and \eref{eq:MPSfullMPS}
are equivalent, but now we have a representation of 
the state which allows a convenient truncation of the 
total Hilbert space, used for the description of a MPS. 
For example, if we introduce a parameter $D$ and 
truncate all effective Hilbert spaces of all sites to the 
dimension $D$, each $A^{[\sigma_{k}]}$-matrix has 
at most the dimension $D \times D$. This reduces the 
resources used to describe a state from $\mathcal{O} (d^{N})$ 
for the full many-body Hilbert space down to 
$\mathcal{O} (ND^{2}d)$. This is linear in the system
size, assuming that the size required for $D$ to 
accurately describe the state grows significantly slower 
than linearly in $N$. This, in fact, turns out to be the case 
for ground state calculations \cite{VerstraetePhysRevB73}. 
Details of this truncation procedure and estimates of the 
resulting error are described in \sref{sub:MPSHilbertSpaceTruncation}.

\subsubsection{Global view and local view}\label{sub:Views}

Matrix product states can be viewed in two alternative ways: a global
view and a local view. Both views are equivalent and both
have their applications. In the global view the state is expressed
as in \eref{eq:MPSprodPsi}, i.e. the effective Hilbert spaces
have been used 'only' to reduce resources. The state is stored in
the $A$-matrices, but the effective basis sets will be contracted
out. This perception has to be handled very careful, because contracting
out the effective basis sets leads to higher costs in resources! In
the local view the state is expressed as in \eref{eq:MPSfullMPS}.
It is called local because there is one special site, the \emph{current}
site, and all other sites are combined in effective orthonormalized
basis sets. Usually, the local view is used iteratively for every
site. In this perception, we need effective descriptions of operators
contributing to the Hamiltonian acting on other sites than the current
site (see \sref{sub:MPSOperatorsBases}).

\subsubsection{Details of the $A$-matrices}\label{sub:MPSDetailsOfA}

The $A$-matrices have some useful properties that hold 
independently of the truncation scheme used to limit the 
effective Hilbert spaces. First of all, we notice that by construction
$dim (\mathcal{H}^{r_{k-1}}) \equiv dim (\mathcal{H}^{l_{k}})$,
otherwise the matrix products in \eref{eq:MPSprodPsi} would
be ill defined. Based on this, we can find another interpretation of
the $A$-matrices in the local view. The part of the chain to the 
left of site $k$ (where $k$ is far from the ends for simplicity) 
is described by the effective basis $\ket{l_{k}}$, which is built 
of truncated $A$-matrices:
\begin{eqnarray}
	\ket{l_{k}}	& = &	\sum_{\sigma_{1} , \dots , \sigma_{k-1}}
								\left( A^{[\sigma_{1}]} \dots A^{[\sigma_{k-1}]} \right)_{1l_{k}}
								\ket{\sigma_{1}} \dots \ket{\sigma_{k-1}}
								\nonumber \\
					& = &	\sum_{\sigma_{k-1}} \sum_{l_{k-1}} \underbrace{
								\sum_{\sigma_{1} , \dots , \sigma_{k-2}}
								\left( A^{[\sigma_{1}]} \dots A^{[\sigma_{k-2}]} \right)_{1l_{k-1}}
								\ket{\sigma_{1}} \dots \ket{\sigma_{k-2}}
								}_{\ket{l_{k-1}}}
								A_{l_{k-1} , l_{k}}^{[\sigma_{k-1}]} \ket{\sigma_{k-1}}
								\nonumber \\
					& = &	\sum_{\sigma_{k-1} , l_{k-1}}
								A_{l_{k-1} l_{k}}^{[\sigma_{k-1}]}
								\ket{l_{k-1}} \ket{\sigma_{k-1}}
								\punkt \label{eq:MapLeftBasis}
\end{eqnarray}
The $A^{[\sigma_{k-1}]}$-matrix maps the effective
left basis $\ket{l_{k-1}}$ together with the local $\ket{\sigma_{k-1}}$
basis onto the effective left basis $\ket{l_{k}}$! The same argument
applied on the effective right basis of site $k$ leads to the transformation
of $\ket{r_{k+1}}$ and $\ket{\sigma_{k+1}}$ onto $\ket{r_{k}}$
via the $A^{[\sigma_{k+1}]}$-matrix:
\begin{equation}
	\ket{r_{k}} = \sum_{\sigma_{k+1} , r_{k+1}}
					A_{r_{k} r_{k+1}}^{[\sigma_{k+1}]}
					\ket{\sigma_{k+1}} \ket{r_{k+1}}
					\punkt \label{eq:MapRightBasis}
\end{equation}
So far, this may be any transformation, but in order 
to deal with properly orthonormal basis sets, we may 
impose unitarity on the transformation (see below).

The $A$-matrices towards the ends of the chain have 
to be discussed separately. The use of open boundary 
conditions implies that we have a 1-dimensional 
effective state space to the left of site one and the right 
of site $N$, respectively, both representing the empty 
state. This implies that $\dim ( \mathcal{H}^{l_{1}} ) = 1 = \dim ( \mathcal{H}^{r_{N}} )$. 
Moving inwards from the ends of the chain, the effective 
Hilbert spaces acquire dimension $d^{1} , d^{2} , \dots$ 
until they become larger than $D$ and need to be 
truncated. Correspondingly, the dimension of matrix 
$A^{[\sigma_{k}]}$ is $D_{k-1} \times D_{k}$, where 
$D_{k} = \min ( d^{k} , d^{N-k} , D )$. There is no 
truncation needed if $\dim ( \mathcal{H}^{l_{k}} ) * d = \dim ( \mathcal{H}^{r_{k}} )$
or $\dim ( \mathcal{H}^{r_{k}} ) * d = \dim ( \mathcal{H}^{l_{k}} )$. 
In these cases we simply choose $A_{( l_{k} \sigma_{k} ) r_{k}} = \unity$
and $A_{l_{k} ( r_{k} \sigma_{k} )} = \unity$, respectively.

Summarizing, the $A$-matrices have two functions. If site $i$ is
the current site in \eref{eq:MPSfullMPS}, the $A^{[\sigma_{i}]}$-matrices
represent the state, i.e. its coefficients specify the linear combination
of basis states $\ket{l_{k}}$, $\ket{\sigma_{k}}$ and $\ket{r_{k}}$.
On the other hand, if not the current site, the $A$-matrices are
used as a mapping to build the effective orthonormal basis for the
current site, as we describe next:

\paragraph{Orthonormal basis sets}

In the local view, the whole system is described by the $A$-matrices
of the current site $k$ in the effective left basis, the effective
right basis, and the local basis of site $k$. A priori, the basis
states form an orthonormal set only for the local basis set, but we 
may ask for the effective basis sets $\ket{l}$ and $\ket{r}$ 
\footnote{From now on the index $k$ is only displayed when several sites are
involved. For the current site or in the case when only one $A$-matrix
is considered the index will be dropped.}
to be orthonormal, too, i.e. require them to obey:
\begin{eqnarray}
	\braket{l'}{l} & = & \delta_{l'l} \komma \nonumber \\
	\braket{r'}{r} & = & \delta_{r'r} \punkt \label{eq:MPSOrtho}
\end{eqnarray}
This immediately implies the following condition on the $A^{[\sigma_{j}]}$-matrices,
using \eref{eq:MapLeftBasis} and \eref{eq:MapRightBasis}
(for a derivation, see \sref{sec:DerivationOrtho}):
\begin{eqnarray}
	\sum_{\sigma_{j}}
	{ A^{[\sigma_{j}]} }^{\dagger} A^{[\sigma_{j}]}
	& = & \unity \qquad \mathrm{for} \, j<k \komma \nonumber \\
	\sum_{\sigma_{j}}
	A^{[\sigma_{j}]} { A^{[\sigma_{j}]} }^{\dagger}
	& = & \unity \qquad \mathrm{for} \, j>k \punkt \label{eq:MPSOrthoA}
\end{eqnarray}
The orthonormality \eref{eq:MPSOrtho} for both the left- \emph{and} 
right basis states holds only for the current site. For the other sites there is always
only one orthonormal effective basis.

\paragraph{Graphical representation}

\begin{figure}[h]
	\begin{centering}
	\includegraphics{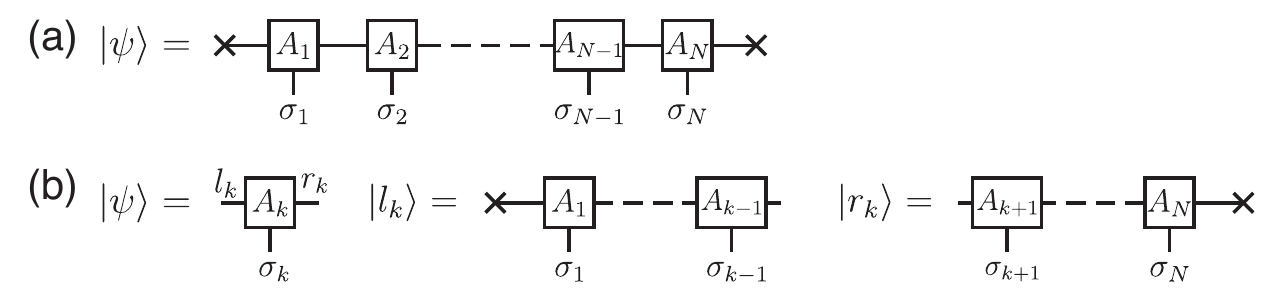}
	\par
	\end{centering}
	\caption{Graphical representation of a matrix product state in the (a) global view and (b) local view.}
	\label{fig:GraphicalRepresentation}
\end{figure}

Matrix product states can be depicted in a convenient graphical representation
(see \fref{fig:GraphicalRepresentation}). In this representation,
$A$-matrices are displayed as boxes and $A^{[\sigma_{k}]}$
is replaced by $A_{k}$ for brevity. Indices correspond to links from
the boxes. The left link connects to the effective left basis, the
right link to the right one, and the link at the bottom to the local
basis. Sometimes indices are explicitly written on the links to emphasize
the structure of the sketch. Connected links denote a summation over
the indices (also called contraction) of the corresponding $A^{[\sigma]}$-matrices.
At the boundaries of the chain, a cross is used to indicate the vacuum
state.

\subsubsection{Orthonormalization of effective basis states}\label{sub:SwitchingCurrentSite}

We now describe how an arbitrary MPS state can be 
rewritten into a form where its local view with respect to 
a given site has orthonormal left- and right basis states. 
It should be emphasized that this really just amounts to 
a reshuffling of information among the state's $A$-matrices 
without changing the state itself, by exploiting the freedom 
that we always can insert any $X^{-1}X = \unity$ at any 
position in the matrix product state without altering it.

\begin{figure}[h]
	\begin{centering}
	\includegraphics{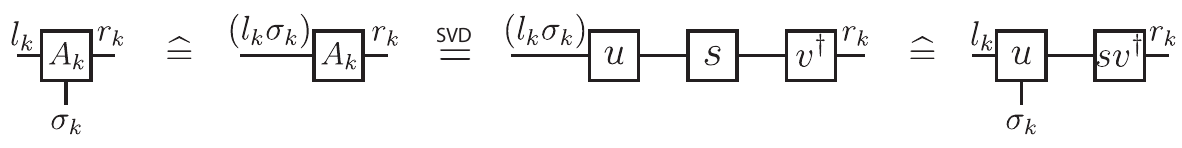}\par
	\end{centering}
	\caption{Singular value decomposition of the $A$-matrices}
	\label{fig:OrthoBasis01}
\end{figure}

Assume site $k$ to be the current site and assume that 
it has an orthonormal left basis (the latter is automatically 
fulfilled for $k=1$). We need a procedure to ensure that, 
when the current site is switched to site $k+1$, this site, 
too, will have an orthonormal left basis. (This is required 
for the orthonormality properties used in the proof in 
\sref{sec:DerivationOrtho}. A similar procedure can be 
used to ensure that site $k-1$ has an orthonormal right 
basis provided $k$ has such a basis.) For this purpose 
we use the singular value decomposition (SVD, see 
\sref{sec:Singular-value-decomposition}) for which we 
have to rewrite $A_{l_{k} r_{k}}^{[\sigma_{k}]}$ by 
\emph{fusing} the indices $l_{k}$ and $\sigma_{k}$:
\begin{equation}
	A_{l_{k} r_{k}}^{[\sigma_{k}]}
	\widehat{=} A_{\left( l_{k} \sigma_{k} \right) r_{k}} 
	= \sum_{m,n} u_{\left( l_{k} \sigma_{k} \right) m} s_{mn} \left( v^{\dagger} \right)_{n r_{k}}
	\widehat{=} \sum_{m} u_{l_{k}m}^{[\sigma_{k}]} \left(sv^{\dagger} \right)_{m r_{k}}
	\komma\label{eq:SVDonA}
\end{equation}
where $m$, $n$ and $r_{k}$ have the same index range 
(see \fref{fig:OrthoBasis01}). Specifically, $u$ fulfills
\begin{equation}
	\unity = u^{\dagger} u
	= \sum_{\left(l_{k} \sigma_{k} \right)} u_{\left( l_{k} \sigma_{k} \right) , m'}^{*} u_{\left( l_{k} \sigma_{k} \right) , m}
	\komma\label{eq:Uortho}
\end{equation}
which is equivalent to the orthonormality condition \eref{eq:MPSOrthoA}
for the $A^{[\sigma_{k}]}$-matrices.

\begin{figure}[h]
	\begin{centering}
	\includegraphics{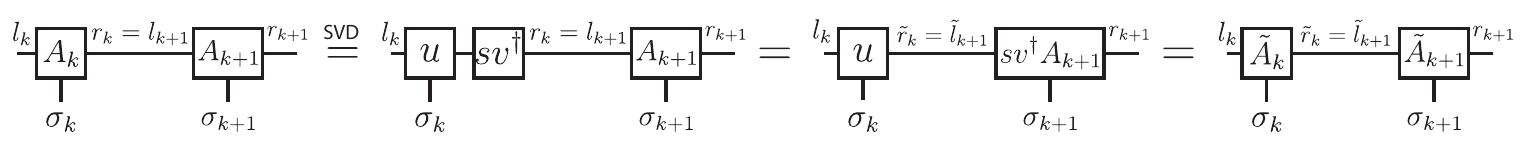}\par
	\end{centering}
	\caption{Rearrangement of the $A$-matrices to switch the current site from site $k$ to $k+1$.}
	\label{fig:OrthoBasis02}
\end{figure}

As $u$ replaces $A^{[\sigma_{k}]}$ and $sv^{\dagger}$
is contracted onto $A^{[\sigma_{k+1}]}$, this leaves the
overall state unchanged (for a graphical depiction see \fref{fig:OrthoBasis02}):
\begin{eqnarray}
	A^{[\sigma_{k}]} A^{[\sigma_{k+1}]}
		& = &	\sum_{\left( r_{k} = l_{k+1} \right)} A_{l_{k} r_{k}}^{[\sigma_{k}]} A_{l_{k+1} r_{k+1}}^{[\sigma_{k+1}]}
					= \sum_{\left(r_{k} = l_{k+1}\right)} \sum_{m} u_{l_{k}m}^{[\sigma_{k}]} \left(sv^{\dagger} \right)_{m r_{k}} A_{l_{k+1} r_{k+1}}^{[\sigma_{k+1}]}
					\nonumber \\
		& = &	u^{[\sigma_{k}]} \left(sv^{\dagger} A_{k+1} \right)^{[\sigma_{k+1}]}
					\equiv \tilde{A}^{[\sigma_{k}]} \tilde{A}^{[\sigma_{k+1}]}
					\punkt \label{eq:UpdateLR}
\end{eqnarray}

\begin{figure}[h]
	\begin{centering}
	\includegraphics{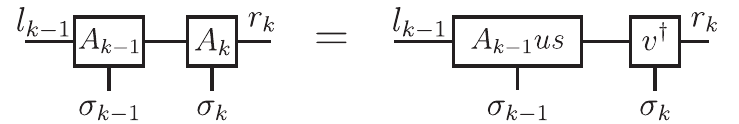}\par
	\end{centering}
	\caption{Orthonormal effective right basis for site $k-1$.}
	\label{fig:OrthoBasis03}
\end{figure}

Site $k+1$ now has an orthonormal effective left basis. A similar
procedure works for the effective right basis, see \fref{fig:OrthoBasis03}.
To obtain an orthonormal effective left basis for the current site
$k$, we start with the first site, update $A^{[\sigma_{1}]}$
and $A^{[\sigma_{2}]}$, move to the next site, update
$A^{[\sigma_{2}]}$ and $A^{[\sigma_{3}]}$,
and so on until site $k-1$. For an orthonormal effective right basis, 
we start from site $N$ and apply an analogous procedure in the 
other direction.

If the state $\ket{\psi}$  is in the local description of site $k$ with
orthonormal basis sets $\ket{l_{k}}$, $\ket{\sigma_{k}}$ and $\ket{r_{k}}$, 
it is now very easy to change the current site to site $k \pm 1$, 
with corresponding new orthonormal basis sets $\ket{l_{k \pm 1}}$,
$\ket{\sigma_{k \pm 1}}$, $\ket{r_{k\pm1}}$. Suppose we want to
change the current site from site $k$ to site $k+1$. 
Following the procedure described above, site $k+1$
already has an orthonormal right basis and all sites left of site
$k$ fulfill the orthonormality condition. All that is left to do,
is to update site $k$ and $k+1$ to obtain an orthonormal left basis
for site $k+1$. This is called a \emph{switch} of the current site
from site $k$ to $k+1$. The switch from site $k$ to site $k-1$
is done analogously.

\subsubsection{Hilbert space truncation}\label{sub:MPSHilbertSpaceTruncation}

A central ingredient in the variational optimization of the 
ground state (see \sref{sub:MPSVarEnergyMinimisation} 
below) is the truncation of the effective Hilbert spaces 
associated with a given $A$-matrix. The strategy for 
truncating the effective Hilbert spaces is completely 
analogous to the original DMRG formulation 
\cite{DukelskyEPL43}. The DMRG truncation scheme 
is based on discarding that part of the Hilbert space 
on which a certain density matrix has sufficiently small 
weight. There are two ways to obtain an appropriate 
reduced density matrix: two-site DMRG
\cite{WhitePhysRevLett69,SchollwockRevModPhys77} 
and one-site DMRG \cite{SchollwockRevModPhys77}. 
The crucial difference between the two is that one-site 
DMRG is strictly variational in the sense that the energy 
is monotonically decreasing with each step,, whereas 
in two-site DMRG the energy may (slightly) increase in 
some steps, but with the advantage that the cutoff 
dimension can be chosen dynamically in each step.

\paragraph{Two-site DMRG}

\begin{figure}[h]
	\begin{centering}
	\includegraphics{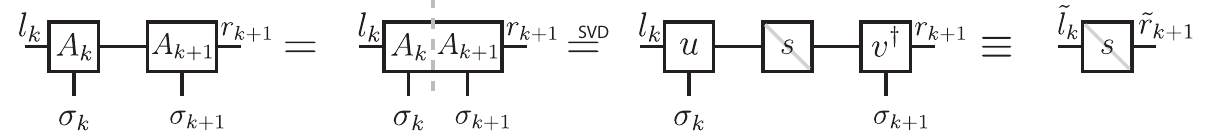} \par
	\end{centering}
	\caption{Procedure for site update within two-site DMRG. 
	The grey line under the $s$ indicates that $s$ is the diagonal
	matrix of singular values.}
	\label{fig:TwoSiteDMRG}
\end{figure}

Two-site DMRG arises when variationally optimizing 
two sites at a time. We consider two current sites, 
say $k$ and $k+1$, and we may choose the cutoff 
dimension site-dependent: $D \rightarrow D_{k} \equiv dim ( \mathcal{H}^{l_{k}} )$.
Following \sref{sub:SwitchingCurrentSite}, we assume 
site $k$ to have an orthonormal left basis and site $k+1$ 
to have an orthonormal right basis. After contracting the 
indices connecting $A^{[\sigma_{k}]}$ and $A^{[\sigma_{k+1}]}$ 
(see \fref{fig:TwoSiteDMRG}), the state is described by 
$A_{l_{k} r_{k+1}}^{[\sigma_{k} \sigma_{k+1}]}$. In this 
description we may optimize the ground state locally by 
variationally minimizing the ground state energy with 
respect to $A_{l_{k} r_{k+1}}^{[\sigma_{k} \sigma_{k+1}]}$ 
(see \sref{sub:MPSVarEnergyMinimisation}). Afterwards, 
we need to decompose $A_{l_{k} r_{k+1}}^{[\sigma_{k} \sigma_{k+1}]}$ 
into $A^{[\sigma_{k}]}$ and $A^{[\sigma_{k+1}]}$ again. 
This can be accomplished via singular value decomposition 
(see \sref{sec:Singular-value-decomposition}) by fusing 
the indices $l_{k} , \sigma_{k} \rightarrow\left( l_{k} \sigma_{k} \right)$ 
and $r_{k+1} , \sigma_{k+1} \rightarrow \left( r_{k+1} \sigma_{k+1} \right)$
(see \fref{fig:TwoSiteDMRG}) to obtain 
$A_{l_{k} r_{k+1}}^{[\sigma_{k} \sigma_{k+1}]} = \sum_{i} u_{l_{k} i}^{[\sigma_{k}]} s_{i} \left( v^{\dagger} \right)_{i r_{k+1}}^{[\sigma_{k+1}]}$,
where $i=1 \dots \min ( dD_{k} , dD_{k+2} )$. Using the 
column unitarity of $u$ and the row unitarity of $v^{\dagger}$
(see \sref{sec:Singular-value-decomposition}), we rewrite the state as 
\begin{eqnarray}
	\ket{\psi}	& = &	\sum_{l_{k}r_{k+1}\sigma_{k}\sigma_{k+1}}
								\left( \sum_{i} u_{l_{k} i}^{[\sigma_{k}]} s_{i} \left(v^{\dagger} \right)_{i r_{k+1}}^{[\sigma_{k+1}]} \right)
								\ket{l_{k}} \ket{\sigma_{k}} \ket{\sigma_{k+1}} \ket{r_{k+1}}
								\nonumber \\
					& = &	\sum_{i} s_{i}
								\underbrace{\left( \sum_{l_{k} \sigma_{k}} u_{l_{k} i}^{[\sigma_{k}]} \ket{l_{k}} \ket{\sigma_{k}} \right)}_{| \tilde{l}_{i} \rangle}
								\underbrace{\left( \sum_{r_{k+1} \sigma_{k+1}} \left(v^{\dagger} \right)_{i r_{k+1}}^{[\sigma_{k+1}]} \ket{\sigma_{k+1}} \ket{r_{k+1}} \right)}_{\ket{\tilde{r}_{i}}}
								\nonumber \\
					& = &	\sum_{i} s_{i} | \tilde{l}_{i} \rangle \ket{\tilde{r}_{i}}
								\komma \label{eq:PsiBond}
\end{eqnarray}
where the new set of basis states $| \tilde{l}_{i} \rangle$ and $\ket{\tilde{r}_{i}}$
is orthonormal with $\langle \tilde{l}_{i'}| \tilde{l}_{i} \rangle = \delta_{i'i}$
and $\braket{\tilde{r}_{i'}}{\tilde{r}_{i}} = \delta_{i'i}$. This 
representation of the state may be seen as residing on the bond
between $k$ and $k+1$, with effective orthonormal basis sets for
the parts of the system to the left and right of the bond. Reduced
density matrices for these parts of the system, obtained by tracing
out the respective complementary part, have the form:
\begin{eqnarray}
	\rho^{\left[ L \right]}	=  \sum_{i} s_{i}^{2} | \tilde{l}_{i} \rangle \langle \tilde{l}_{i} |
	\komma \qquad
	\rho^{\left[ R \right]}	& = & \sum_{i} s_{i}^{2} \ket{\tilde{r}_{i}} \bra{\tilde{r}_{i}}
	\punkt \label{eq:RhoBond}
\end{eqnarray}
The standard DMRG truncation scheme amounts to 
truncating $\rho^{\left[ L \right]}$ and $\rho^{\left[ R \right]}$ 
according to their singular values $s_{i}$. We could 
either keep all singular values greater than a certain 
cutoff, thereby specifying a value for $D_{k+1}$ 
between $1$ and $\min \left(dD_{k} , dD_{k+2} \right)$,
or alternatively choose $D_{k}=D$ to be site-independent 
for simplicity. This step makes the method not strictly 
variational, since we discard some part of the Hilbert 
space which could increase the energy. It turns out 
that this potential increase of energy is negligible in 
practice. We can obtain a measure for the information 
lost due to truncation by using the von Neumann entropy 
$S = - \tr \left( \rho \ln \rho \right)$, given by
\begin{equation}
	\varepsilon \equiv - \sum_{i>D} s_{i}^{2} \ln \left(s_{i}^{2}\right)\komma\label{eq:LostEntropy}
\end{equation}
where $\sum s_{i}^{2} = 1$ due to the normalization of $\ket{\psi}$.

\paragraph{One-site DMRG}

One-site DMRG arises when variationally optimizing 
one site at a time. In contrast to two-site DMRG, 
one-site DMRG does not easily allow for dynamical 
truncation during the calculation. (It is possible in 
principle to implement the latter,  but if one decides to 
use dynamical truncation, it would be advisable to do 
so using two-site DMRG.) The truncation is fixed by 
the initial choice of $D$, but it is still possible to determine 
an estimate on the error of this truncation by analyzing 
the reduced density matrix. Starting from an expression 
for the full density matrix in the local view (current site 
$k$ with orthonormal effective basis sets)
\begin{eqnarray}
	\rho = \ket{\psi} \bra{\psi} & = &	\left( \sum_{l r \sigma} A_{l r}^{[\sigma]} \ket{l} \ket{\sigma} \ket{r}\right)
													\left( \sum_{l' r' \sigma'} { A_{l' r'}^{[\sigma']} }^{*} \bra{l'} \bra{\sigma'} \bra{r'} \right)
													\nonumber \\
										& = &	\sum_{l r \sigma l' r' \sigma'} A_{lr}^{[\sigma]} { A_{l' r'}^{[\sigma']} }^{*}
													\ket{l} \bra{l'} \ket{\sigma} \bra{\sigma'} \ket{r} \bra{r'}
													\komma \label{eq:MPSRhoLocal}
\end{eqnarray}
we trace out the effective right basis and obtain a reduced density
matrix for the current site and the left part of the system:
\begin{equation}
	\rho^{\left[l_{k+1} \right]} = \sum_{l r \sigma l' \sigma'} A_{l r}^{[\sigma]} { A_{l' r}^{[\sigma']} }^{*}
										\ket{l} \bra{l'} \ket{\sigma} \bra{\sigma'}
										\punkt \label{eq:MPSRhoLocalLeft1}
\end{equation}
This reduced density matrix carries the label $l_{k+1}$ because it 
corresponds precisely to the density matrix $\ket{l_{k+1}} \bra{l'_{k+1}}$. 
So if we switch the current site from site $k$ to site $k+1$, we can check 
the error of the truncation of $\mathcal{H}^{l_{k+1}}$. Fusing the indices
$l$ and $\sigma$, we obtain
\begin{eqnarray}
	\fl
	\rho^{\left[ l_{k+1} \right]}	& = &	\sum_{l r \sigma l' \sigma'} 
													A_{\left(l \sigma \right) r} A_{\left(l' \sigma' \right) r}^{*}
													\ket{\left(l \sigma \right)} \bra{\left(l' \sigma' \right)}
													= \sum_{l r \sigma l' \sigma'}
													A_{\left(l \sigma \right) r} \left(A^{\dagger} \right)_{r \left(l' \sigma' \right)}
													\ket{\left(l \sigma \right)} \bra{\left(l' \sigma' \right)}
													\nonumber \\
	\fl
										& = &	\sum_{l \sigma l' \sigma'}
													\left(AA^{\dagger} \right)_{\left(l \sigma \right) \left(l' \sigma' \right)}
													\ket{\left(l \sigma \right)} \bra{\left(l' \sigma' \right)}
													\punkt \label{eq:MPSRhoLocalLeft2}
\end{eqnarray}
We do not need to diagonalize the coefficient matrix $AA^{\dagger}$ to 
obtain the largest weights in the density matrix, because we get its 
eigenvalues as a byproduct of the following manipulations anyway \cite{SchollwockRevModPhys77}. 
To switch the current site we need to apply a singular value decomposition 
(see \sref{sub:SwitchingCurrentSite}) and obtain $A = usv^{\dagger}$ 
(this is not the usual $A$-matrix, but the index- fused form). This directly 
yields $AA^{\dagger} = usv^{\dagger} vsu^{\dagger} =u s^{2} u^{\dagger}$, 
which corresponds to the diagonalization of $\rho^{\left[ l_{k+1} \right]}$, 
implying that the weights of the density matrix are equal to $s^{2}$. Of course
this works also for the right effective basis. With such an expression,
we can check whether the effective Hilbert space dimension $D$ of
$\mathcal{H}^{l_{k+1}}$ is too small or not. For example, we could
ask for the smallest singular value $s_{D}$ to be at least $n$ orders
of magnitude smaller than the largest one $s_{1}$, i.e. the respective
weights in the density matrix would be $2n$ orders of magnitude apart.
If the singular values do not decrease that rapidly, we have to choose
a greater $D$.

\subsubsection{Scalar product}\label{sub:MPSScalarProduct}

\begin{figure}[h]
	\begin{centering}
	\includegraphics{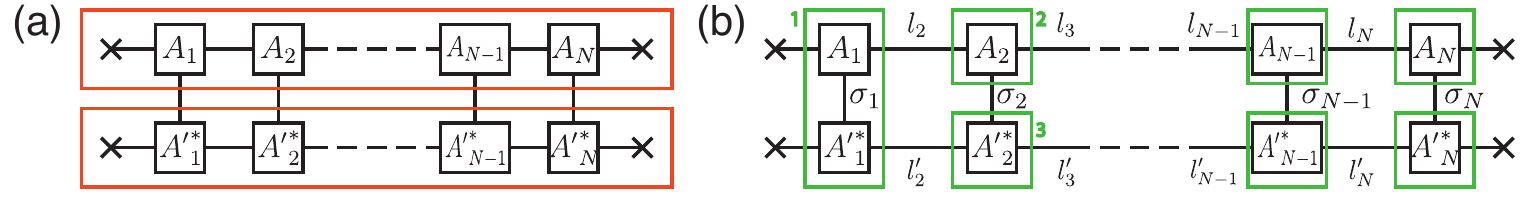} \par
	\end{centering}
	\caption{Scalar product, computed in two different
	orders. (a) First all $A$-matrices of $\ket{\psi}$ and $\ket{\psi'}$
	are contracted and then contraction over the local indices is carried out.
	b) First, for site one, we contract
	over the local indices of $A_{1}$ and $A'_{1}$. Then we contract
	over the effective index between $A_{1}$ and $A_{2}$ and afterwards
	over the indices between the resulting object and $\left(A'_{2}\right)^{*}$.
	Proceeding over the whole chain yields the scalar product.}
	\label{fig:ScalarProduct}
\end{figure}

The scalar product of two states $\ket{\psi}$ and $\ket{\psi'}$
is one of the simplest operations we can perform with matrix product
states. It is calculated most conveniently in the global view because then we do
not need to care about orthonormalization of the $A$-matrices:
\begin{eqnarray}
	\fl
	\braket{\psi'}{\psi}	& = &	\bra{\sigma'_{1}} \dots \bra{\sigma'_{N}}
											\sum_{\sigma'_{1} \dots \sigma'_{N}} \left( A'^{[\sigma'_{1}]} \dots A'^{[\sigma'_{N}]} \right)^{*}
											\sum_{\sigma_{1} \dots \sigma_{N}} \left(A^{[\sigma_{1}]} \dots A^{[\sigma_{N}]} \right)
											\ket{\sigma_{1}} \dots \ket{\sigma_{N}}
											\nonumber \\
	\fl
								& = &	\sum_{\sigma_{1} \dots \sigma_{N}}
											\left( A'^{[\sigma_{1}]} \dots A'^{[\sigma_{N}]} \right)^{*}
											\left(A^{[\sigma_{1}]} \dots A^{[\sigma_{N}]} \right)
											\komma \label{eq:MPSScalarProduct}
\end{eqnarray}
using the orthonormality of the local basis 
$\braket{\sigma'_{k}}{\sigma_{l}} = \delta_{kl}\delta_{\sigma'_{k}\sigma_{k}}$.
In principle the order in which these contractions are carried out
is irrelevant, but in practice it is possible to choose an order in which this summation
over the full Hilbert space is carried out very efficiently by exploiting 
the one-dimensional structure of the matrix product state (see \fref{fig:ScalarProduct}
for a graphical explanation). 
For details on the numerical costs, see \sref{sec:NumericalCosts}. 
In method (a), after contracting all $A$-matrices of 
$\ket{\psi}$ and $\ket{\psi'}$, we have to perform a 
contraction over the full Hilbert space, i.e.  a $1 \times d^{N}$ 
matrix is multiplied with a $d^{N} \times1$ matrix. This contraction 
is of order $\mathcal{O} \left( d^{N} \right)$, which is completely
unfeasible for practical purposes. In method (b) the most 
'expensive' contraction is in the middle of the chain, say 
at site $k$, and it is of order $\mathcal{O}\left(dD^{3}\right)$. 
Here the $A$-matrices are viewed as three-index objects 
$A_{l_{k}r_{k}\sigma_{k}}$ with dimension $D \times D \times d$. 
All sites left of site $k$ are represented by a $D \times D$ 
matrix, say $L_{l'_{k}}^{l_{k}}$. Contracting 
this with the matrix at site $k$ yields the object 
$\sum_{l_{k}} L_{l'_{k}}^{l_{k}} A_{l_{k} r_{k} \sigma_{k}}$, 
which has dimensions $D \times D \times d$, and since the sum 
contains $D$ terms, the overall cost is $\mathcal{O} \left( dD^{3} \right)$. 
Thus, in practice, method (b) is rather efficient and renders 
such calculations feasible in practice.

\paragraph{Partial product}

\begin{figure}[h]
	\begin{centering}
	\includegraphics{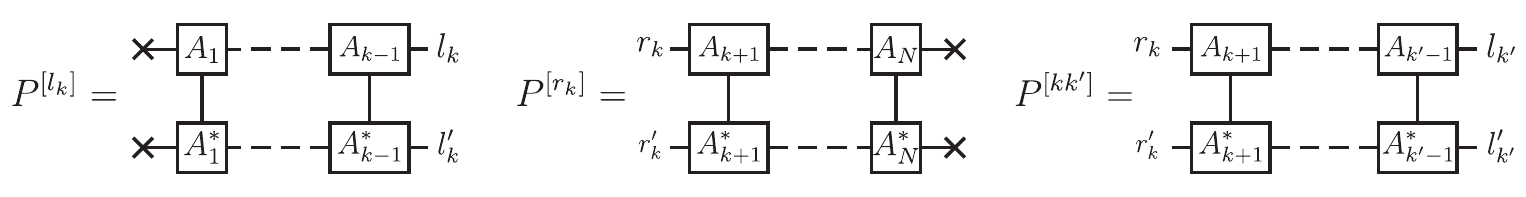} \par
	\end{centering}
	\caption{Partial products associated with site $k$.}
	\label{fig:PartialProducts}
\end{figure}

Sometimes it is required to calculate a product over only a part of the
matrix product state. This is done the same way as the scalar product
\begin{eqnarray}
\fl	\left( P^{\left[ L_{k} \right]} \right)_{l^{\phantom{,}}_{k} l'_{k}}
	& \equiv &
	\sum_{\sigma_{1} \dots \sigma_{k-1}}
	\left( A^{[\sigma_{1}]} \dots A^{[\sigma_{k-1}]} \right)_{l'_{k}}^{*}
	\left( A^{[\sigma_{1 }]} \dots A^{[\sigma_{k-1}]} \right)_{l^{\phantom{,}}_{k}}
	\komma \label{eq:PartialLeft} \\
\fl	\left(P^{\left[ R_{k} \right]} \right)_{r^{\phantom{,}}_{k} r'_{k}}
	& \equiv &
	\sum_{\sigma_{k+1} \dots \sigma_{N}}
	\left( A^{[\sigma_{k+1}]} \dots A^{[\sigma_{N}]} \right)_{r'_{k}}^{*}
	\left( A^{[\sigma_{k+1}]} \dots A^{[\sigma_{N}]} \right)_{r^{\phantom{,}}_{k}}
	\komma \label{eq:PartialRight} \\
\fl	\left(P^{\left[ kk' \right]} \right)_{r^{\phantom{,}}_{k} r'_{k} , l^{\phantom{,}}_{k'} l'_{k'}}
	& \equiv &
	\sum_{\sigma_{k+1} \dots \sigma_{k'-1}}
	\left( A^{[\sigma_{k+1}]} \dots A^{[\sigma_{k'-1}]} \right)_{r'_{k^{\phantom{,}}} l'_{k'}}^{*}
	\left( A^{[\sigma_{k+1}]} \dots A^{[\sigma_{k'-1}]} \right)_{r^{\phantom{,}}_{k^{\phantom{,}}} l^{\phantom{,}}_{k'}}
	\punkt \label{eq:PartialBetween}
\end{eqnarray}
Notice that $P^{\left[ L_{k} \right]}$ and $P^{\left[ R_{k} \right]}$
are matrices in the indices $l_{k}$ and $r_{k}$, respectively
(see \fref{fig:PartialProducts}). In fact, they correspond to
the overlap matrices $\braket{l'_{k}}{l_{k}}$ and $\braket{r'_{k}}{r_{k}}$,
respectively.

\subsubsection{Reduced density matrix}\label{sub:MPSReducedDensityMatrix}

\begin{figure}[h]
	\begin{centering}
	\includegraphics{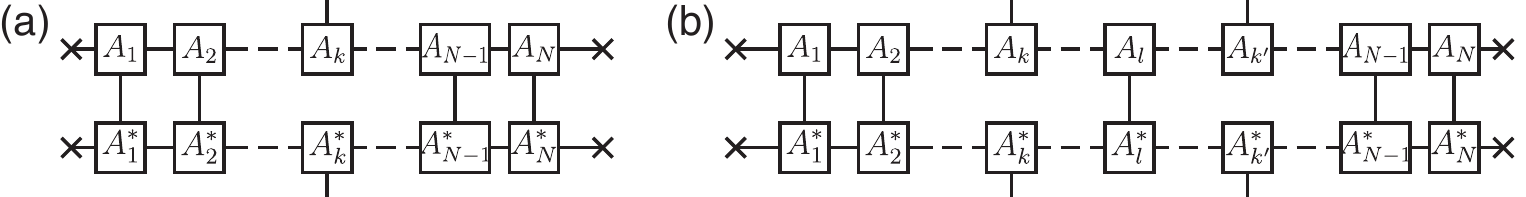} \par
	\end{centering}
	\caption{Reduced density matrix (a) $\rho_{\left\{ k\right\} }$
		for site $k$ and (b) $\rho_{\left\{ kk'\right\} }$ for sites $k$
		and $k'$, where $k<l<k'$.}
	\label{fig:ReducedDensityMatrix}
\end{figure}

The pure density matrix given by the matrix product state 
$\ket{\psi}$ is defined as $\rho=\ket{\psi}\bra{\psi}$. To 
describe only a part of the system, we need  to calculate 
the reduced density matrix. Let $I$ be a set of sites and 
$\sigma_{\mathrm{s}} = \left\{ \sigma_{k \in I} \right\}$ a 
fused index for their local states. Tracing out all other sites 
with combined index $\sigma_{\mathrm{b}} = \left\{ \sigma_{k \notin I} \right\}$ 
we obtain
\begin{equation}
	\rho_{I} = \sum_{\sigma^{\phantom{,}}_{1} \dots \sigma^{\phantom{,}}_{N} \sigma'_{1} \dots \sigma'_{N}}
	\delta_{\sigma^{\phantom{,}}_{\mathrm{b}} \sigma'_{\mathrm{b}}}
	\left(A^{[\sigma'_{1}]} \dots A^{[\sigma'_{N}]} \right)^{*}
	\left(A^{[\sigma^{\phantom{,}}_{1}]} \dots A^{[\sigma^{\phantom{,}}_{N}]} \right)
	\ket{\sigma_{\mathrm{s}}} \bra{\sigma'_{s}}
	\punkt \label{eq:ReducedDensityMatrix}
\end{equation}
This is a completely general expression, but in the cases 
where $I = \left\{ k \right\}$ or $I = \left\{ k,k' \right\}$ it 
reduces to (see \fref{fig:ReducedDensityMatrix})
\begin{eqnarray}
\fl	\rho_{\left\{ k \right\} }	& = &	P^{\left [L_{k} \right]}
												\left( A^{[\sigma_{k}]} \otimes { A^{[\sigma'_{k}]} }^{*} \right)
												P^{\left[ R_{k} \right]}
												\ket{\sigma_{k}}\bra{\sigma'_{k}}
												\komma \label{eq:ReducedDensityMatrixOne} \\
\fl	\rho_{\left\{ kk' \right\} }	& = &	P^{\left [L_{k} \right]}
												\left( A^{[\sigma_{k}]} \otimes { A^{[\sigma'_{k}]} }^{*} \right)
												P^{\left[ kk' \right]}
												\left( A^{[\sigma_{k'}]} \otimes { A^{[\sigma'_{k'}]} }^{*} \right)
												P^{\left[ R_{k'} \right]}
												\ket{\sigma_{k}} \ket{\sigma_{k'}} \bra{\sigma'_{k}} \bra{\sigma'_{k'}}
												\punkt \label{eq:ReducedDensityMatrixTwo}
\end{eqnarray}
A similar strategy can be used to calculate the density matrices 
needed for the main text, by contracting out the $\sigma_k$'s for 
all sites except those involved in the clusters $A$, $B$ or 
$A \cup B$. In fact, \eref{eq:ReducedDensityMatrixTwo} gives 
$\hat{\rho}^{A \cup B}$ for two clusters of size one at sites $k$ 
and $k'$.

\subsubsection{Operators in an effective basis}\label{sub:MPSOperatorsBases}

\begin{figure}[h]
	\begin{centering}
		\includegraphics{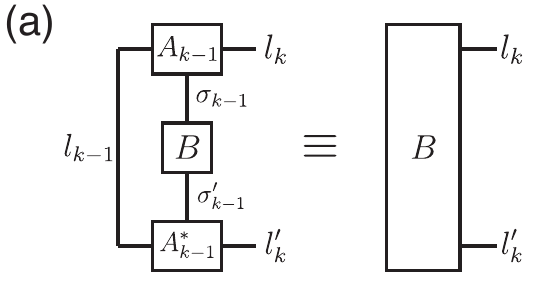}
		\includegraphics{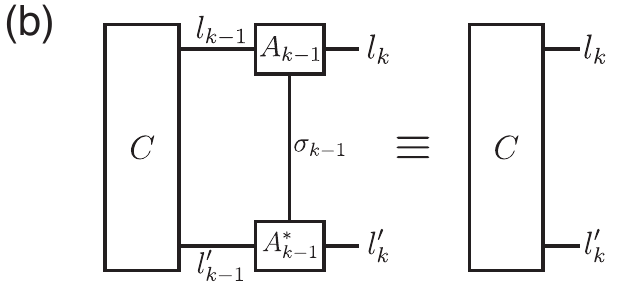}
		\par
	\end{centering}
	\caption{The $k$-left-representation of 
		(a) the operator $B$, obtained from its 
		$\left(k-1\right)$-local-representation and 
		(b) the operator $C$, obtained from its 
		$\left(k-1\right)$-left-representation.}
	\label{fig:OperatorToLeft}
\end{figure}

Let $k$ be the current site with orthonormal effective basis sets
$\ket{l_{k}}$ and $\ket{r_{k}}$. Consider an operator $B$, which
acts on the local basis of site $k-1$ only, with matrix elements
$B_{\sigma'_{k-1} \sigma_{k-1}} = \bra{\sigma'_{k-1}} B \ket{\sigma_{k-1}}$.
We call this the $(k-1)$\emph{-local-representation}
of $B$. To represent $B$ in the effective left basis of site $k$,
called the $k$\emph{-left-representation} of $B$, we use the 
transformation properties of $A^{[\sigma_{k-1}]}$ (see \fref{fig:OperatorToLeft}), 
\begin{eqnarray}
\fl	\bra{l'_{k}} B \ket{l_{k}}	& = &	\left( \bra{l'_{k-1}} \bra{\sigma'_{k-1}}
													\sum_{l'_{k-1} \sigma'_{k-1}}
													{ A_{l'_{k-1} l'_{k}}^{[\sigma'_{k-1}]} }^{*} \right)
													B_{\sigma'_{k-1} \sigma_{k-1}}
													\left( \sum_{l_{k-1} \sigma_{k-1}}
													A_{l_{k-1} l_{k}}^{[\sigma_{k-1}]}
													\ket{l_{k-1}} \ket{\sigma_{k-1}} \right)
													\nonumber \\
\fl										& = &	\sum_{l_{k-1} \sigma'_{k-1} \sigma_{k-1}}
													{ A_{l_{k-1} l'_{k}}^{[\sigma'_{k-1}]} }^{*}
													A_{l_{k-1} l_{k}}^{[\sigma_{k-1}]}
													B_{\sigma'_{k-1} \sigma_{k-1}}
													\komma \label{eq:OperatorLocalToLeft}
\end{eqnarray}
where the only condition to derive these results, was that
site $k-1$ has an orthonormal effective left basis.
Similarly, if the $(k-1)$-left-representation of an operator
$C$ is known, its $k$-left-representation can be obtained via (see
\fref{fig:OperatorToLeft})
\begin{equation}
	\bra{l'_{k}} C \ket{l_{k}} = 
	\sum_{l_{k-1} l'_{k-1} \sigma_{k-1}}
	{ A_{l'_{k-1} l'_{k}}^{[\sigma_{k-1}]} }^{*}
	A_{l_{k-1} l_{k}}^{[\sigma_{k-1}]}
	C_{l'_{k-1} l_{k-1}}
	\punkt \label{eq:OperatorLeftToLeft}
\end{equation}

\begin{figure}[h]
	\begin{centering}
	\includegraphics{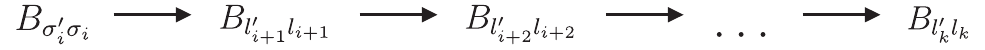} \par
	\end{centering}
	\caption{Iterative calculation of
		the $k$-left-description of an operator $B$, given in the $i$-local-description,
		by \eref{eq:OperatorLocalToLeft} and \eref{eq:OperatorLeftToLeft}
		for any $k>i$.}
	\label{fig:OperatorLeftToLeftIteratively}
\end{figure}

\Eref{eq:OperatorLocalToLeft} and \eref{eq:OperatorLeftToLeft}
can be used iteratively to transcribe the $i$-local-representation
of $B$ into its $k$-left-representation for any $k>i$ (see \fref{fig:OperatorLeftToLeftIteratively}).
This reasoning also applies to the right site of site $k$ and so
it is possible to obtain a description of any local operator on any
site.

\begin{figure}[h]
	\begin{centering}
	\includegraphics{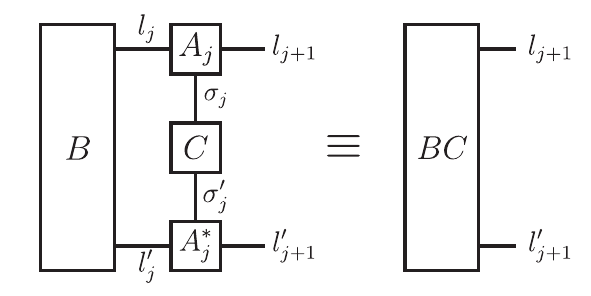}
	\par
	\end{centering}
	\caption{The $(j+1)$-left-representation
		of the operators $C$, given in the $j$-local-representation, and
		$B$, given in the $j$-left-representation.}
	\label{fig:OperatorsLeftLocalToLeft}
\end{figure}

To obtain a description of a pair of local operators acting on 
different sites, we have to transcribe them step by step. Let 
site $k$ be the current site with orthonormal effective basis sets 
and $B,C$ two operators acting locally on site $i$ and $j$ 
respectively ($i<j<k$). First we obtain the $j$-left-representation
of $B$, namely $B_{l'_{j}l_{j}}$, as described above. Then both operators
are transformed together into the $\left(j+1\right)$-left-representation
(see \fref{fig:OperatorsLeftLocalToLeft}),
\begin{equation}
	\bra{l'_{j+1}} \left(BC\right) \ket{l_{j+1}} = 
	\sum_{l_{j} l'_{j} \sigma_{j} \sigma'_{j}}
	{ A_{l'_{j} l'_{j+1}}^{[\sigma'_{j}]} }^{*}
	A_{l_{j} l_{j+1}}^{[\sigma_{j}]}
	B_{l'_{j} l_{j}}
	C_{\sigma'_{j} \sigma_{j}}
	\komma \label{eq:OperatorsLeftLocalToLeft}
\end{equation}
which in turn can be transformed iteratively into the desired $k$-left-representation
of the operators $B$ and $C$.

\subsubsection{Local operators acting on $\ket{\psi}$}\label{sub:MPSOperatorsEvaluation}

Any combination of operators can be calculated directly in the global
view or in the local view via the effective descriptions introduced
in the previous section.

\paragraph{Global view}

\begin{figure}[h]
	\begin{centering}
	\includegraphics{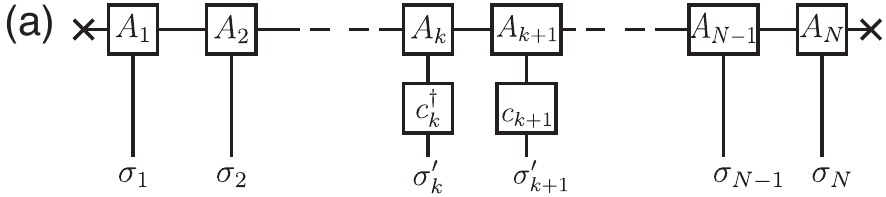}
	\includegraphics{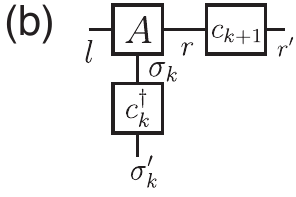}
	\par
	\end{centering}
	\caption{The nearest neighbour hopping term $c_{k}^{\dagger}c_{k+1}$
		acting on $\ket{\psi}$ in (a) the global view
		and (b) the local view.}
	\label{fig:OperatorAction}
\end{figure}

The operators, known in the local basis of the site they are acting
on, are contracted directly with the corresponding $A$-matrix. For
example, the formula for a nearest neighbour hopping term $c_{k}^{\dagger} c_{k+1}$
(see \fref{fig:OperatorAction}) reads as
\begin{eqnarray}
	\fl
	c_{k}^{\dagger} c_{k+1} \ket{\psi}	& = &	\sum_{\sigma_{1} \dots \sigma_{N}}
																\left( \sum_{\sigma'_{k}} \left(c_{k}^{\dagger} \right)_{\sigma'_{k} \sigma_{k}} \right)
																\left( \sum_{\sigma'_{k+1}} \left(c_{k+1} \right)_{\sigma'_{k+1} \sigma_{k+1}} \right)
																\left( A^{[\sigma_{1}]} \dots A^{[\sigma_{N}]} \right)
																\nonumber \\
													&  &		\ket{\sigma_{1}} \dots \ket{\sigma_{k-1}} \ket{\sigma'_{k}} \ket{\sigma'_{k+1}} \ket{\sigma_{k+2}} \dots \ket{\sigma_{N}}
																\punkt \label{eq:OperatorsGlobal}
\end{eqnarray}

\paragraph{Local view}

Let $k$ be the current site with orthonormal effective basis sets.
If we want to evaluate operators acting on other sites than the
current site $k$, we need an effective description of these operators
in one of the effective basis sets of site $k$ to contract
these operators with the $A$-matrix of the current site. For example,
to calculate the action of the nearest neighbour hopping term $c_{k}^{\dagger} c_{k+1}$
on $\ket{\psi} = A_{lr}^{[\sigma_{k}]} \ket{l} \ket{\sigma_{k}} \ket{r}$,
we need $(c_{k}^{\dagger})_{\sigma_{k}' \sigma_{k}}$ and $(c_{k+1})_{r' r}$
to obtain (see \fref{fig:OperatorAction})
\begin{equation}
	c_{k}^{\dagger} c_{k+1} \ket{\psi} = 
	\sum_{r \sigma_{k}}
	\left( \sum_{\sigma_{k}'} \left( c_{k}^{\dagger} \right)_{\sigma_{k}' \sigma_{k}} \right)
	\left( \sum_{r'} \left( c_{k+1} \right)_{r' r} \right)
	A_{l r}^{[\sigma_{k}]} \ket{l} \ket{\sigma_{k}'} \ket{r'}
	\punkt \label{eq:OperatorsLocal}
\end{equation}

\subsubsection{Expectation values}\label{sub:MPSExpectationValues}

\begin{figure}[h]
	\begin{centering}
	\includegraphics{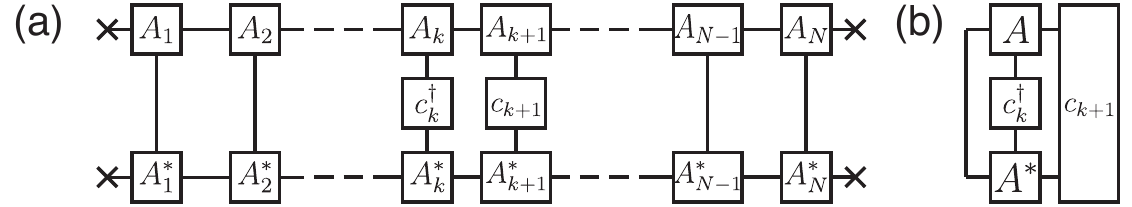} \par
	\end{centering}
	\caption{The expectation value of the nearest
		neighbour hopping $c_{k}^{\dagger}c_{k+1}$ in (a) the global view
		and (b) the local view.}
	\label{fig:ExpectationValues}
\end{figure}

Expectation values are merely the scalar product between the state
with itself including the action of an operator and can be easily
worked out in both the global and the local view (see \fref{fig:ExpectationValues}).
Since both methods are equivalent, the local variant is much more
efficient as it involves much less matrix multiplications. However,
it requires careful orthonormalization of the remainder of the $A$-matrices.
The iterative scheme, introduced in \sref{sec:MPSVariationalOptimizationScheme},
allows for that and works in the local picture.

\subsection{Variational optimization scheme}\label{sec:MPSVariationalOptimizationScheme}

\begin{figure}[h]
	\begin{centering}
	\includegraphics{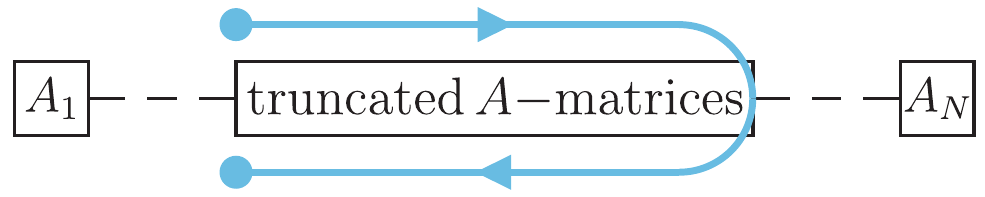}
	\par
	\end{centering}
	\caption{One complete sweep.}
	\label{fig:Sweep}
\end{figure}

The basic techniques introduced in the previous sections are 
the building blocks for DMRG sweeps, an iterative scheme 
to determine the ground state in the usual DMRG sense. 
This scheme starts at some site as current site, for example 
the first site where truncation occurs, and minimizes the energy 
of $\ket{\psi}$ with respect to that site. Afterwards the current 
site is shifted to the next site, and the energy of $\ket{\psi}$ 
with respect to that site is minimized. This is repeated until the 
last site where truncation occurs is reached and the direction 
of the switches is reversed. When the starting site is reached 
again, one \emph{sweep} has been finished (see \fref{fig:Sweep}). 
These sweeps are repeated until $\ket{\psi}$ converges.

\subsubsection{Energy minimization of the current site}\label{sub:MPSVarEnergyMinimisation}

\begin{figure}[h]
	\begin{centering}
	\includegraphics{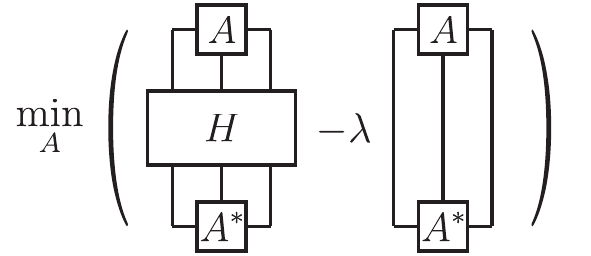} \par
	\end{centering}
	\caption{The minimization problem expressed in the current site.}
	\label{fig:MinimizationCurrentSite}
\end{figure}

In order to find the ground state of the system we have to 
minimize the energy $E = \bra{\psi} H \ket{\psi}$ of the 
matrix product state $\ket{\psi}$ with the constraint that the 
norm of $\ket{\psi}$ must not change. Introducing $\lambda$ 
as Lagrange multiplier to ensure proper normalization, we 
arrive at the problem of determining
\begin{equation}
	\underset{\ket{\psi}}{\min} \left( \bra{\psi} H \ket{\psi} - \lambda \braket{\psi}{\psi} \right) \punkt \label{eq:Minimization}
\end{equation}
In the sweeping procedure introduced above, the current 
site is changed from one site to the next and the energy 
is minimized in each local description. Thus, we need 
\eref{eq:Minimization} in terms of the parameters of the 
current site. Let us describe how to do this for the case of 
one-site DMRG, where the $A$-matrices are optimized 
one site at a time. (The procedure for two-site DMRG is 
entirely analogous, except that it involves combining $A$-matrices 
of two neighboring sites by fusing their indices to obtain a 
combined two-site $A$-matrix, see \sref{sub:MPSHilbertSpaceTruncation}.)
Inserting \eref{eq:MPSfullMPS} into \eref{eq:Minimization} 
yields (see \fref{fig:MinimizationCurrentSite})
\begin{equation}
	\underset{A^{[\sigma]}}{\min}
	\left( \sum_{l r \sigma l' r' \sigma'}
	{ A_{l' r'}^{[\sigma']} }^{*}
	H_{l' r' \sigma' l r \sigma}
	A_{l r}^{[\sigma]}
	- \lambda \sum_{l r \sigma}
	{ A_{l r}^{[\sigma]} }^{*}
	A_{l r}^{[\sigma]} \right)
	\komma \label{eq:MinimizationCurrentSite}
\end{equation}
where $H_{l' r' \sigma' l r \sigma} = \bra{l'} \bra{\sigma'} \bra{r'} H \ket{l} \ket{\sigma} \ket{r}$
is the Hamiltonian expressed in the two orthonormal effective 
basis sets and the local basis of the current site.

The multidimensional minimization problem \eref{eq:Minimization}
has been transformed to a local minimization problem where one 
$A$-matrix (or two) is optimized at a time and all others are kept 
constant. Such a procedure could, in principle, cause the system 
to get stuck in a local minimum in energy, but experience shows 
that the procedure works well \cite{SchollwockRevModPhys77}, 
especially in the presence of a gap.

To obtain a solution for \eref{eq:MinimizationCurrentSite},
we differentiate the equation with respect to 
${ A_{l' r'}^{[\sigma']} }^{*}$ (this is possible because the 
Hilbert space has an hermitian scalar product) and obtain
\begin{equation}
	0 = \sum_{l' r' \sigma'} H_{l' r' \sigma' l r \sigma} A_{l r}^{[\sigma]} - \lambda A_{l' r'}^{[\sigma']} \punkt \label{eq:MinimizationDiff}
\end{equation}
The matrix elements $H_{l' r' \sigma' l r \sigma}$ may be 
calculated easily using the techniques introduced in 
\sref{sec:MPS} (see \sref{sub:MPSVarSweepingDetails} 
for details). Changing to matrix notation and replacing 
$\lambda$ with $E_{0}$ in anticipation of its interpretation 
as an energy, we obtain an eigenvalue equation:
\begin{equation}
	H A_{l r}^{[\sigma]} \ket{l} \ket{\sigma} \ket{r} = E_{0} A_{l r}^{[\sigma]} \ket{l} \ket{\sigma} \ket{r} \punkt \label{eq:MinimizationEigen}
\end{equation}
The minimization problem reduces to a local eigenvalue
problem, which can be solved by standard techniques. 
The full Hilbert space of the current site has dimension 
$dD^{2}$ and may become large, but it is not necessary 
to determine the full spectrum of $H$, since we are interested 
only in the ground state. The Lanczos algorithm is an 
effective algorithm to achieve exactly that. The advantage 
of this algorithm is that we only have to compute $H\ket{\psi}$, 
which saves much effort. The Lanczos algorithm produces 
as output the ground state eigenvalue and eigenvector. 
The latter gives the desired optimized version of the matrix 
$A^{\sigma}_{lr}$, which then has to be rewritten (with or 
without Hilbert space truncation, as needed) into a form that 
satisfies the orthonormality requirements of the left and right 
basis sets, as described in \sref{sub:SwitchingCurrentSite}.

\subsubsection{Sweeping details}\label{sub:MPSVarSweepingDetails}

Before the actual sweeping may be started we 
have to set up an initial state, prepare a current 
site with orthonormal effective basis sets and 
calculate effective descriptions of operators which 
are part of the Hamiltonian. After this initialization 
we may determine the ground state with respect 
to this current site and shift the current site to the 
next site. That current site again has orthonormal 
effective basis sets due to the switching procedure 
introduced in \sref{sub:SwitchingCurrentSite}, but 
we also need effective representations of the 
operators acting in the Hamiltonian. At this step the 
structure of the matrix product state saves much 
effort, as most of the needed representations are 
already calculated.

\paragraph{Structure of the Hamiltonian terms}

\begin{figure}[h]
	\begin{centering}
	\includegraphics{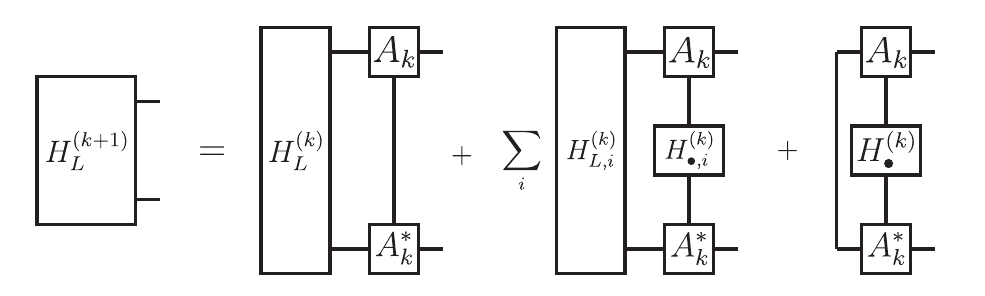} \par
	\end{centering}
	\caption{Iterative calculation of the operator $H_{L}^{(k+1)}$.
		The sum over $i$ indicates that $H_{L\bullet}^{(k)}$
		has the form $\sum_{i} H_{L,i}^{(k)} \otimes H_{\bullet,i}^{(k)}$,
		where $H_{L,i}^{(k)}$ acts only on sites $k' < k$ and $H_{i,\bullet}^{(k)}$
		only on site $k$. The calculation of $H_{R}^{(k-1)}$ works analogously.}
	\label{fig:UpdateHEff}
\end{figure}

The Hamiltonian $H_{l'r'\sigma'lr\sigma}$, 
acting in the space spanned by the states 
$\ket{l}$, $\ket{\sigma}$, $\ket{r}$, breaks 
up into several terms:
\begin{eqnarray}
\fl	H_{l' r' \sigma' l r \sigma}	& = &	\unity_{l' l} \otimes \left(H_{\bullet}\right)_{\sigma' \sigma} \otimes \unity_{r' r}
													+ \left(H_{L}\right)_{l' l} \otimes \unity_{\sigma' \sigma} \otimes \unity_{r' r}
													+ \unity_{l' l} \otimes \unity_{\sigma' \sigma} \otimes \left(H_{R}\right)_{r' r}
													\nonumber \\
										&  &		+ \left(H_{L\bullet}\right)_{l' l \sigma' \sigma} \otimes \unity_{r' r}
													+ \unity_{l' l} \otimes \left(H_{\bullet R} \right)_{r' r \sigma' \sigma}
													+ \left(H_{L\bullet R}\right)_{l' l r' r \sigma' \sigma}
													\komma \label{eq:HamiltonianBreakup}
\end{eqnarray}
where the indices denote on which parts of the 
system the respective term acts on ($L$ and 
$R$ indicate left and right of the current site, 
respectively, $\bullet$ indicates action on the 
current site). Of course, the six terms of 
\eref{eq:HamiltonianBreakup} depend on the
current site $k$: $H_{\bullet}^{(k)}$, $H_{L}^{(k)}$, 
$H_{R}^{(k)}$, $H_{L\bullet}^{(k)}$, $H_{\bullet R}^{(k)}$ 
and $H_{L\bullet R}^{(k)}$. The terms $(H_{L})_{l' l}$ 
and $(H_{R})_{r' r}$ contain all terms which involve 
only sites $k' < k$ and $k' > k$, respectively. The 
iterative structure of the method directly yields the 
following equalities:
\begin{eqnarray}
	H_{L}^{(k+1)} & = & H_{L}^{(k)}+H_{L\bullet}^{(k)}+H_{\bullet}^{(k)} \komma \label{eq:HEffLeftRight} \\
	H_{R}^{(k-1)} & = & H_{\bullet}^{(k)}+H_{\bullet R}^{(k)}+H_{R}^{(k)} \komma \label{eq:HEffRightLeft}
\end{eqnarray}
where the terms on the right hand side are meant to 
be expressed in the effective basis of the operator 
on the left hand site (see \fref{fig:UpdateHEff}).

\paragraph{Initialization}

First of all we need an initial matrix product state, which is most
conveniently chosen to consist of identity transformations at the
ends of the chain (see \sref{sub:MPSDetailsOfA}) and random
$A$-matrices for the rest of the chain. We take the first
site where Hilbert space truncation is applied as current site $k$
and obtain an orthonormal effective right basis (the effective left
basis is already orthonormal) using the orthonormalization
procedure introduced in \sref{sub:SwitchingCurrentSite} starting
from site $N$. Additionally it is convenient, while dealing with
site $N$, to calculate and \emph{store} the operator $H_{R}^{(N-1)}$
(see \eref{eq:HEffRightLeft}) and the effective description of 
all operators of site $N$ which contribute to $H_{\bullet R}^{(k)}$ 
and $H_{L \bullet R}^{(k)}$ in the effective right basis of site $N-1$
(see \sref{sub:MPSOperatorsBases}). This ensures, when 
the sweeping procedure reaches site $N-1$, that
all necessary operators are already calculated. This is repeated from
site $N$ down to site $k+1$, and similarly for the sites $k'<k$
in the other direction. The result of these initialization steps is
that we have a current site $k$ with orthonormal effective basis
sets, effective descriptions of the Hamiltonian terms $H_{L}^{(k)}$
and $H_{R}^{(k)}$ and effective descriptions of all operators
contributing to $H_{L\bullet}^{(k)}$, $H_{\bullet R}^{(k)}$
and $H_{L \bullet R}^{(k)}$. Moreover, with an appropriate
extension to the switching procedure of \sref{sub:SwitchingCurrentSite},
all effective descriptions for other current sites are available for
use when needed in future sweeping steps.

\paragraph{Extended switching procedure}

The switching procedure of \sref{sub:SwitchingCurrentSite} is
applied as before. Additionally, depending on the direction of the
switch, $H_{L}^{(k+1)}$ or $H_{R}^{(k-1)}$
are calculated and stored as well as the operators needed 
for the Hamiltonian \eref{eq:HamiltonianBreakup}. 
This extended switching ensures that for the new 
current site all required operators are calculated, 
if they had been for the old current site.

\paragraph{Complete ground state calculation}

The methods introduced above make the procedure to determine the ground
state very efficient as the global problem is mapped onto many
local problems involving only a few terms to calculate. The iterative
structure of the matrix product states and the effective Hamiltonian
terms strongly increase the efficiency. A full ground state calculation
consists of:
\begin{enumerate}
\item Initialization as described above
\item Full sweeps from site $K$ to site $K'$ and back to site $K$, 
		with sites $K$ and $K'$ the first and last site where the 
		effective Hilbert spaces are truncated.
\item After each sweep $i$ the overlap $\braket{\psi_{i-1}}{\psi_{i}}$
		between the state before and after the sweep is calculated. If the
		matrix product state does not change any more, stop the sweeping.
		A criterion, for example, for when to stop would be to require that
		\begin{equation}
			\frac{\left| \braket{\psi_{i-1}}{\psi_{i}} - \braket{\psi_{i-2}}{\psi_{i-1}} \right|}{\left| \braket{\psi_{i-1}}{\psi_{i}} \right|} \leq \epsilon \komma \label{eq:SweepingStop}
		\end{equation}
		where $\epsilon$ is a small control parameter, typically of order $10^{-10}$.
\end{enumerate}

\paragraph{Numerical costs}

The step with the most impact on the numerical costs of the algorithm
is the calculation of $H\ket{\psi}$ in the Lanczos method. This method
is an iterative scheme using several \emph{Lanczos steps}, of which
usually less than $100$ are needed for one ground state calculation.
Each Lanczos step calculates $H\ket{\psi}$ exactly once. This calculation
basically consists of elementary matrix multiplications, 
see \sref{sec:NumericalCosts} for details on the numerical costs of such calculations.
The six terms introduced in \eref{eq:HamiltonianBreakup} are
not all equally time consuming. Most of them contain identity maps
which do not need to be carried out and thus the term $H_{L\bullet R}$ 
is the most time consuming, requiring operations of order 
$\mathcal{O} ( dD^{2} (2D+d) )$. The total numerical 
cost for the minimization process is
\begin{equation}
	C = N_{\mathrm{Sweep}} \times 2N \times N_{\mathrm{Lanczos}} \times \left( dD^{2} \left (2D+d \right) \right) \komma \label{eq:FullCost}
\end{equation}
where $N_{\mathrm{Sweep}}$ is the number of sweeps, $N$ the chain
length and $N_{\mathrm{Lanczos}}$ the number of Lanczos steps.
In practice the cutoff dimension is significantly higher than the
local Hilbert space dimension $d$ and thus \eref{eq:FullCost} 
is nearly linear in $d$.

\subsection{Abelian symmetries}\label{sec:AbelianSymmetries}

Matrix product states can be easily adapted to 
properly account for conserved quantum 
numbers, representing the global symmetries of 
the Hamiltonian. We will limit ourselves to Abelian 
symmetries, meaning that the irreducible 
representation of the symmetry group is Abelian, 
as these are easily implemented, which is not 
necessarily the case for non-Abelian symmetries 
\cite{McCullochEuroPhysLett57}.

An Abelian symmetry allows a quantum number 
$Q$ to be attached to every state. The property 
that the symmetry is Abelian manifests itself in that 
this quantum number is strictly additive. For two 
states $\ket{Q_{1}}$ and $\ket{Q_{2}}$, the quantum 
number of the direct product of these two states is 
given by $\ket{Q_{1}} \otimes \ket{Q_{2}} = \ket{Q_{1} + Q_{2}}$. 
For example, if the Hamiltonian commutes with the 
number operator for the full system, the quantum 
number $Q$ could represent particle number.

For matrix product states, the introduction of Abelian 
symmetries has the consequence that the $A$-matrix 
$A^{[\sigma]}_{lr}$ may be written as 
$( A^{Q_{\sigma}}_{Q_{l} Q_{r}} )_{\alpha_{l} \beta_{r}}^{\gamma_{\sigma}}$. 
Here $Q_{\sigma}$, $Q_{l}$, $Q_{r}$ are the quantum 
numbers attached to the local, left effective and right 
effective basis, respectively. The index $\alpha_{l}$ 
distinguishes different states $\ket{Q_{l} , \alpha_{l}}$ 
characterized by the same quantum number $Q_{l}$, 
and similarly for $\ket{Q_{r}, \beta_{r}}$ and 
$\ket{Q_{\sigma}, \gamma_{\sigma}}$. If $A$ describes, 
for example, the mapping of the $\ket{l}$-basis of the 
left block together with the local basis to a combined 
(truncated) $\ket{r}$-basis, then the only non-zero blocks 
of the $A$-matrix are those for which $Q_{\sigma} + Q_{l} = Q_{r}$. 
For the current site, the total symmetry $Q_{\mathrm{tot}}$ 
of the full quantum many-body state manifests itself in 
that the corresponding $A$-matrix fulfills 
$Q_{l} + Q_{r} + Q_{\sigma} = Q_{\mathrm{tot}}$.

For the handling of matrix product states quantum 
numbers imply a significant amount of bookkeeping, 
i.e. for every coefficient block we have to store its 
quantum number. The benefit is that we can deal 
with large effective state spaces at reasonable 
numerical cost. The Lanczos algorithm, in particular, 
takes advantage of the block structure. 

Of course, the treatment of Abelian symmetries is 
generic and not limited to only one symmetry. We 
may incorporate as many symmetries as exist for 
a given Hamiltonian, by writing $Q$ as a vector of 
the corresponding quantum numbers.

\subsection{Additional details}\label{sec:VMPSadd}

\subsubsection{Derivation of the orthonormality condition}\label{sec:DerivationOrtho}

The orthonormality condition \eref{eq:MPSOrthoA} is easily
derived by induction. The starting point is condition \eref{eq:MPSOrtho} 
and we limit to the derivation for the left basis. The derivation 
for the right basis is analogous.

The induction argument can be initialized with site $k=1$ 
because its effective left basis is already orthonormal as 
it consists only of the vacuum state. Now, consider the 
case that site $k$ has an orthonormal effective left basis 
and construct the condition for site $k+1$ to have an 
orthonormal effective left basis:
\begin{eqnarray}
\braket{l'_{k+1}}{l_{k+1}} 	& = &	\left( \sum_{l'_{k} \sigma'_{k}} \bra{l'_{k}} \bra{\sigma'_{k}} { A^{\left[ \sigma'_{k} \right]}_{l'_{k} l'_{k+1}} }^{*} \right)
												\left( \sum_{l_{k} \sigma_{k}} A_{l_{k} l_{k+1}}^{\left[ \sigma_{k} \right]} \ket{l_{k}} \ket{\sigma_{k}} \right) 
												\nonumber \\
									& = &	\sum_{l'_{k} l_{k} \sigma'_{k} \sigma_{k}}
												{ A_{l'_{k} l'_{k+1}}^{\left[ \sigma'_{k} \right]} }^{*}
												A_{l_{k} l_{k+1}}^{\left[ \sigma_{k} \right]}
												\underbrace{\braket{l'_{k}}{l_{k}}}_{\delta_{l'_{k}l_{k}}}
												\underbrace{\braket{\sigma'_{k}}{\sigma_{k}}}_{\delta_{\sigma'_{k}\sigma_{k}}}
												= \sum_{l_{k} \sigma_{k}}
												{ A_{l_{k} l'_{k+1}}^{\left[ \sigma_{k} \right]} }^{*}
												A_{l_{k}l_{k+1}}^{\left[\sigma_{k}\right]}
												\nonumber \\
									& = &	\left( \sum_{\sigma_{k}}
												{ A^{\left[ \sigma_{k} \right]} }^{\dagger}
												A^{\left[ \sigma_{k} \right]}
												\right)_{l'_{k+1}l_{k+1}}
												\punkt \label{eq:DeriveOrthoLeft}
\end{eqnarray}
Condition \eref{eq:MPSOrthoA} follows with $\braket{l'_{k+1}}{l_{k+1}}\overset{!}{=}\delta_{l'_{k+1}l_{k+1}}$.

\subsubsection{Singular value decomposition}\label{sec:Singular-value-decomposition}

The singular value decomposition can be seen as a generalization of
the spectral theorem, i.e. of the eigenvalue decomposition. It is
valid for any real or complex $m \times n$ rectangular matrix. Let $M$ 
be such a matrix, then it can be written in a singular value decomposition
\begin{equation}
M = U S V^{\dagger} \komma \label{eq:USVt}
\end{equation}
where $U$ is a $m \times m$ unitary matrix, $S$ a $m\times n$
matrix with real, nonnegative entries on the diagonal and zeros off the
diagonal, and V a $n \times n$ unitary matrix. The numbers on the
diagonal of $S$ are called \emph{singular values}, and there are $p=min\left(n,m\right)$
of them. The singular values are unique, but $U$ and $V$ are not, in general. It is
convenient to truncate and reorder these matrices in such a fashion that their dimension are
$m \times p$ for $U$, $p\times p$ for $S$ (with the singular values ordered in a non-increasing
fashion) and $n \times p$ for $V$ (i.e. $p \times n$ for $V^\dagger$). 
A consequence of this truncation is that $U$ or $V$ is no longer quadratic and
unitarity is not defined for such matrices. This property is replaced
by \emph{column unitarity} (orthonormal columns) of $U$ and \emph{row
unitarity} (orthonormal rows) for $V^{\dagger}$ - no matter which
one is no longer quadratic. In this article all singular value decompositions
are understood to be ordered in this fashion.

\subsubsection{Numerical costs of index contractions}\label{sec:NumericalCosts}

The numerical costs of matrix multiplications and index contractions
of multi-index objects depend on the dimension of both the resulting
object and of the contracted indices. In the case of matrix multiplications
this is quite simple. Consider a $n \times m$ matrix $M_{1}$ multiplied
by a $m \times p$ matrix $M_{2}$. The result is a $n \times p$
matrix $M$:
\begin{equation}
	M_{ij} = \sum_{k=1}^{m} \left( M_{1} \right)_{ik} \left( M_{2} \right)_{kj}
	\punkt \label{eq:NCMatrixProduct}
\end{equation}
Evidently, each of the $n*p$ matrix elements $M_{ij}$ requires a
sum over $m$ products of the form $\left( M_{1} \right)_{ik} \left( M_{2} \right)_{kj}$.
Thus the process for calculating $M_{1}M_{2}$ is of order $\mathcal{O} \left( nmp \right)$.

The numerical costs of multi-index objects are obtained analogously.
Consider two multi-index objects, $M_{1}$ with indices $i_{1} , \dots , i_{n}$
and dimensions $p_{1} \times \dots \times p_{n}$ and $M_{2}$ with indices
$j_{1} , \dots , j_{m}$ and dimensions $q_{1} \times \dots \times q_{m}$.
If we contract the indices $i_{1}$ and $i_{2}$ of $M_{1}$ with
the indices $j_{1}$ and $j_{2}$ of $M_{2}$ (assuming that $p_{1} = q_{1}$
and $p_{2} = q_{2}$), we obtain the multi-index object $M$:
\begin{equation}
M_{i_{3} \dots i_{n} j_{3} \dots j_{m}} = \sum_{k=1}^{p_{1}} \sum_{l=1}^{p_{2}} 
														\left( M_{1} \right)_{kli_{3} \dots i_{n}}
														\left( M_{2} \right)_{klj_{3} \dots j_{m}}
														\punkt \label{eq:MultiIndexProduct}
\end{equation}
Thus for every entry of $M$, $p_{1}$ times $p_{2}$ multiplications
have to be done, so that the process is of order 
$\mathcal{O} \left( \left( p_{3} \dots p_{n} \right) \left( p_{1} p_{2} \right) \left( q_{3} \dots q_{m} \right) \right)$.

\end{document}